\documentclass[runningheads]{llncs}
\usepackage{makeidx} 
\usepackage{multirow}
\usepackage{graphicx}
\usepackage{amsmath}
\usepackage{amssymb}
\usepackage{amsfonts}
\usepackage{url}
\usepackage{rotating}
\usepackage[vlined,boxed,commentsnumbered,ruled,linesnumbered]{algorithm2e}
\usepackage{caption}
\usepackage{subcaption}
\usepackage{lscape}
\usepackage{bm}
\usepackage{braket} 

\usepackage{amsthm}
\usepackage{float}

\usepackage{booktabs} 

\usepackage[colorlinks, citecolor=blue]{hyperref} 

\usepackage{listings}
\usepackage{color}

\usepackage{pgfplots, pgfplotstable}
\usepgfplotslibrary{fillbetween}
\usetikzlibrary{patterns}

\allowdisplaybreaks[4] 
\usepackage{soul}
\usepackage{autobreak}
\usepackage{pifont}
\usepackage{rotfloat}
\usepackage{ifsym}
\usepackage{bbding}
\usepackage{amsmath} 
\usepackage{mathtools}
\usepackage{lineno}

\usepackage{algpseudocode}
\usepackage{amssymb} 
\usepackage{booktabs}   
\usepackage{makecell}   
\usepackage{threeparttable}

\usepackage{arydshln}
\usepackage{amsthm}
\theoremstyle{definition}
\newtheorem{pro_}{Property} 

\begin{document}
\title{Algebraic Cryptanalytic Extraction on Hard-Label Neural Networks}

\author{Zirui Chen\inst{1} \and
Shi Tang\inst{2} \and
Zhengchao Gao\inst{2} \and
Yongjia Su\inst{2} \and \\
Lingyue Qin\inst{1} \and
Xiaoyang Dong\inst{1}}
\authorrunning{Z. Chen et al.}

\institute{Tsinghua University, Beijing, P. R. China \\
\email{chenzr25@mails.tsinghua.edu.cn}, \email{\{qinly,xiaoyangdong\}@tsinghua.edu.cn}\\
\and
Shandong University, Qingdao, P. R. China \\
\email{\{shi.tang,chao\_qwq,yongjia.su\}@mail.sdu.edu.cn}
}
\titlerunning{}

\maketitle         

\begin{abstract}


Although the state-of-the-art model extraction attack on the hard-label Fully-connected Neural Network (FCN) by Carlini {\em et al.} at EUROCRYPT 2025 has polynomial-time complexity in theory, its dual-point clustering relies on singular value decomposition (SVD) with a time complexity of $\mathcal{O}(n^2 (d^{(k)})^3)$, resulting in huge runtime in practice. 
To address this computational bottleneck, this work transforms Carlini {\em et al.}'s geometric-view hard-label attack into an algebraic framework, and proposes two efficient clustering methods: Normal Rank Check (NRC) and Approximate Signature Vector (ASV). 
The NRC and ASV methods replace Carlini {\em et al.}'s heavy SVD-based rank checking with simple rank checking or inner-product operations, reducing the clustering complexity to $\mathcal{O}(n (d^{(k)})^3)$ on average. Furthermore, this paper presents the first model extraction attack against hard-label max-pooling Convolutional Neural Networks (CNNs) by combining the ASV method with the kernel-centric clustering scheme instead of the neuron-centric clustering, which fully exploits the property of weight sharing in convolutions and fills a cryptanalysis gap. Experiments on FCNs and the max-pooling LeNet-5 demonstrate that our NRC/ASV methods drastically cut clustering time, and improve the overall efficiency in the model extraction. 
\end{abstract}

\section{Introduction}
\label{sect:Introduction}

Deep Neural Networks (DNNs) map inputs to outputs via a function defined by their architecture, weights, and biases. Extracting these weights and biases from a black-box DNN (given only its inputs and outputs) has been a longstanding challenge spanning three decades \cite{blum1988training,fefferman1994reconstructing}. As early as 2005, Lowd and Meck introduced the first adversarial reverse engineering attack for this problem \cite{lowd2005adversarial}, with recent years seeing continued innovations from industry and academia \cite{batina2019csi,jagielski2020high,oliynyk2023know,rolnick2020reverse,tramer2016stealing}.
At CRYPTO 2020, 
Carlini, Jagielski, and Mironov achieved the seminal parameter extraction attack on the ReLU-based fully connected neural networks (FCNs) with a cryptographic differential attack \cite{DBLP:conf/crypto/CarliniJM20}. This approach recovers the unsigned weights using a polynomial number of raw output queries, but relies on exponential time of guessing for sign recovery  \cite{DBLP:conf/crypto/CarliniJM20}.
At EUROCRYPT 2024, Canales-Mart{\'{\i}}nez {\em et al.} proposed the {\em neuron wiggle technique} to recover neuron signs in polynomial time \cite{DBLP:conf/eurocrypt/CanalesMartinezCHRSS24}. Building on this, Foerster {\em et al.} developed an end-to-end attack for practical models by combining methods from Carlini {\em et al.} and Canales-Mart{\'{\i}}nez {\em et al.} at NeurIPS 2024 \cite{foerster2024beyond}. At EUROCRYPT 2026, Liu {\em et al.} further extended parameter extraction to the deep layers with more novel techniques \cite{DBLP:conf/eurocrypt/LiuSELBP26}.  In 2026, the parameter extraction attacks are proposed against the Convolutional Neural Networks (CNNs) with max pooling \cite{DBLP:conf/crypto/ChenTGSQD26,liu2026model-cnn} and average pooling \cite{cnn_average_pooling}.  Besides, various model extraction attacks have emerged for different networks, such as PReLU networks \cite{chen2025delving},  non-linear activations networks \cite{DBLP:conf/crypto/AsselineauDFM26,qi2026various}, as well as in the physical leakage settings \cite{batina2019csi,horvath2024sok,ches_extraction}.

Another direction focuses on ReLU-based DNNs in the hard-label scenario, where attackers only access final classification labels ({\em e.g.}, ``dog'' or ``car''). Yi Chen {\em et al.} first proposed an exponential-time extraction method at ASIACRYPT 2024 \cite{DBLP:conf/asiacrypt/ChenDGSWW24}. At EUROCRYPT 2025,  Carlini {\em et al.} achieved a breakthrough by presenting a polynomial-time, polynomial-query attack by analyzing the geometric properties of decision boundaries \cite{DBLP:conf/eurocrypt/CarliniCHRS25}. Canales-Mart{\'{\i}}nez later proposed the hard-label {\em SOE} to recover the signs at LATINCRYPT 2025 \cite{DBLP:conf/latincrypt/CanalesMartinezS25}. At CRYPTO 2026, Ito, Miura, and Todo 
proposed a cross-layer extraction strategy to handle always-active neurons \cite{DBLP:conf/crypto/ItoMT26}. Sun {\em et al.} proposed the first extraction attack on hard-label CNNs with average pooling \cite{cnn_average_pooling}, which was later improved by \cite{gongzheng_cnn}, while the attacks on max-pooling CNNs \cite{DBLP:journals/iacr/ChenTGSQD26,liu2026model-cnn} require raw output queries, making the hard-label attacks against max-pooling CNNs an open problem.

\subsubsection*{Our Contributions.} 
A major computational hurdle of Carlini {\em et al.}'s hard-label attack \cite{DBLP:conf/eurocrypt/CarliniCHRS25} arises during the signature recovery phase, which requires clustering the collected dual points into consistent groups. 
The authors explicitly acknowledge this limitation, noting that their proof-of-concept implementation avoids the full computational burden of the clustering in \cite[Section 7.1]{DBLP:conf/eurocrypt/CarliniCHRS25}:
\begin{quote}
    ``\emph{Our proof-of-concept implementation disregards the runtime of the $n^2$ pairwise clustering $t_{\text{cluster}}$ needed to identify dual points that are mutually consistent. Although this method is known to work in principle (we validated it post-hoc) and  runs in polynomial time, 
 applying it naively (i.e., without many potential optimizations) to a real attack would involve a $(1\text{ million})^2$ time complexity, requiring weeks of computation.}''
\end{quote}
Our contributions come from two aspects and (partial) black-box experiments:
\begin{itemize}
    \item {\bf Improved  Carlini {\em et al.}'s Hard-label Attack:} We first convert Carlini {\em et al.}'s attack \cite{DBLP:conf/eurocrypt/CarliniCHRS25} from their geometric view to our algebraic view, which inspires two more efficient  clustering methods named as {\em Normal Rank Check} (NRC) and {\em Approximate Signature Vector} (ASV) methods.     
    The first method NRC computes the rank of only 4 normal vectors of the decision boundaries of  two dual points in layer $k$ to determine the consistency, while  Carlini {\em et al.} compute the rank of about $2d^{(k)}$ vectors of the same length  \cite{DBLP:conf/eurocrypt/CarliniCHRS25}. 
     
    The second method ({\em i.e.,} ASV) further enhances this efficiency: in unitary balanced networks, it theoretically bounds the expected directional distance between the derived ASV $\vec{v}$ and the corresponding neuron weight $\vec{\omega}$ by $1/d^{(k)}$. For FCNs with high weight dimension $d^{(k)}$, experiments have also confirmed that the directions of $\vec{v}$ and $\vec{\omega}$ are nearly parallel  (identical or opposite), consistent with the theoretical results. 
    Specifically, we first compute the ASV $\vec{v}$ for each dual point. If two dual points correspond to the same neuron,  
    then their ASVs $\vec{v}_1$ and $\vec{v}_2$ will be nearly parallel to $\vec{\omega}$,  simultaneously. Therefore, the directions of  $\vec{v}_1$ and $\vec{v}_2$ will also be nearly parallel, which could be detected by a simple inner product in $\mathcal{O}(d^{(k)})$.  
    
    Suppose there are $n$ dual points for layer $k$,  Carlini {\em et al.}'s clustering \cite{DBLP:conf/eurocrypt/CarliniCHRS25} is to compute the rank of $2d^{(k)}$ vectors for any two dual points through SVD, which has a total time complexity of $\mathcal{O}(n^2 (d^{(k)})^3)$. In contrast, the time complexities for both NRC or ASV methods are $\mathcal{O}(n(d^{(k)})^3+n^2d^{(k)})$ in the worst case, or $\mathcal{O}(n(d^{(k)})^3)$ in the average case. 
    In our experiment with 2,000 dual points, the clustering time is $2^{6.09}$ seconds (about 1 minute) with our ASV method, while  Carlini {\em et al.}'s method \cite{DBLP:conf/eurocrypt/CarliniCHRS25} requires about $2^{21.54}$ seconds (847 hours), {\em i.e.}, the time is significantly reduced by a factor of $2^{15}$. More comparisons are given in Table  \ref{tab:clustering_comparison}. In practical attacks, the NRC and ASV methods could be combined to achieve efficiency and robustness.

\item {\bf Kernel-Centric ASV in Hard-label Attack on the  Max-pooling CNN:} Based on the ASV method, we introduce the first hard-label attack on max-pooling CNNs,  filling a cryptanalysis gap. We define the dual points ({\em i.e.}, dual RPCPs or dual PSPs) based on the critical points introduced by Zirui Chen {\em et al.} 
\cite{DBLP:journals/iacr/ChenTGSQD26}. 
In Carlini {\em et al.}'s attack \cite{DBLP:conf/eurocrypt/CarliniCHRS25} on FCNs and Sun {\em et al.}'s attack \cite{cnn_average_pooling} on CNNs, the dual points clustering methods are both {\em neuron-centric}, {\em i.e.}, two clustered dual points should correspond to the same neuron. 
When applying the {\em neuron-centric} method to the max-pooling CNN, one has to collect many dual RPCPs or dual PSPs to expect that at least two fall into the same neuron or max pooling. 
However, unlike FCNs, in layer $k$ of a single-channel CNN, 
the convolutional kernel weight to be recovered is the same one for all neurons. Therefore, there is a natural question that {\em can we cluster different dual RPCPs and dual PSPs corresponding to different neurons or max poolings within the same layer $k$ into the same category, and use them to determine the unique convolutional kernel weight?} This paper answers the question by proposing the {\em Kernel-Centric ASV} method with a {\em kernel-centric} (instead of Carlini {\em et al.}'s {\em neuron-centric}) clustering method to solve the unique convolutional kernel weight. 


\end{itemize}

\subsubsection{Black-box and Partial Black-box Experiments.} 
\begin{itemize}
  
  \item{\bf Black-box Signature Recovery on 64-64$ \times$4-10 FCN.} The experiment targets on 64-64$ \times$4-10 FCN, identical to the ``Tiny'' model from the source code provided by Carlini {\em et al.} \cite{DBLP:conf/eurocrypt/CarliniCHRS25}, but trained on MNIST dataset. Due to the high computational cost of Carlini {\em et al.}'s method \cite{DBLP:conf/eurocrypt/CarliniCHRS25}, they only perform a proof-of-concept white-box signature recovery, {\em i.e.}, relying on white-box access to gradients or neuron indices across all three phases (dual point finding, clustering, and signature computing). For this network, we treat all the three phases as a black box and extract the signatures of the first two layers and 26 neurons of layer 3. The results and the comparison with the experiment \cite{DBLP:conf/eurocrypt/CarliniCHRS25} on the same platform are given in the first three rows of Table~\ref{tab:Total Experiments}. Carlini {\em et al.}'s method needs 0.93 hours to extract the 1st layer, while our method only needs 0.04 hours, reducing the runtime by 94.6\%. For the 2nd layer, about 300, 000 dual points are needed to cluster, and our method needs 4.22 hours, while Carlini {\em et al.}'s method cannot give a result within 1 month. 

\item{\bf Black-box Extractions on Hard-label Max-pooling CNNs.} For the experiments on CNNs, we recover the full convolutional block of a (2+1) CNN and a (2+2) LeNet-5 with max pooling in the hard-label setting. The bottom four rows of Table~\ref{tab:Total Experiments} compare our results with Sun \emph{et al.} (target average pooling) \cite{cnn_average_pooling} and Chen \emph{et al.} (requires raw outputs) \cite{DBLP:journals/iacr/ChenTGSQD26}.

\item {\bf Partial Black-box Signature Recovery on 3072-256{$\times$}3-64-10 FCN.} 
For this experiment, only the {\em dual point finding} phase relies on white-box gradient access. The other two phases are still black-box. We use 100,000 dual points to recover the signatures of the first two layers in Table \ref{tab:Total_3072FCN_Experiments}. 
Due to the  high computational cost of Carlini {\em et al.}'s method \cite{DBLP:conf/eurocrypt/CarliniCHRS25},
we estimate its runtime by multiplying the total required comparisons (obtained via white-box access) by the average comparison time measured on a subset of dual points.
\end{itemize}
The source codes for all the experiments can be found via
\begin{center}
    XXX
\end{center}
\begin{table}\scriptsize
\centering
\captionsetup{labelfont=bf}
\caption{{\bf Black-box Experiments on FCN and Max-pooling LeNet-5 in S1 Setting.} 
All notations are summarized in Tab. \ref{tab:notations} in {\sf Supp.} \ref{sect:def_notations}. and the architecture of LeNet-5 is given in {\sf Supp.} \ref{supp:model structure}. 
The accuracy term $(\varepsilon,0)$ is defined in Sect. \ref{subsect:adversarial goal and assumptions}, and $\max \lvert \theta - \hat{\theta} \rvert$ is the maximum extraction error of model parameters.}
\label{tab:Total Experiments}

\renewcommand{\arraystretch}{0.8} 
\setlength{\tabcolsep}{2pt}

\newcommand{\ggap}{\\[0.6pt]} 
\resizebox{\textwidth}{!}{
\begin{threeparttable}
    
    \begin{tabular}{l c c c c c c c} 
    \toprule
    \makecell[l]{\textbf{Models}}
    & \makecell[c]{\textbf{Architecture}$^\dag$\\ 
    $d^{(k)}$--$d_f^{(k)}$--$d^{(k+1)}$}
    & \textbf{Kernel}
    & \textbf{Method}
    & \makecell[c]{\textbf{Runtime}\\hours}
    & \textbf{Queries}
    & $(\varepsilon, 0)$
    & $\max \lvert \theta - \hat{\theta} \rvert$ 
    \\
    \midrule
    \multirow{10}{*}{\makecell[l]{FCN \\ 64-64$\times$4-10}}
    
    & \multirow{3}{*}{\makecell[l]{$F_1$: 64--64}}
    & --
    & ASV
    & 0.05 
    & $2^{24.30}$ & $2^{-39.38}$ & $2^{-44.20}$
    \\
    \cmidrule{3-8} 
  
    & 
    & --
    & \cite{DBLP:conf/eurocrypt/CarliniCHRS25}
    & 0.92 
    & $2^{24.30}$ & $2^{-39.14}$ & $2^{-44.17}$
    \\
    \cmidrule{2-8} 

    & \multirow{3}{*}{\makecell[l]{$F_2$: 64--64}}
    & --
    & NRC
    & 4.22 
    & $2^{30.76}$ & $2^{-36.76}$ & $2^{-41.30}$ 
    \\
    \cmidrule{3-8}
    & 
    & --
    & \cite{DBLP:conf/eurocrypt/CarliniCHRS25}
    &$\approx$1478.55$^\blacklozenge$ 
    & $\approx 2^{30.76}$ & -- & -- 
    \\
    \cmidrule{2-8}

    & \multirow{4}{*}{\makecell[l]{$F_3$: 64--64}}
    & --
    & NRC 
    & 6.58 
    & $2^{32.18}$ & \makecell[c]{26/64$^\lozenge$ \ggap  recovered} & $2^{-36.02}$ 
    \\
    \cmidrule{3-8}
    & 
    & --
    &\cite{DBLP:conf/eurocrypt/CarliniCHRS25}
    & $\approx$1079.65$^\blacklozenge$ 
    & $\approx 2^{30.76}$ & \makecell[c]{25/64$^\lozenge$ \ggap recovered} & -- 
    \\    
    \midrule[0.9pt] 
    
    (2+1) CNN
    & \makecell[l]{
        $C_1$: 1024--784--196 \ggap
        $C_2$: 196--100--25
      }
    & \makecell[c]{
        (1,1,5,5) \ggap
        (1,1,5,5)
      }
    & \makecell[c]{
        Kernel-Centric \ggap
        ASV
      }
    &1.46 
    & $2^{28.35}$ & $2^{-31.41}$ & $2^{-38.74}$
    \\
    \midrule
    
    LeNet-5 
    & \makecell[l]{
        $C_1$: 1024--4704--1176 \ggap
        $C_2$: 1176--1600--400
      }
    & \makecell[c]{
        (1,6,5,5) \ggap
        (6,16,5,5)
      }
    &\makecell[c]{
        Kernel-Centric \ggap
        ASV
      }
    &12.65 
    & $2^{30.96}$ & $2^{-27.89}$ & $2^{-36.18}$
    \\
    \midrule

    LeNet-5 \cite{cnn_average_pooling}$^{\spadesuit}$& $C_2$: 1176--1600--400 
    & (6,16,5,5) 
    & \makecell[c]{
        Neuron-Centric \ggap
        Clustering
      }
    &190.84$^\ddag$ 
    & $2^{32.34}$ & -- & $2^{-17.75}$
    \\
    \midrule
    LeNet-5 \cite{DBLP:journals/iacr/ChenTGSQD26}$^{\clubsuit}$
    & \makecell[l]{
        $C_1$: 1024--4704--1176 \ggap
        $C_2$: 1176--1600--400
      }
    & \makecell[c]{
        (1,6,5,5) \ggap
        (6,16,5,5)
      }
    & --
    &42.11$^\ddag$ 
    & $2^{27.29}$ & $2^{-23.38}$ & $2^{-27.31}$
    \\
    \bottomrule
    \end{tabular}
    
    \begin{tablenotes}[flushleft] 
        \item[$\dag$:] $F_i$ is the FCN layer $i$; 
        $C_i$ is the CNN layer $i$, defined in  Tab. \ref{tab:notations}. 
        \item[$\blacklozenge$:]  
        Due to the  high computational cost of Carlini {\em et al.}'s method \cite{DBLP:conf/eurocrypt/CarliniCHRS25}, the complexity of only $F_1$ layer extraction is given. For $F_2$ layer, we estimate their runtime as follows (also given in Sect. \ref{Experiments}): 
We simulate the clustering process for all $300,000$ dual points using white-box neuron indices, assuming each SVD-based consistency check is correct. This full-scale clustering requires a total of $2,653,799,211$ comparison. We then evaluate the actual time cost on a smaller subset of $3,000$ dual points, which takes $3,585.770$ seconds to perform $1,790,440$ comparison. According to Sect.~\ref{subsect:Carlini's limitation}, the estimated runtime for Carlini \emph{et al.}'s method to cluster the entire dual points set is about $\frac{3,585.770\text{ s}}{1,790,440 \times 3,600 \text{ s/h}} \times 2,653,799,211= 1,476$ hours. Together with the costs of dual point finding and signature computing,  the extraction time for $F_2$ is about 1,478.55 hours. Similarly, for $F_3$, $300,000$ dual points are clustered with about $1077$ hours.

        \item[$\lozenge$:] In $F_3$, in order to make practical experimental comparison, only about 26 neurons' signatures are recovered.   
        \item[$\spadesuit$:] Sun {\em et al.} \cite{cnn_average_pooling} target the LeNet-5 with {\em Average Pooling} in hard-label setting.
        \item[$\clubsuit$:] Chen {\em et al.} \cite{DBLP:journals/iacr/ChenTGSQD26} target LeNet-5 with {\em Max Pooling} in raw-output setting.
        \item[$\ddag$:] The figures are borrowed from \cite{cnn_average_pooling} and \cite{DBLP:journals/iacr/ChenTGSQD26}, respectively. 
    \end{tablenotes}
\end{threeparttable}
}
\end{table}

\begin{table}\scriptsize
\centering
\captionsetup{labelfont=bf}
\caption{\bf Partial Black-box Experiments on {\em 3072-FCN} in S1 Setting.} 
\label{tab:Total_3072FCN_Experiments}
\renewcommand{\arraystretch}{0.8} 
\setlength{\tabcolsep}{2pt}

\newcommand{\ggap}{\\[0.9pt]} 
\resizebox{\textwidth}{!}{
\begin{threeparttable}
    
    \begin{tabular}{l l c c c c c} 
    \toprule
    \makecell[l]{\textbf{Models}}~~~~~~~~~~~~~~
    & \makecell[c]{\textbf{Architecture}\\ 
    $d^{(k)}$--$d^{(k+1)}$}
    & \textbf{Method}
    & \makecell[c]{\textbf{Runtime}\\hours}
    & \textbf{Queries}
    & $(\varepsilon, 0)$
    & $\max \lvert \theta - \hat{\theta} \rvert$ 
    \\
    \midrule
    \multirow{6}{*}{\makecell[l]{FCN \\ 3072-256$\times$3-64-10}}
    
    & \multirow{3}{*}{\makecell[l]{$F_1$: 3072--256}}
    & ASV
    & 6.36 
    & $2^{22.77}$ & $2^{-17.74}$ & $2^{-24.56}$
    \\
    \cmidrule{3-7} 
  
    & 
    & \cite{DBLP:conf/eurocrypt/CarliniCHRS25}
    & $\approx$825845.75$^\blacklozenge$ 
    & $\approx2^{22.77}$ & -- & --
    \\
    \cmidrule{2-7} 

    & \multirow{3}{*}{\makecell[l]{$F_2$: 256--256}}
    & ASV
    & 4.17 
    & $2^{22.77}$ & \makecell[c]{238/256 \ggap  recovered} & $2^{-23.19}$ 
    \\
    \cmidrule{3-7}
    & 
    & \cite{DBLP:conf/eurocrypt/CarliniCHRS25}
    & $\approx$93568.23$^\blacklozenge$  
    & $\approx2^{22.77}$ & \makecell[c]{238/256 \ggap  recovered} & -- 
    \\
    \bottomrule
    \end{tabular}
    \begin{tablenotes}[flushleft]
        \item[$\blacklozenge$:] Similarly,  the runtime is estimated by multiplying the total comparisons (obtained via white-box access) by the average comparison time measured on a subset of dual points. For $F_1$ layer, the full-scale clustering for $100,000$ dual points requires $1,296,384,990$ comparison. We evaluate on $200$ dual points, taking $42,371.518$ seconds to perform $18,476$ comparison. According to Sect.~\ref{subsect:Carlini's limitation}, the estimated runtime for Carlini \emph{et al.}'s method to cluster is about $\frac{42,371.518\text{ s}}{18,476\times 3,600 \text{ s/h}} \times 1,296,384,990= 825,843$ hours. Similarly, for $F_2$ layer, $100,000$ dual points are clustered with about $93567$ hours. 
    \end{tablenotes}
\end{threeparttable}
}
\end{table}

\section{Preliminaries}
\label{sect:Preliminaries}

All notations are summarized in Tab. \ref{tab:notations} in {\sf Supp.} \ref{sect:def_notations}. 
A Deep Neural Network (DNN) with logit output is a function $\mathcal{F}_\theta$ parameterized by $\theta$ that takes inputs $\in \mathbb{R}^{d^{(1)}}$ and returns values  $\in \mathbb{R}^{d^{(r+1)}}$. 
In the hard-label setting, DNN returns the most likely class label ({\em e.g.}, ``cat'' or ``dog''), instead of the logit values. 
    
\begin{definition}
    {\bfseries\upshape (Hard-label DNN)}
   A hard-label DNN takes an input $X^{(1)} \in \mathbb{R}^{d^{(1)}}$, processes it through the DNN, and applies a function $z$ on the resulting logit vector to returns a hard label:
\begin{equation}
    \mathcal{H}_\theta 
    = (z \circ \mathcal{F}_\theta)(X^{(1)}) = z(\mathcal{F}_\theta(X^{(1)}))=\mathop{\arg\max}\limits_{i} \mathcal{F}_\theta(X^{(1)})_i
\end{equation}
\end{definition}
Note that if there are ties, {\em i.e.}, multiple elements of $\mathcal{F}_\theta(X^{(1)})$ have the same maximum, the hard-label is the smallest index of these equal elements \cite{DBLP:conf/eurocrypt/CarliniCHRS25,DBLP:conf/asiacrypt/ChenDGSWW24}.

\subsection{Fully-connected Neural Network (FCN)} 
The FCN is a sequence of FCN layers, and each layer includes a linear function $f^{(k)}: \mathbb{R}^{d^{(k)}}\mapsto \mathbb{R}^{d^{(k+1)}}$ ($k\geq 1$), and 
a component-wise ReLU function $\sigma$: 
\begin{equation}
    \mathcal{F}_{\theta}=f^{(r+1)}\circ \sigma\circ f^{(r)}\circ \sigma\circ\cdots f^{(2)}\circ \sigma\circ f^{(1)}, 
\end{equation}
where $f^{(k)}:\mathbb{R}^{d^{(k)}} \rightarrow \mathbb{R}^{d^{(k+1)}}$ is an affine transformation:
        $$f^{(k)}(X)=A^{(k)}X^{(k)}+B^{(k)},$$ 
    where $X^{(k)}\in\mathbb{R}^{d^{(k)}}$ is the input vector. The weight matrix $A^{(k)}\in \mathbb{R}^{d^{(k+1)}\times d^{(k)}}$ and the bias vector $B^{(k)}\in \mathbb{R}^{d^{(k+1)}}$ are composed of floating-point numbers, which are the model parameters. The unsigned weight  $A^{(k)}$ and bias $B^{(k)}$ are named {\em signature} \cite{DBLP:conf/eurocrypt/CanalesMartinezCHRSS24}. 
Given input $X^{(1)}\in \mathbb{R}^{d^{(1)}}$, the ReLU function $\sigma$ in layer $k$ is also interpreted as the matrix 
\begin{equation}\label{eqn:I}
    I^{(k)}=\operatorname{diag}(\tau^{(k)}_0, \tau^{(k)}_1, \dots, \tau^{(k)}_{d^{(k)}-1}), 
\end{equation}
where $\tau^{(k)}_i\in \{0,1\},~ 0\leq i < d^{(k)}$ are determined by the input $X^{(1)}$. 
\begin{definition}{\bfseries\upshape(Linear Neighborhood)}
    Given an input  $x\in  \mathbb{R}^{d^{(1)}}$, the matrices $A^{(k)},~B^{(k)},~I^{(k)}, ~(1\leq k\leq r+1)$ are all fixed. The linear neighborhood of $x$ is defined as the subset $\chi \subset \mathbb{R}^{d^{(1)}}$, so that all  $x'\in \chi$, the same $A^{(k)},~B^{(k)},~I^{(k)}$ will be  applied to compute the output of the FCN. 
\end{definition}
FCN has been proved to be piecewise linear function \cite{DBLP:conf/crypto/CarliniJM20,DBLP:conf/eurocrypt/CanalesMartinezCHRSS24}, {\em i.e.}, for $x$ in the linear neighborhood $\chi$ of $X^{(1)}$, 
the FCN is reduced to 
\begin{equation}\small
    \label{eqn:FCN_linear}
    \begin{array}{lll}
     \mathcal{F}_{\theta}(x) &=&  A^{(r+1)} \left(I^{(r)}\left(A^{(r)} \cdots \left(I^{(1)}\left(A^{(1)}x + B^{(1)}\right)\right) \cdots + B^{(r)}\right) \right)+ B^{(r+1)} \\
        &=& A^{(r+1)}I^{(r)}A^{(r)} \cdots I^{(2)}A^{(2)}I^{(1)}A^{(1)}x + \beta =\Gamma x + \beta.
    \end{array}
\end{equation}

\subsection{Convolutional Neural Network (CNN)} 
In {$(m+n)$-deep} CNN, the function $\mathcal{F}_{\theta}: \mathbb{R}^{d^{(1)}} \rightarrow \mathbb{R}^{d^{(m+n+2)}}$ is composed of two sequential blocks: the first block (Convolutional Block) consists of $m$ CNN layers, and each layer includes  a convolutional function $f_c^{(k)}$, a ReLU $\sigma_c^{(k)}$, and a pooling function $\rho^{(k)}$ ($1\leq k\leq m$); the second block consists of $n$ FCN layers:
    \begin{equation}
        \label{eqn:cnn_f}
        \mathcal{F}_{\theta}=\underbrace{f^{(n+1)} \circ \sigma^{(n)} \circ \cdots \circ \sigma^{(1)} \circ f^{(1)}}_{\mathclap{\text{Fully-connected Block}}} \,
            \circ \underbrace{\rho^{(m)} \circ \sigma_c^{(m)} \circ f_c^{(m)} \circ \cdots \circ \rho^{(1)} \circ \sigma_c^{(1)} \circ f_c^{(1)}}_{\mathclap{\text{Convolutional Block}}}.
    \end{equation}
At CRYPTO 2026, Zirui Chen {\em et al.} \cite{DBLP:journals/iacr/ChenTGSQD26} gave an algebraic view of CNN. 
Given $X^{(1)}\in \mathbb{R}^{d^{(1)}}$, the CNN layer  ``$\rho^{(k)} \circ \sigma_c^{(k)} \circ f_c^{(k)}$'' is translated into 
\begin{equation}
  \begin{array}{ll}
\rho^{(k)} \circ \sigma_c^{(k)} \circ f_c^{(k)} &=P^{(k)}I^{(k)}(A^{(k)}X^{(k)}+ B^{(k)}),
  \end{array}
\end{equation}
where convolutional kernal matrix $A^{(k)}$ and the bias $B^{(k)}$ are the only trained parameters in CNN which will be extracted,  $I^{(k)}$ is the same to Eq. \eqref{eqn:I}, and $P^{(k)}$ is the so-call Boolean max-pooling matrix \cite{DBLP:journals/iacr/ChenTGSQD26}. More details on those matrices are given in \textsf{Supp.} \ref{sect:def_notations}, and a toy example is given  Fig. \ref{fig:toy_example_diagram} in \textsf{Supp.}  \ref{sec:toy_example_cnn}.  
CNN has been also proved to be piecewise linear \cite{DBLP:journals/iacr/ChenTGSQD26}, {\em i.e.}, for $x$ in a linear neighborhood of $X^{(1)}$, the CNN layers and FCN layers of CNN ``collapse'' into an affine transformation. Assume that the function of the $m$ CNN layers is $ \mathcal{F}_{1}$, and the function of the $n$ FCN layers is $\mathcal{F}_{2}$, then 
\begin{equation}\label{eqn:cnn_f1_linear}
  \begin{array}{ll}
y&=\mathcal{F}_{1}(x) = P^{(m)}(I^{(m)}(A^{(m)}  \cdots (P^{(1)}(I^{(1)}(A^{(1)}x+B^{(1)})))\cdots+B^{(m)}))\\
       & = P^{(m)}I^{(m)}A^{(m)}  \cdots P^{(2)}I^{(2)}A^{(2)}P^{(1)}I^{(1)}A^{(1)}x+ \beta_1 = \Gamma_1 \cdot x+\beta_1.
  \end{array}
\end{equation}
Together with Eq. \eqref{eqn:FCN_linear},  the full CNN ``collapses'' into
\begin{equation}\label{eqn:affine_transformation}
  \begin{array}{ll}
\mathcal{F}_{\theta}(x) =\Gamma_2  (\Gamma_1 \cdot x+\beta_1)+ \beta_2 = \Gamma_2 \cdot \Gamma_1
\cdot  x + \Gamma_2 \beta_1 + \beta_2.
\end{array}
\end{equation}

\begin{definition} \label{def:layer_merging_intro}
    {\bfseries\upshape (Layer Merging)}
    Given $(m+n)$-deep CNN network, suppose that we have complete knowledge of the first $k-1$ layers of the convolutional block, and we are currently recovering layer $k$. Let $F_x^{k-1}$ and $G_x^{k+1}$ represent, respectively, the fully recovered and non-recovered parts of the CNN.
        $$\mathcal{F}_{\theta} = \underbrace{f^{(n+1)} \circ \cdots \circ \sigma^{(1)}\circ f^{(1)} \circ 
        \rho^{(m)} \circ \cdots \circ f_c^{(k+1)}}_{G_x^{k+1}}
        \circ \rho^{(k)} \circ \sigma_c^{(k)} \circ f_c^{(k)} \circ 
        \underbrace{\rho^{(k-1)} \circ \cdots \circ f_c^{(1)}}_{F_x^{k-1}}.$$
   Given $X^{(1)} \in  \mathbb{R}^{d^{(1)}}$, $G_x^{k+1}$ and $F_x^{k-1}$ become $G_x^{k+1}(X^{(k+1)})=
    G^{k+1}X^{(k+1)}+B^{k+1}$ and $F_x^{k-1}(X^{(1)})=F^{k-1}X^{(1)}+B^{k-1}$, where the matrices $F^{k-1}\in \mathbb{R}^{d^{(k)}\times d^{(1)}}$, $B^{k-1}\in \mathbb{R}^{d^{(k)}}$, $G^{k+1}\in \mathbb{R}^{d^{(n+2)}\times d^{(k+1)}}$, $B^{k+1}\in \mathbb{R}^{d^{(n+2)}}$,   respectively. The definition of layer merging for FCN is similarly defined.  
\end{definition}


\begin{definition} 
    {\bfseries\upshape (Neuron State and  Critical Point \cite{DBLP:conf/crypto/CarliniJM20})}
    Let $\mathcal{V}(\eta;X^{(1)})$ denote the value that neuron $\eta$ takes with $X^{(1)} \in \mathbb{R}^{d^{(1)}}$ before applying its corresponding ReLU. If $\mathcal{V}(\eta;X^{(1)})>0$ (respectively, $\mathcal{V}(\eta;X^{(1)})<0$), the neuron state of $\eta$ is activated (respectively, inactivated). In FCN, if $\mathcal{V}(\eta;X^{(1)})=0$, the neuron state is critical, and $X^{(1)}$ is a critical point.
\end{definition}

\begin{definition} 
    {\bfseries\upshape (ReLU-Pooling Critical Point, RPCP \cite{DBLP:journals/iacr/ChenTGSQD26})}
    \label{def:RPCP} In CNN, a RPCP is an input $X^{(1)} \in \mathbb{R}^{d^{(1)}}$ that makes the input of a ReLU of $\sigma^{(k)}_c$  0, and by varying the input in its tiny vicinity, the output of CNN changes non-linearly. 
\end{definition} 

\begin{pro_}\label{pro:rpcp}\cite{DBLP:journals/iacr/ChenTGSQD26}
   If $X^{(1)}$ is a RPCP corresponding to the $t$-th neuron in layer $k$, then $A^{(k)}_t X^{(k)} + B_t^{(k)}=0$, and the inputs of other neurons in the same Local Reception Field of Pooling (LRF-P) are smaller than 0. 
\end{pro_}

\begin{definition} 
    {\bfseries\upshape (Pooling Switching Point, PSP \cite{DBLP:journals/iacr/ChenTGSQD26})}
    \label{def:PSP}
    In CNN, a PSP is an input $X^{(1)} \in \mathbb{R}^{d^{(1)}}$ such that two inputs within the local receptive field of a pooling layer (LRF-P) achieve the same maximum value. By varying the inputs in the tiny vicinity of $X^{(1)}$, the output of CNN changes non-linearly. 
\end{definition}

\begin{pro_}\label{pro:psp}\cite{DBLP:journals/iacr/ChenTGSQD26}
   Given PSP $X^{(1)}$ for the $i$-th and $j$-th neuron in layer $k$, then $A^{(k)}_{i} X^{(k)} + b^{(k)} = A^{(k)}_{j} X^{(k)} + b^{(k)}$, and the inputs of  other neurons in the same LRF-P are smaller than it. 
\end{pro_}

\subsection{Adversarial Goals and Assumptions}
\label{subsect:Adversarial Goals and Assuamptions}
We make the following assumptions on the capabilities of the attacker:
\begin{itemize}
\item \textbf{Architecture knowledge.}  The attacker has the full knowledge of the target neural network's architecture, while the parameters are treated as secret.  
\item \textbf{Full-domain inputs.} The attacker can query arbitrary inputs from $\mathbb{R}^{d^{(1)}}$.
\item \textbf{Precise Computation.} The neural network is specified and evaluated using a sufficiently high precision floating-point arithmetic. 
\item \textbf{ReLU Activations.} All activation functions $\sigma^{(k)}$ are the ReLU function.
\item \textbf{Hard-label Outputs.} Given input $X^{(1)}$, the model returns the label of the most likely class, which corresponds to the maximum of the raw output.
\end{itemize}

\subsubsection{Adversarial Goal.} 
\label{subsect:adversarial goal and assumptions}
The goal is to achieve an $(\varepsilon,\xi)$-functionally equivalent parameter extraction, rather than replicating the exact original parameters
\cite{DBLP:conf/crypto/CarliniJM20,DBLP:conf/eurocrypt/CarliniCHRS25}. 
\begin{definition}
    {\bfseries\upshape ($(\varepsilon,\xi)$-Functional Equivalence \cite{DBLP:conf/crypto/CarliniJM20})}
    Two neural networks $\mathcal{F}_{\theta}$ and $\mathcal{F}_{\hat{\theta}}$ are $(\varepsilon,\xi)$-functional equivalent on the input dataset $S$ if 
    $$Pr_{x\in S}(\lvert \mathcal{F}_{\theta}(x)-\mathcal{F}_{\hat{\theta}}(x) \rvert \leq \varepsilon) \geq 1-\xi.$$    
\end{definition}
In our analysis, we specifically focus on the $(\varepsilon,0)$-functional equivalence given by  Carlini {\em et al.} \cite{DBLP:conf/crypto/CarliniJM20} which represents an upper bound on the maximum error of the output, and adopt their error bounds propagation method 
to it. 

\section{Algebraic Analysis of Carlini {\em et al.}'s Hard-label Attack}
\label{sect:EC25}

\begin{figure}[htbp]
	\centering
    \begin{minipage}[b]{0.48\linewidth} 
		\centering
		\includegraphics[width=0.9\linewidth]{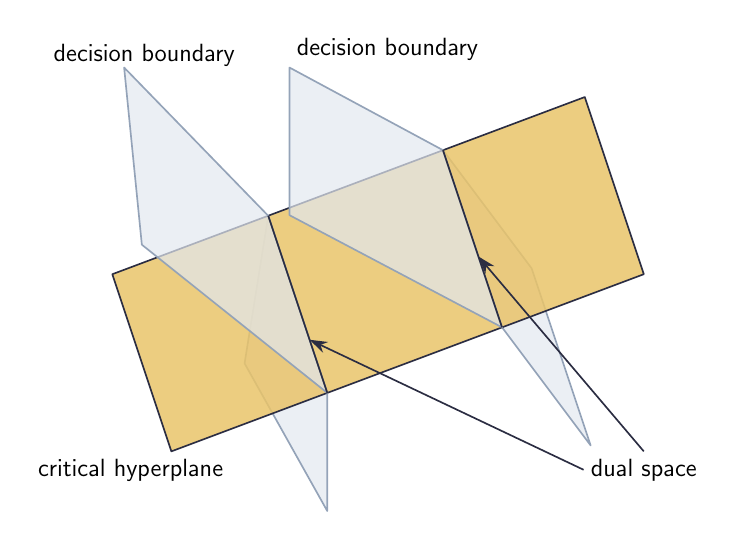}
		\caption{Intersection of critical hyperplane and dual space with decision boundaries}
		\label{fig:Consistent_intersection_spaces}
	\end{minipage}
    \vrule width 0.6pt 
	\hspace{1pt}   
	\begin{minipage}[b]{0.48\linewidth} 
		\centering
		\includegraphics[width=0.9\linewidth]{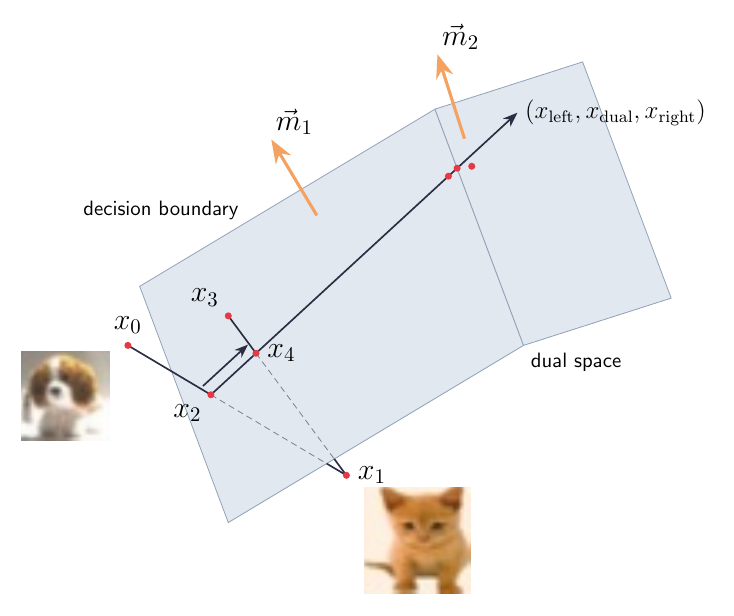}
		\caption{Carlini {\em et al.}'s method for finding dual spaces \cite{DBLP:conf/eurocrypt/CarliniCHRS25}}
		\label{fig:finding_intersection_spaces}
	\end{minipage}
\end{figure}

In the hard-label setting, 
the only change of the model output that the adversary can observe is label switching. Hence, the boundary between pairs of class labels, which is called the {\em decision boundary}, can be detected. 
At EUROCRYPT 2025 \cite{DBLP:conf/eurocrypt/CarliniCHRS25}, Carlini {\em et al.} observed that the decision boundary of a ReLU network is composed of piecewise-linear regions and it bends precisely when intersecting a {\em critical hyperplane} due to a single neuron's state change (see Fig. \ref{fig:Consistent_intersection_spaces}). This critical hyperplane consists of the critical points that are otherwise undetectable in this setting. 
Note that points on the two adjacent decision boundary patches share the same neuron states, except for the state of the neuron corresponding to the critical hyperplane. 
As shown in Fig. \ref{fig:Consistent_intersection_spaces}, in the input space, the intersection of the two $(d^{(1)}-1)$-dimension decision boundary patches yields a $(d^{(1)}-2)$-dimension subspace (called the {\em dual space}), which is entirely contained in the $(d^{(1)}-1)$-dimension critical hyperplane (the yellow region).

\subsection{Algebraic View on  Carlini {\em et al.}'s Signature Recovery}
\label{subsec:Carlini's algebraic view on DNN}
\begin{definition} \label{def:decision boundary}
    {\bfseries\upshape (Decision Boundary)}
    The decision boundary consists of points  $X^{(1)} \in \mathbb{R}^{d^{(1)}}$ so that two logit outputs of the FCN reach the same maximum value.
\end{definition}
Suppose the first $k-1$ layers are extracted. By Def. \ref{def:layer_merging_intro}, for any input $X^{(1)}$, the forward mapping matrices $F^{k-1}$, $B^{k-1}$ are known, yielding $X^{(k)}=F^{k-1}X^{(1)}+B^{k-1}$. 
If $X^{(1)}$ lies on the decision boundary where the network's 
the $i$-th and $j$-th outputs, $\mathcal{F}_i(X^{(1)})$ and $\mathcal{F}_j(X^{(1)})$, achieve the same maximum value, 
we have: 
\begin{equation}\small
    \label{eq:decision boundary}
    \mathcal{F}_i(X^{(1)})-\mathcal{F}_j(X^{(1)})=(G_i^{k+1}-G_j^{k+1}) I^{(k)} ( A^{(k)}X^{(k)} + B^{(k)} ) + B_i^{k+1}-B_j^{k+1}=0,
\end{equation}
where $G_i^{k+1}$ and $B_i^{k+1}$ are the $i$-th rows of matrices $G^{k+1}$ and $B^{k+1}$, respectively. 
{Let $\vec{m}=(G_i^{k+1}-G_j^{k+1}) I^{(k)} A^{(k)} F^{k-1} \in \mathbb{R}^{d^{(1)}}$, which is the normal vector of the decision boundary in the input space of the network.} 

\begin{definition}
    \label{def:critical hyperplane}
    {\bfseries\upshape (Critical Hyperplane)}
    For the $t$-th neuron in layer $k$, its critical hyperplane consists of this neuron's critical points $X^{(1)} \in \mathbb{R}^{d^{(1)}}$.  
\end{definition}
If $X^{(1)}$ is a critical point corresponding to the $t$-th neuron in layer $k$, then 
\begin{equation}
    \label{eq:critical point}
    A^{(k)}_t X^{(k)} + B_t^{(k)} = 0.
\end{equation}
At this point, the $t$-th diagonal entry of the activation matrix $I^{(k)}$ can be either 1 or 0 without altering the output, as the neuron always outputs 0. Consequently the corresponding activation matrices can be $I_{+}^{(k)}$ or $I_{-}^{(k)}$, differing only at the $t$-th diagonal entry ($1$ in $I_{+}^{(k)}$ and $0$ in $I_{-}^{(k)}$). Then, we have:
\begin{align}
G^{k+1} (I^{(k)}_{+}-I^{(k)}_{-}) ( A^{(k)}X^{(k)} + B^{(k)} ) =0.
\end{align}

\begin{definition} \label{def:dual points}
    {\bfseries\upshape (Dual Points)}
    Dual points are located at the intersection of a critical hyperplane and the decision boundary patches on both sides. Consequently, dual points fulfill the properties of both decision boundary points and critical hyperplane points ({\em i.e.}, Eqs. \eqref{eq:decision boundary} and \eqref{eq:critical point}). 
\end{definition}

If $X^{(1)}$ is a dual point corresponding to the $t$-th neuron in layer $k$, and is located on the decision boundary between class $i$ and class $j$, it satisfies:
\begin{align}\label{eq:dual_1}
    (G_i^{k+1} - G_j^{k+1}) (I^{(k)}_{+} - I^{(k)}_{-}) ( A^{(k)}X^{(k)} + B^{(k)} ) = 0. 
\end{align}

\begin{definition} \label{def:dual space}
    {\bfseries\upshape (Dual Space)}
    A dual space is the subspace spanned by dual points corresponding to the same neuron and the same decision boundary.
\end{definition}

Both a decision boundary and a critical hyperplane are governed by a single constraint (Eqs. \eqref{eq:decision boundary} and \eqref{eq:critical point} respectively), making them the $(d^{(1)}-1)$-dimension subspaces in the input space. Consequently, the dual space, constrained by both equations simultaneously, is $(d^{(1)}-2)$-dimension.

By employing the \emph{Dual Points Finding} algorithm proposed by Carlini {\em et al.} in \cite[Section 4]{DBLP:conf/eurocrypt/CarliniCHRS25}, upon collecting a dual point $x_{\text{dual}}$, as shown in Fig. \ref{fig:finding_intersection_spaces}, we simultaneously obtain two adjacent points $x_{\text{left}}$ and $x_{\text{right}}$. While $x_{\text{dual}}$ lies precisely on the intersection of the decision boundary and the critical hyperplane, $x_{\text{left}}$ and $x_{\text{right}}$ reside on the two distinct decision boundary patches on either side of this critical hyperplane. From these points, we can extract the local normal vectors to the adjacent decision boundary patches at inputs $x_{\text{left}}$ and $x_{\text{right}}$, 
denoted as $\vec{m}_\ell$ and $\vec{m}_r$, respectively as follows. 
According to  \cite{DBLP:conf/asiacrypt/ChenDGSWW24,DBLP:conf/eurocrypt/CarliniCHRS25}, starting from a point on the decision boundary ({\em e.g.}, $x_{\text{left}}$), we take a small step $\alpha_0$ along a reference basis direction $\vec{e_0}$ to step off the boundary, and then perform a binary search along another orthogonal basis direction $\vec{e_i}$ to find the distance $\alpha_i$ required to return to the decision boundary. By computing the ratio $\frac{\alpha_i}{\alpha_0}$, we obtain the relative coordinate weights of the normal vector. Iterating this procedure across all $d^{(1)}$ dimensions allows us to completely reconstruct the directions of $\vec{m}_\ell$ up to an unknown scalar factor, {\em i.e.},
\begin{equation}
    \label{eq:input space normal vector}
   \widehat{\vec{m}}_\ell = (1, -\frac{\alpha_0}{\alpha_1}, \dots, -\frac{\alpha_0}{\alpha_{d^{(1)}-1}}).
\end{equation}

For arbitrary points $X^{(1)}$ and ${X'}^{(1)}$ sampled from the same dual space, their difference effectively eliminates the constant terms of Eq. \eqref{eq:dual_1}:
\begin{equation}
    \label{eq:dual space difference}
    (G_i^{k+1} - G_j^{k+1}) (I^{(k)}_{+} - I^{(k)}_{-}) A^{(k)} F^{k-1}(X^{(1)} - {X'}^{(1)}) = 0. 
\end{equation}
Note that the term $I_{+}^{(k)} - I_{-}^{(k)}$ yields a diagonal matrix with a single non-zero entry (value 1) at the $(t,t)$-th position, effectively acting as a selection operator. Let the row vector $G_i^{k+1} - G_j^{k+1} = (g_0, g_1, \dots, g_{d^{(k+1)}-1}) \in \mathbb{R}^{d^{(k+1)}}$, then, 
\begin{align} \label{eq:gl-gr}
    (G_i^{k+1} - G_j^{k+1}) (I^{(k)}_{+} - I^{(k)}_{-}) A^{(k)} = g_t A^{(k)}_t.
\end{align}
Let $\vec{\delta}=X^{(1)} - {X'}^{(1)}$. Then, Eq. (\ref{eq:dual space difference}) can be simplified as:
\begin{equation}
    \label{eq:dual space equation}
    g_t A^{(k)}_t F^{k-1} \vec{\delta} = 0, 
\end{equation}
where $g_t A^{(k)}_t$ represents the weight vector of the target $t$-th neuron, scaled by an unknown constant $g_t$. 
\paragraph{Remark.} In the hard-label setting, the model extraction attack can still be regarded as a differential attack, where the difference between inputs from the same dual space is studied as it propagates through the sequential cryptographic/DNN operations. For a given input difference $\vec{\delta}$, the corresponding output difference---specifically defined as the difference between the logit gaps $\mathcal{F}_i - \mathcal{F}_j$ in Eq. \eqref{eq:decision boundary} of the two respective inputs---deterministically evaluates to $0$.

\subsubsection{Is a Single Dual Space Sufficient for One Neuron Weight Recovery?}
By sampling $n$ ($n$ is slightly larger than $d^{(k)}$) different dual points within the same dual space,  compute  $\vec{\delta}_{j}=X^{(1)}-X'^{(1)}$ ($0\leq j\leq n-1$), and construct a system of equations of Eq. \eqref{eq:dual space equation}: $
    {(F^{k-1} \vec{\delta}_{j})}^{\sf T} g_t {A^{(k)}_t}^{\sf T} = 0.$
To obtain a unique non-trivial solution (up to a scalar) for this linear system, 
the $\{F^{k-1} \vec{\delta}_{j}\}_{0\leq j\leq n-1}$ must attain a rank of $d^{(k)}-1$.
Note that $F^{k-1} \vec{\delta}_{j}$ is the difference between $X^{(k)}$ and $X'^{(k)}$. Since both $X^{(k)}$ and $X'^{(k)}$ are sampled from the same dual space, they are simultaneously subject to the constraints of the decision boundary (Def. \ref{def:decision boundary}) and the critical hyperplane (Def. \ref{def:critical hyperplane}). Geometrically, this intersection forms at most a $(d^{(k)}-2)$-dimension subspace (assuming all neurons in layer $k-1$ are activated, it is a $(d^{(k)}-2)$-dimension subspace). Hence, the matrix formed by any subset of the differential vectors $\{X^{(k)} - X'^{(k)}\}$ derived from a single dual space has a rank of at most $d^{(k)}-2$. Hence, we need at least two dual spaces corresponding to the same neuron to recover its weight vector.

\subsubsection{Clustering Dual Points from Different Dual Spaces.}
Given a set of dual points collected thus far, the next step is to cluster them into disjoint groups, where each group corresponds to a distinct target neuron.
To achieve this, Carlini \emph{et al.} \cite[Section 5.2]{DBLP:conf/eurocrypt/CarliniCHRS25} introduced the \emph{ISCONSISTENT} algorithm, which verifies whether two dual spaces $D_1$ and $D_2$ correspond to the same neuron by analyzing the rank of their union.
Let $F^{k-1}$ and ${F'}^{k-1}$ denote the corresponding local forward mapping matrices associated with $D_1$ and $D_2$, respectively. Suppose we sample differential vectors $\vec{\delta}_{0}, \dots, \vec{\delta}_{s_1-1}$ from $D_1$, and $\vec{\delta}'_{0}, \dots, \vec{\delta}'_{s_2-1}$ from $D_2$ ($s_1$ and $s_2$ are slightly larger than $d^{(k)}$). We stack each set of vectors:
$S = \begin{pmatrix}
        \vec{\delta}_{0}, \dots, \vec{\delta}_{s_1-1}
    \end{pmatrix}^{\sf T}$, 
$S'= \begin{pmatrix}
        \vec{\delta}'_{0}, \dots, \vec{\delta}'_{s_2-1}
    \end{pmatrix}^{\sf T}$.
The problem of determining consistency is thus reduced to checking whether the following homogeneous system of equations admits a non-trivial solution:
\begin{align} \label{eqn: check consistency}
    \begin{pmatrix}
    S {F^{k-1}}^{\mathsf T}\\
    S' {{F'}^{k-1}}^{\mathsf T}
    \end{pmatrix}
    g_t {A_{t}^{(k)}}^{\mathsf{T}} = 0.
\end{align}
Let $\Lambda^{(k-1)}_1$ and $\Lambda^{(k-1)}_2$ denote the sets of active neurons at layer $k-1$ for the dual points in $D_1$ and $D_2$, respectively.
And let $M_{SF}$ denote the coefficient matrix multiplying $g_t {A_{t}^{(k)}}^{\mathsf{T}}$ in Eq.~(\ref{eqn: check consistency}).
As discovered by Carlini {\em et al.} \cite[Section 4.3.2]{DBLP:conf/crypto/CarliniJM20}, the rank of the matrix is usually not full since the ReLU functions suppress negative outputs, {\emph i.e.}, some columns of $M_{SF}$ are zero column vectors. Following the criterion established in \cite[Section 5.2]{DBLP:conf/eurocrypt/CarliniCHRS25}:
\begin{itemize}
    \item If $D_1$ and $D_2$ belong to the same neuron, ${rank}(M_{SF}) < \lvert \Lambda^{(k-1)}_1 \cup \Lambda^{(k-1)}_2 \rvert$.
    \item If $D_1$ and $D_2$ belong to different neurons, ${rank}(M_{SF}) = \lvert \Lambda^{(k-1)}_1 \cup \Lambda^{(k-1)}_2 \rvert$.
\end{itemize}
By evaluating this rank condition, all sampled dual points are clustered.

\subsection{Limitations in Black-box Setting}
\label{subsect:Carlini's limitation}


As claimed by Carlini \emph{et al.} in \cite[Section 7.1]{DBLP:conf/eurocrypt/CarliniCHRS25}, the clustering step presents a bottleneck in implementations.  In contrast, the other phases of the attack were optimized and empirically viable in the black-box setting. According to \cite{DBLP:conf/eurocrypt/CarliniCHRS25}, when using a 256-core server with GPU support to recover a 3072-256{$\times$}3-64-10 FCN (referred to as {\em 3072-FCN}) with four hidden layers, trained on CIFAR10, finding the dual points (accelerated by computing white-box gradients symbolically rather than via black-box binary search) and unifying the corresponding dual spaces (neglecting clustering) took approximately 16 hours, while the sign recovery phase was heavily parallelized, requiring only 8.5 hours. Consequently, the clustering becomes the bottleneck of the black-box model extraction attack.
The procedure involves two main computational costs:
\begin{enumerate}
    \item \textbf{Pairwise Comparisons:} Clustering $n$ collected dual points requires $\mathcal{O}(n^2)$ independent invocations of the {\em ISCONSISTENT} algorithm.
    \item \textbf{Rank Computation:} Each {\em ISCONSISTENT} call computes the rank of the $(s_1 +s_2) \times d^{(k)}$ projected matrix $M_{SF}$ according to Eq.~(\ref{eqn: check consistency}), where $s_1$ and $s_2$ are slightly larger than $d^{(k)}$. Consequently, performing Singular Value Decomposition (SVD) on this matrix costs $\mathcal{O}((d^{(k)})^3)$.
\end{enumerate}
Multiplying these factors, the time complexity of Carlini \emph{et al.}'s clustering  \cite{DBLP:conf/eurocrypt/CarliniCHRS25}  is: 
\begin{equation}\label{eq:carlini_cluster_time}
    \mathcal{O}(n^2  (d^{(k)})^3).
\end{equation}
For deeper layers, recovering near-dead neurons typically requires millions of dual points. Therefore, the black-box extraction becomes extremely time-consuming.

\section{Improving Clustering by the Normal Vectors 
}
\label{sect:our clustering method}

\subsection{Compute $\vec{G}_{\ell}$ and $\vec{G}_r$ up to a scalar}
Given a dual point $x$ corresponding to the $t$-th neuron in layer $k$. Recall Eq. (\ref{eq:gl-gr}) and  denote the row vectors
\begin{equation} \label{eq:G_vectors}
\begin{aligned}
    \vec{G}_{\ell} &= (G_i^{k+1} - G_j^{k+1}) I_{+}^{(k)} A^{(k)} = \sum_{s\in \Lambda^{(k)}} g_s A^{(k)}_s + g_tA^{(k)}_t, \\
    \vec{G}_r &= (G_i^{k+1} - G_j^{k+1}) I_{-}^{(k)} A^{(k)} =
    \sum_{s\in \Lambda^{(k)}} g_s A^{(k)}_s.
\end{aligned}
\end{equation}
where $\Lambda^{(k)}$ denotes the index set of active neurons in layer $k$ for the dual point $x$. Then the difference between the row vectors $\vec{G}_{\ell}$ and $\vec{G}_r$ is the effective weight vector of the target neuron $t$, scaled by a constant factor $g_t$, {\em i.e.},
$\vec{G}_{\ell} - \vec{G}_r = g_tA_t^{(k)}$.
{\em Note that the vectors $\vec{G}_{\ell}$ and $\vec{G}_r$ also correspond to the normal vectors of the decision boundary patches on the left (activated) and right (inactivated) sides of the critical hyperplane, respectively.} 
Since we are operating in the hard-label setting, the exact magnitudes of the gradients $\vec{G}_{\ell}$ and $\vec{G}_r$ are inaccessible. 
Specifically, according to Def.~\ref{def:decision boundary} and Eq. \eqref{eq:G_vectors}, the normal vector to the decision boundary ({\em e.g.}, on the left patch) $\vec{m}_{\ell} = (G_i^{k+1} - G_j^{k+1}) I_{+}^{(k)} A^{(k)}F^{k-1}$, then 
\begin{equation}
    \label{eq:Gl system}
    {F^{k-1}}^{\sf T} {\vec{G}_{\ell}}^{\sf T} = {\vec{m}_{\ell}}^{\sf T}.
\end{equation}
This constructs a linear system with an unknown vector ${\vec{G}_{\ell}}^{\sf T}$. Recall from Eq.~\eqref{eq:input space normal vector} that,  the normal vectors $\vec{m}_{\ell}$ and $\vec{m}_r$ can be recovered up to a scalar multiple. Consequently, we can recover ${\vec{G}_{\ell}}^{\sf T}$ up to a scalar multiple from ${\vec{m}_{\ell}}^{\sf T}$ by solving this system. 
Let $\Lambda^{(k-1)}$ denote the index set of active neurons at layer $k-1$. For a vector $\vec{v}$, $\vec{v}[\Lambda^{(k-1)}]$ denotes the subvectors indexed by $\Lambda^{(k-1)}$. We will discuss how to solve the ${\vec{G}_{\ell}}^{\sf T}$ under 3 cases in the following section. 

\subsection{Normal Rank Check: Improved Rank Check by Normal Vectors}\label{sect:rank_check}

\subsubsection{Case 1: $d^{(k)}=\dim{({\vec{G}_{\ell}}^{\sf T})}=\operatorname{rank}(F^{k-1}) = \lvert \Lambda^{(k-1)} \rvert$.} In this case, the full vector $\vec{G}_{\ell}^{\sf T}$ in Eq.~\eqref{eq:Gl system} is recovered up to a scalar. 
By normalizing the solution, we extract its unit normal row vector: $\vec{n}_{\ell} = \frac{\vec{G}_{\ell}}{\|\vec{G}_{\ell}\|}$. A similar procedure is applied to the right patch to compute $\vec{n}_r = \frac{\vec{G}_r}{\|\vec{G}_r\|}$.
Consequently, according to Eq. \eqref{eq:G_vectors}, $g_t A^{(k)}_t$ can be expressed as a linear combination of the observed unit normal vectors:
\begin{equation}
    \label{eq:difference between normals}
    g_t A^{(k)}_t = \|\vec{G}_{\ell}\| \cdot \vec{n}_{\ell} - \|\vec{G}_r\|\cdot \vec{n}_r, 
\end{equation}
which 
indicates that the target weight vector $g_tA^{(k)}_t$ lies within the 2-dimensional plane $P$ spanned by $\vec{n}_{\ell}$ and $\vec{n}_r$, denoted as $P = span(\vec{n}_{\ell}, \vec{n}_r)$.
Given another dual point belonging to the same neuron, similarly compute the normal vectors of its adjacent decision boundary patches, denoted as $\vec{n}'_{\ell}$ and $\vec{n}'_r$.
Consequently, the corresponding target weight vector also satisfies $g_tA^{(k)}_t \in \mathrm{span}(\vec{n}'_{\ell}, \vec{n}'_r)$, which we denote as plane $P'$.
If the two dual points correspond to the same neuron, $g_tA^{(k)}_t$ resides in both $P$ and $P'$, {\em i.e.}, these two planes must intersect along at least a 1-dimensional line (which corresponds to the direction of $A^{(k)}_t$). By constructing a combined matrix $N=(\vec{n}_{\ell}^{\sf T}, \vec{n}_r^{\sf T}, \vec{n}_{\ell}^{\prime \sf T}, \vec{n}_r^{\prime \sf T})^{\sf T} \in \mathbb{R}^{4 \times d^{(k)}}$, the consistency between the two dual points can be evaluated via the rank of $N$:
\begin{itemize}
    \item If the two dual points belong to the same neuron, then $\mathrm{rank}(N) < 4$;
    \item If the two dual points belong to different neurons, then $\mathrm{rank}(N) = 4$.
\end{itemize}
In practice, we employ Singular Value Decomposition (SVD) to compute the minimum singular value $\sigma_{\min}$ of $N$. 
Then the rank condition is converted into verifying whether $\sigma_{\min}$ is sufficiently close to zero. 
Given a small threshold $\tau$, if $\sigma_{\min} < \tau$, the two dual points are assumed to belong to the same neuron.

\subsubsection{Case 2: $d^{(k)}>\operatorname{rank}(F^{k-1}) = \lvert \Lambda^{(k-1)} \rvert$.}  In this case,  we can recover ${\vec{G}_{\ell}}^{\sf T}[\Lambda^{(k-1)}]$ 
 from Eq.~\eqref{eq:Gl system}  up to a scalar multiple. 
 Given the active sets $\Lambda^{(k-1)}$ and $\Lambda^{\prime (k-1)}$ for dual points $x$ and $x'$, we compute and normalize its normal vectors on their respective active sets ({\em e.g.}, $\vec{n}_{\ell} = \frac{\vec{G}_{\ell}[\Lambda^{(k-1)}]}{\lVert \vec{G}_{\ell}[\Lambda^{(k-1)}] \rVert}$). 
When comparing the two dual points, we extract their shared active subset $\bar{\Lambda}^{(k-1)} = \Lambda^{(k-1)} \cap \Lambda^{\prime (k-1)}$, slice the raw vectors according to $\bar{\Lambda}^{(k-1)}$ and re-normalize them:
\begin{equation}
\label{eq:nl and nr}
    \vec{n}_{\ell} = \frac{\vec{G}_{\ell}[\bar{\Lambda}^{(k-1)}]}{\lVert \vec{G}_{\ell}[\bar{\Lambda}^{(k-1)}]\rVert}, \quad 
    \vec{n}_r = \frac{\vec{G}_r[\bar{\Lambda}^{(k-1)}]}{\lVert \vec{G}_r[\bar{\Lambda}^{(k-1)}]\rVert}.
\end{equation}
Then we perform 
the rank checking as {Case 1}. Note that 
the check requires $|\bar{\Lambda}^{(k-1)}| \geq 4$. Otherwise, the matrix $N \in \mathbb{R}^{4 \times |\bar{\Lambda}^{(k-1)}|}$ necessarily has rank smaller than $4$. We therefore exclude pairs with $|\bar{\Lambda}^{(k-1)}| < 4$ from the consistency check.

\subsubsection{Case 3: $d^{(k)}> \lvert \Lambda^{(k-1)}  \rvert> \operatorname{rank}(F^{k-1})$.} For Eq.~\eqref{eq:Gl system}, if $\operatorname{rank}(F^{k-1})<|\Lambda^{(k-1)}|$, it admits multiple solutions. 
We use the least-squares solver to obtain its minimum-norm solution, denoted by $\vec{h}$. Let $\vec{h}'$ be any other solution and define $\vec{z}=\vec{h}'-\vec{h}$. Since both vectors $\vec{h}$ and $\vec{h'}$ satisfy Eq.~\eqref{eq:Gl system}, we have
\begin{equation}
    (F^{k-1})^{\sf T}\vec{z} = (F^{k-1})^{\sf T}(\vec{h}'-\vec{h}) = \vec{0}.
\end{equation}
Hence, $\vec{z}\in\ker((F^{k-1})^{\sf T})$. 
For any scalar $\alpha$, $\vec{h}+\alpha\vec{z}$ is also a solution. If $\vec{h}^{\sf T}\vec{z}\neq 0$, choosing
$\alpha=-\frac{\vec{h}^{\sf T}\vec{z}}{\|\vec{z}\|^2}
$
gives
$\|\vec{h}+\alpha\vec{z}\|^2=\|\vec{h}\|^2-\frac{(\vec{h}^{\sf T}\vec{z})^2}{\|\vec{z}\|^2}<\|\vec{h}\|^2$,
which contradicts the fact that $\vec{h}$ is the minimum-norm solution.
Consequently, $\vec{h}^{\sf T}\vec{z} = 0$, {\em i.e.}, $\vec{h}$ is orthogonal to this kernel space.

For kernel space $\ker((F^{k-1})^{\sf T})$, let
$q=|\Lambda^{(k-1)}| - \operatorname{rank}(F^{k-1})$
denote its dimension. Thus, $q$ represents the number of directions that cannot be recovered from Eq.~\eqref{eq:Gl system}.
Since the original $\vec{G}_{\ell}^{\sf T}$ is also a solution, the difference between $\vec{G}_{\ell}^{\sf T}$ and $\vec{h}$ must lie in this kernel space. Therefore, there exists $\vec{z}_0\in\ker((F^{k-1})^{\sf T})$ such that
$\vec{G}_{\ell}^{\sf T}=\vec{h}+\vec{z}_0$.
Since $\vec{h}$ is orthogonal to $\vec{z}_0$, $\vec{h}$ is exactly the projection of $\vec{G}_{\ell}^{\sf T}$ obtained by discarding its kernel-space component $\vec{z}_0$. 
Therefore, when $q$ is small and
$\lVert\vec{z}_0\rVert/ \lVert\vec{G}_{\ell}^{\sf T}\rVert$ is small, only a small number of directions and a small component of $\vec{G}_{\ell}^{\sf T}$ are lost. In this case, $\vec{h}$ still provides a good approximation to the direction of $\vec{G}_{\ell}^{\sf T}$ and can be used in the subsequent comparison.

\subsubsection{Experiments on Consistency Verification Using the Minimum Singular Value $\sigma_{\min}$.}
Our experiment targets the same  {\em 3072-FCN} model used by Carlini {\em et al.}  \cite{DBLP:conf/eurocrypt/CarliniCHRS25} and also given in Sect. \ref{subsect:Carlini's limitation}.  
We sample 10,000 dual points across the input space. For each layer, we filter the points from the previous layers. Then evaluate two dual points by extracting their corresponding $\vec{n}_\ell$ and $\vec{n}_r$ from Eq.~\eqref{eq:nl and nr}, 
and calculate the minimum singular value $\sigma_{\min}$ of $N=(\vec{n}_{\ell}^{\sf T}, \vec{n}_r^{\sf T}, \vec{n}_{\ell}^{\prime \sf T}, \vec{n}_r^{\prime \sf T})^{\sf T} \in \mathbb{R}^{4 \times d^{(k)}}$. 
\begin{figure}[htbp]
    \centering
    \includegraphics[width=0.9\textwidth]{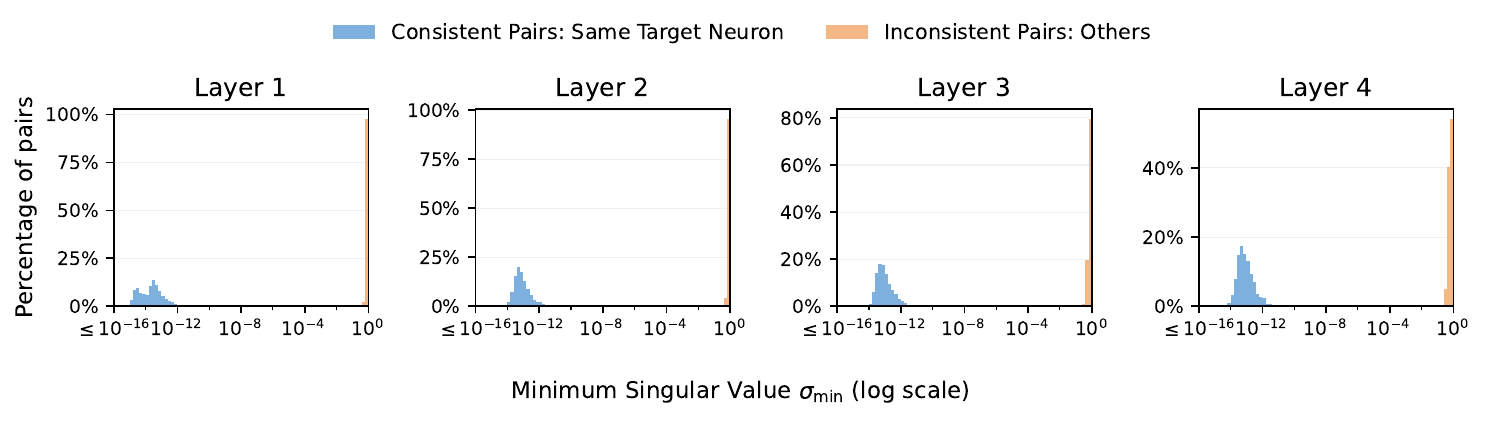}
    \caption{Distribution of $\sigma_{\min}$ for same-target-neuron and other dual-point pairs}
    \label{fig:sigma_min_distribution}
\end{figure} 
\vspace{-0.5em}

As illustrated in Fig.~\ref{fig:sigma_min_distribution}, the  $\sigma_{\min}$ from pairs of dual points corresponding to the {\em same} neuron in the target layer are all smaller than $10^{-8}$, while that from other pairs are close to 1. 
This separation on distribution of $\sigma_{\min}$ demonstrates the metric provides a reliable criterion for clustering dual points. 
To quantify the practical effectiveness of the consistency check, we evaluate its precision and recall during clustering:
\begin{align*}
    \mathrm{Precision}= \frac{\mathrm{TP}}{\mathrm{TP}+\mathrm{FP}},
\quad
\mathrm{Recall}=\frac{\mathrm{TP}}{\mathrm{TP}+\mathrm{FN}}.
\end{align*}
A ``True Positive (TP)'' occurs when both dual points corresponding to the same neuron in the target layer are identified as consistent. 
A ``False Positive (FP)'' arises when two dual points from different neurons, or the same neuron in another layer, are mistakenly clustered together. 
Conversely, a ``False Negative (FN)'' indicates two dual points belonging to the same target-layer neuron fail to be clustered together. 
``Precision'' measures how reliably the consistency check avoids grouping unrelated dual points and maintains cluster purity, while ``Recall'' evaluates how effectively it recognizes true target-neuron pairs, thereby reducing missed consistent pairs.
\begin{itemize}
    \item In Layer 1, every dual point meets Case 1 condition, and therefore the $\vec{G}_{\ell}^{\sf T}$ and $\vec{G}_{r}^{\sf T}$ are fully recovered. All dual points are clustered with 99.98\% Precision and 100 \% Recall.
    
    \item In Layer 2, the $\vec{G}_{\ell}^{\sf T}[\Lambda^{(k-1)}]$ and $\vec{G}_{r}^{\sf T}[\Lambda^{(k-1)}]$ are uniquely recoverable 
    with no rank deficiency (Case 2). 
    Only $0.0005\%$ of the pairs meet $|\bar{\Lambda}^{(k-1)}| < 4$. Excluding these pairs, the consistency check achieves $99.96\%$ Precision and $100\%$ Recall.
    
    \item In Layer 3, let $q$ denote the rank deficit $q = |\Lambda^{(k-1)}-\operatorname{rank}(F^{k-1})|$. Among the remaining candidate points, only 1.44\% have $q>0$ (Case 3). We retain all candidates, using minimum-norm solutions for rank-deficient systems, and obtain $99.95\%$ precision and $99.36\%$ recall.
    
    \item In Layer 4, $14.23\%$ have $q>0$, Applying the same procedure without rank-deficiency filtering yields $100\%$ precision and $88.55\%$ recall.
\end{itemize}

\subsection{Faster Consistency Checking: Approximate Signature Vector}
This section introduces a more efficient method called {\em Approximate Signature Vector} ({\em ASV}) method using unit normal row vectors $ \vec{n}_{\ell}, \vec{n}_r$ derived in Sect. \ref{sect:rank_check}

\begin{definition} \label{def:asv}
 {\bfseries\upshape (Approximate Signature Vector, ASV)} Given $ \vec{n}_{\ell}, \vec{n}_r$, the  {approximate signature vector (ASV) $\vec{v}$} is defined as:
\begin{align}
\label{eq:signature vector}
\vec{v}=\vec{n}_{\ell}-(\vec{n}_{\ell} \cdot \vec{n}_r^{\sf T})\vec{n}_r.
\end{align}
Let $\vec{\omega} = g_t A^{(k)}_t$. 
Then, $\vec{n}_{\ell} = \frac{1}{\|\vec{G}_{\ell}\|}(\vec{\omega} + \|\vec{G}_r\|\vec{n}_r)$. Since $\vec{n}_r\vec{n}_r^{\sf T}=1$, Eq. \eqref{eq:signature vector} becomes
\begin{align}
\label{eq:signature vector 2}
\vec{v}
= \frac{1}{\|\vec{G}_{\ell}\|}(\vec{\omega}-(\vec{\omega} \cdot \vec{n_r}^{\sf T})\vec{n_r}).
\end{align}

\end{definition}


\subsubsection{Directional Distance between ASV $\vec{v}$ and Neuron Weight $\vec{\omega}$.} 
Eq. \eqref{eq:signature vector 2} demonstrates that the ASV $\vec{v}$ deviates from the exact scaled weight $\vec{\omega}$ solely by the term $(\vec{\omega} \cdot \vec{n}_r^{\sf T})\vec{n}_r$.
To measure the precision of this directional alignment, we employ the cosine similarity $\cos \varphi_{\vec{v},\vec{\omega}} = \frac{\vec{v}\cdot \vec{\omega}^{\sf T}}{\|\vec{v}\|\|\vec{\omega}\|}$, and introduce the directional distance ${\sf DGap}(\vec{v},\vec{\omega})$ of $\vec{v},\vec{\omega}$ by the following cosine distance 
\begin{equation}\label{eq:direction_gap}
 {\sf DGap}(\vec{v},\vec{\omega})= 1-\lvert \cos \varphi_{\vec{v},\vec{\omega}} \rvert=    1-\frac{\lvert \vec{v}\cdot \vec{\omega}^{\sf T}\rvert}{\|\vec{v}\|\cdot \|\vec{\omega}^{\sf T}\|}. 
\end{equation}

Following the assumption that the network is a unitary balanced network, which is also used in the analysis on the expected signal-to-noise ratio of {\em Neuron Wiggle} by Canales-Mart{\'{\i}}nez {\em et al.}  \cite[Appendix A]{DBLP:conf/eurocrypt/CanalesMartinezCHRSS24} at EUROCRYPT 2024, we propose the following proposition. 
\begin{proposition}\label{pro:expectation_dgap}
Assuming the network is a unitary balanced network, and each neuron weight is a randomly chosen unit vector in $d^{(k)}$-dimensional space.  We have 
\begin{equation}\label{eq:expectation_dgap}
\mathbb{E}\bigl[{\sf DGap}(\vec{v},\vec{\omega})\bigr] \le \frac{1}{d^{(k)}},
\qquad \lim_{d^{(k)}\to\infty}\frac{1}{d^{(k)}} = 0.
\end{equation}
\end{proposition}
\begin{proof}
See {\sf Supp}.~\ref{supp:decoupled_error_proof}. 
\end{proof}
Proposition \ref{pro:expectation_dgap} shows that the expectation $\mathbb{E}[{\sf DGap}(\vec{v},\vec{\omega})]$
approaches zero, and hence the ASV $\vec{v}$ and the  neuron weight $\vec{w}$ are nearly parallel in direction, when the weight dimension increases.  

\subsubsection{White-box Experiment for Verification of ${\sf DGap}(\vec{v},\vec{\omega})\rightarrow 0$.}

\begin{figure}
    \centering
    \includegraphics[width=0.9\textwidth]{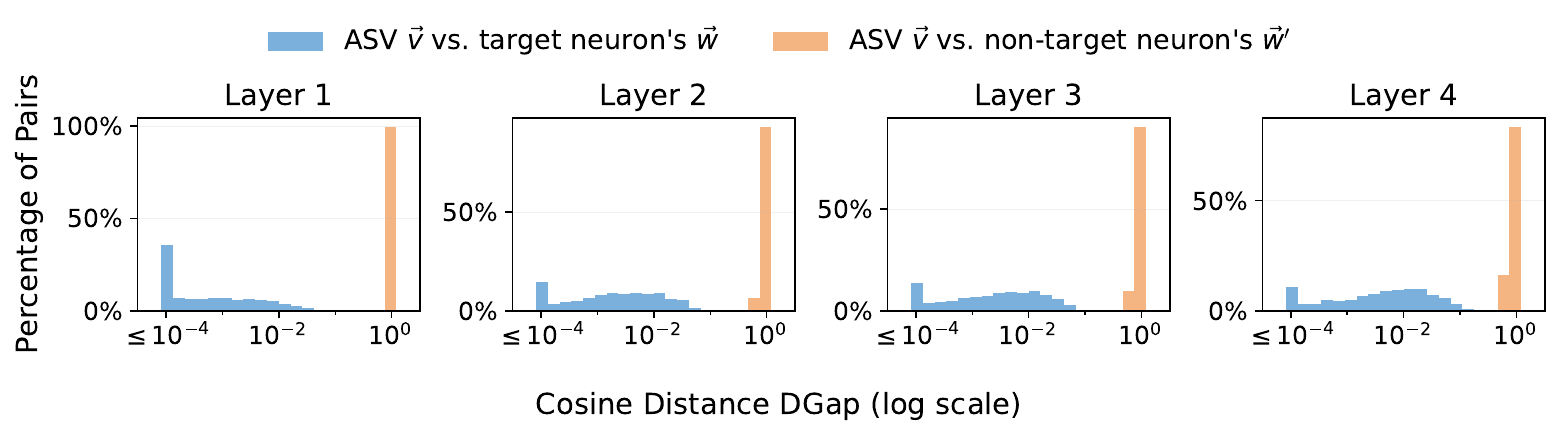}
    \caption{Distribution of  ({\sf DGap}) between ASV \((\vec{v})\) and  neuron weights \((\vec{w})\) }
    \label{fig:combined_cosine_dists_histogram}
\end{figure}

For the {\em 3072-FCN}, we randomly sample 10,000 dual points. For each point, compute its ASV $\vec{v}$ and ${\sf DGap}$ between $\vec{v}$ and the ground-truth weights of all neurons within the same layer. As shown in Fig. \ref{fig:combined_cosine_dists_histogram}, the resulting distributions exhibit a clear separation: the cosine distance between an ASV and its corresponding target neuron $\vec{\omega}$ tightly clusters near $0$, while the distances to  other non-target neurons $\vec{\omega}'$ consistently concentrate approach $1$. 

\subsubsection{Consistency Check using ASVs.}

If two dual points belong to the same neuron, the cosine similarities between their ASVs (denoted as $\vec{v}_1$ and $\vec{v}_2$) and the ground-truth weight $\vec{\omega}$ both approach $0$. According to the triangle inequality for spatial angles \cite[Chapter I.1]{bridson2013metric}, 
\begin{equation}
    0 \leq \varphi_{\vec{v}_1,\vec{v}_2} \leq \varphi_{\vec{v}_1,\vec{\omega}} + \varphi_{\vec{v}_2,\vec{\omega}}.
\end{equation}
Since both $\varphi_{\vec{v}_1,\vec{\omega}}$ and $\varphi_{\vec{v}_2,\vec{\omega}}$ approach $0$, $\varphi_{\vec{v}_1,\vec{v}_2}$ will also approach $0$.
\begin{pro_}\label{pro:asv} {\bfseries\upshape (Parallel  ASVs)}  \em
    If two dual points belong to the same neuron, their corresponding approximate signature vectors $\vec{v}_1$ and $\vec{v}_2$ will be nearly identical or opposite (both are nearly collinear with the underlying weight vector $\vec{\omega}$). Therefore, their consistency score is bounded by a small threshold $\tau$:
\begin{equation}
\label{eq:cosine_similarity}
\text{consistency score} = {\sf DGap}(\vec{v_1},\vec{v_2}) =  1-\lvert \cos \varphi_{\vec{v}_1,\vec{v}_2} \rvert = 1- \frac{\lvert \vec{v}_1 \vec{v}_2^{\sf T} \rvert}{\|\vec{v}_1\| \|\vec{v}_2\|}<\tau.
\end{equation} 
\end{pro_}

\subsubsection{Immunity to Activation State Blindness.}
One potential concern in the hard-label setting is the inability to directly observe the activation states of the patches $X^{(1)}_{left}$ and $X^{(1)}_{right}$. Given a dual point, we can extract two distinct normal vectors from its adjacent patches as $\vec{n}_1$ and $\vec{n}_2$, but we cannot ascertain which corresponds to the activated normal $\vec{n}_{\ell}$ and which to the inactivated normal $\vec{n}_r$. 
To bypass this ambiguity, we compute two ASVs for each dual point: 
\begin{equation}
    \vec{v} = \vec{n}_1 - (\vec{n}_1 \cdot \vec{n}_2^{\sf T})\vec{n}_2, \quad \vec{v}' = \vec{n}_2 - (\vec{n}_2 \cdot \vec{n}_1^{\sf T})\vec{n}_1.
\end{equation}
Given two distinct dual points, we obtain two pairs of ASVs: $(\vec{v}_1, \vec{v}'_1)$ and $(\vec{v}_2, \vec{v}'_2)$. To determine if these two dual points belong to the same neuron, we evaluate the consistency score ${\sf DGap}(\cdot, \cdot)$ across all possible cross-combinations. The final consistency score $\mathcal{S}$ between the two dual points is defined as the minimum value among these combinations:
\begin{equation}
    \mathcal{S} = \min \Big\{{\sf DGap}(\vec{v}_1, \vec{v}_2), {\sf DGap}(\vec{v}_1, \vec{v}'_2), {\sf DGap}(\vec{v}'_1, \vec{v}_2), {\sf DGap}(\vec{v}'_1, \vec{v}'_2) \Big\}.
\end{equation}
Since the ASVs closely approximate the true target weight, the correct pairing will minimize the score $\mathcal{S}$, effectively resolving the activation state blindness.

\subsubsection{Experiments on Consistency Verification.}
\begin{figure}[htbp]
    \centering
    \includegraphics[width=0.9\textwidth]{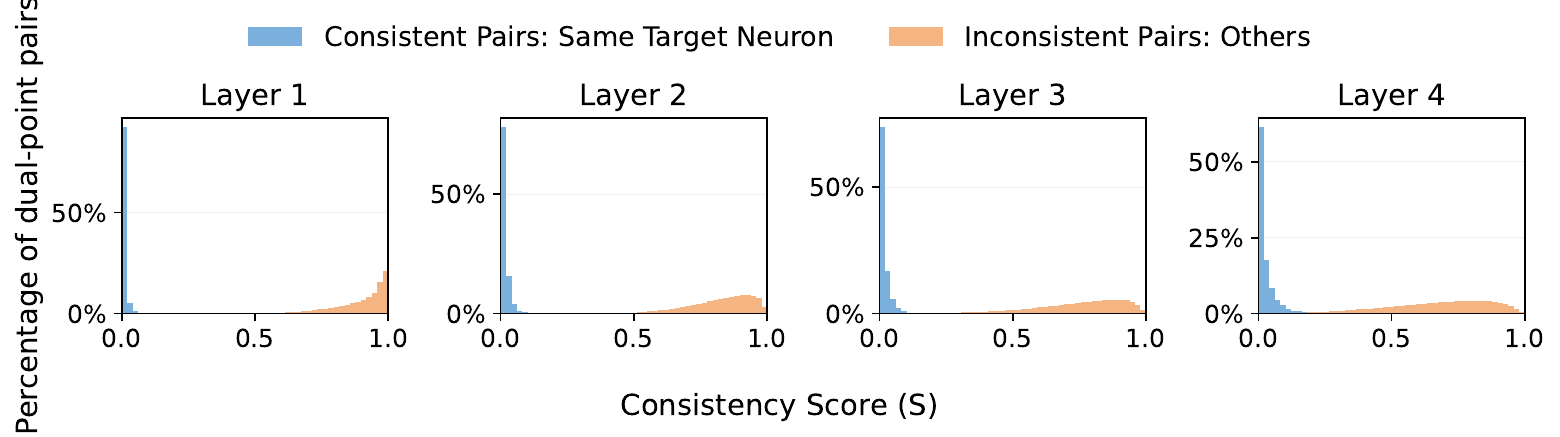}
    \caption{Distribution of the Consistency Scores $\mathcal{S}$ of ASV pairs from the same versus different neurons across DNN layers}
    \label{fig:combined_consistency_distribution}
\end{figure}

To verify the practical performance, we conducted an experiment on {\em 3072-FCN}. As illustrated in Fig. \ref{fig:combined_consistency_distribution}, evaluating the consistency score $\mathcal{S}$ on pairs of dual points corresponding to the \emph{same} neuron yields a significantly different distribution compared to pairs originating from \emph{different} neurons. This separation on distribution demonstrates that the ASV-based metric provides a reliable criterion for clustering dual points.

We observe a performance threshold regarding the shared active dimensions: when the number of shared active dimensions ($\bar{\Lambda}^{(k-1)}$ defined in Eq. \eqref{eq:nl and nr} in Sect. \ref{sect:rank_check}) is small ($|\bar{\Lambda}^{(k-1)}|<10$), the clustering precision is severely degraded (20.03\% for Layer 3 and 25.81\% for Layer 4). Conversely, when the shared dimensions $|\bar{\Lambda}^{(k-1)}|$ exceed this threshold, the precision leaps to 98.92\% and 94.60\%, respectively. 
Therefore, only the pairs of dual points with $|\bar{\Lambda}^{(k-1)}| \geq 10$ will be tested by the the ASV consistency check, and about 91.10\% and 78.58\% dual points belonging to Layers 3 and 4 are  clustered after the ASV consistency check, and the remain dual points are discarded. 

\paragraph{Remark on The Choice of the Small Threshold $\tau$.} 
As shown in Fig. \ref{fig:combined_consistency_distribution}, the small threshold $\tau$ serves as a separation boundary between the distributions of consistent and inconsistent pairs. An excessively large $\tau$ admits false positives, 
whereas an overly strict $\tau$ yields false negatives.  
In practice, $\tau$ can be determined empirically. We first experiment on a small subset of dual points.  
By computing the consistency score $\mathcal{S}$ for all pairs within this subset, we analyze their distribution.
Similar to the phenomenon shown in Fig.~\ref{fig:combined_consistency_distribution}, but without access to ground-truth neuron indices, the scores naturally form a distinct bimodal distribution.
This natural separation allows us to select a threshold $\tau$ that cleanly divides the two groups. For example, the 
Fig.~\ref{fig:combined_consistency_distribution} indicates a sound $\tau=0.2$. 
Then we apply this $\tau$ to cluster the full dataset of dual points.


\subsection{The Steps and Complexities of NRC and ASV Clustering}
Both NRC and ASV share the following steps: 
\begin{itemize}
    \item {\bf Step 1. Filtering and Normal Extraction.} Given a collection of dual points, first filter out the points of previous layers. For each remaining point, compute $\vec{n}_{\ell}$ and $\vec{n}_r$ of its adjacent decision boundary patches. For ASV method,  calculate the ASV $\vec{v} = \vec{n}_{\ell} - (\vec{n}_{\ell} \cdot \vec{n}_r^{\sf T})\vec{n}_r$.

\item {\bf Step 2. Seed Selection and Testing.} 
Randomly select a unclustered dual point from the candidate set, and evaluate it with all other points using NRC or ASV methods.  

\item {\bf Step 3. Cluster and Remove.} 
All dual points that yield a minimum singular value  $\sigma_{\min}<\tau_{\mbox{\tiny NRC}}$ for NRC or consistency score $\mathcal{S}< \tau_{\mbox{\tiny ASV}}$ for ASV  
are grouped into the same cluster as the seed. Once clustered, these points are strictly removed from the candidate pool. We then iteratively repeat Step 2, selecting a new seed from the shrinking pool, until all dual points are successfully partitioned.

\end{itemize}

\subsubsection{Complexity.} 
The NRC and ASV methods have similar complexities: 
\begin{itemize}
    \item \textbf{Complexity of Step 1:} Computing the normal vectors requires solving local linear equations, taking $\mathcal{O}((d^{(k)})^3)$ operations per dual point. For $n$ dual points, this preprocessing strictly bounds to $\mathcal{O}(n(d^{(k)})^3)$.
    
    \item \textbf{Complexity of Step 2 and 3:} For NRC, the cost of applying SVD to 4 vectors is $\mathcal{O}(d^{(k)})$. For ASV, the vector inner product costs $\mathcal{O}(d^{(k)})$.
    Therefore, both of these time complexities are strictly lower than the rank check given by Carlini {\em et al.} in Eq. \eqref{eqn: check consistency}, which has a time complexity of $\mathcal{O}((d^{(k)})^3)$.  
   
    For NRC and ASV, the worst-case matching scales as $\mathcal{O}(n^2 d^{(k)})$. 
    In average case, we assume the $n$ points are uniformly distributed across $d^{(k)}$ hidden neurons, each neuron corresponds to a cluster of approximately $n/d^{(k)}$ points. The clustering procedure effectively proceeds in $d^{(k)}$ rounds with 
    $\sum_{i=0}^{d^{(k)}-1} \left( n - i \frac{n}{d^{(k)}} \right) = \mathcal{O}(n d^{(k)})$ pairwise comparisons.
    Since each comparison takes $\mathcal{O}(d^{(k)})$ operations for NRC or ASV, this lowers the  average-case time complexity to $\mathcal{O}(n d^{(k)} \cdot d^{(k)}) = \mathcal{O}(n(d^{(k)})^2)$. 

\end{itemize}
Consequently, 
our method successfully reduces the clustering complexity from Carlini {\em et al.}'s original 
 $\mathcal{O}(n^2(d^{(k)})^3)$ \cite{DBLP:conf/eurocrypt/CarliniCHRS25} to  $\mathcal{O}(n(d^{(k)})^3+n (d^{(k)})^2)\approx \mathcal{O}(n(d^{(k)})^3)$ in average case or $\mathcal{O}(n(d^{(k)})^3+n^2 d^{(k)})$ in worst case by NRC or ASV .

\subsection{Comparison and Combination of the NRC and ASV Methods}\label{sect:comparison_combination_NRC_ASV} 
Both the NRC and ASV methods derive $\vec{n}_\ell$ and $\vec{n}_r$ from normal vectors already recorded during the {\em Dual Points Finding} phase \cite{DBLP:conf/eurocrypt/CarliniCHRS25}, neither method incurs additional model queries during clustering. Carlini {\em et al.}'s method \cite{DBLP:conf/eurocrypt/CarliniCHRS25} adopts heavy SVD and NRC employs a very lightweight SVD. While,  ASV method relies on efficient vector inner products rather than complex SVDs. 

\subsubsection{Dual Points Sharing the Same Decision Boundary.}  ASV method resolves a critical vulnerability in both the NRC and Carlini {\em et al.}'s \cite{DBLP:conf/eurocrypt/CarliniCHRS25} methods: if two dual points share the same decision boundary patch, they inherently share the same normal vector ({\em e.g.}, $\vec{n}_r = \vec{n}_{\ell}'$). Although these two points belong to different neurons, the shared normal vector reduces the matrix rank, causing the consistency check to fail. 
For NRC and Carlini {\em et al.}'s method \cite{DBLP:conf/eurocrypt/CarliniCHRS25}, we must identify if any pair of their normal vectors is nearly identical. In contrast, the computed ASVs directly approximate the target weight vectors in direction, it inherently eliminates the need for these redundant pairwise checks. Consequently, the ASV-based clustering is faster than NRC.

\subsubsection{Conditions of ASV and the Combination of NRC and ASV.} 
According to Proposition \ref{pro:expectation_dgap}, the ${\sf DGap}(\vec{v},\vec{\omega})$ of the ASV method becomes larger when $d^{(k)}$ decreases, which may lead to ineffective distinguishing between consistent dual point pairs and inconsistent pairs. 
Also ASV becomes ineffective when the the number of shared active neurons is extremely small ({\em e.g.}, $|\bar{\Lambda}^{(k-1)}|<10$). Depending on the specific network architecture, extreme cases may arise when the network is relatively narrow. 

For example, in our experiment on the relatively narrow network, a 64-64$\times$4-10 FCN, after trained on the MNIST dataset, 
{the average {\sf DGap} between ASV $\vec{v}$ and its underlying neuron weight $\vec{w}$ rise to around $0.055$-$0.066$ in Layers 2 and 3, compared to around $0.008$ for the wide {\em 3072-FCN} network in Table~\ref{tab:dgap_statistics} in {\sf Supp.} \ref{supp:DGap_verification_on_FCN}, leading to a higher false positive rate for the ASV method.} Favorable conditions arise at Layer 1, where ASVs are computed directly from the input-space normals $\vec{m}_\ell$ and $\vec{m}_r$, avoiding errors from recovering normals through $F^{k-1}$, and at Layer 4, where there are no deeper hidden neurons whose dual points could be mistakenly clustered into the target layer. 

In general, ASV is particularly suitable for networks with high-dimensional weight spaces according to Proposition \ref{pro:expectation_dgap}. 
For example, {targeting the {\em 3072-FCN}, ASV clusters 100,000 dual points and successfully recovers 100\% and 92\% of the target neurons for the first two layers (achieving partial black-box signature recovery in Table \ref{tab:Total_3072FCN_Experiments}), identical to absolutely correct results using white-box clustering in Carlini {\em et al.}'s proof-of-concept experiment \cite{DBLP:conf/eurocrypt/CarliniCHRS25}.}
The NRC method does not rely on the high-dimensional weight vectors and provides a robust consistency check applicable to a wide range of narrow or wide architectures.

In  narrow network like 64-64$\times$4-10, NRC  maintains a precision of over 99\%. Therefore, in practical experiment on 64-64$\times$4-10, a combined method of ASV-NRC is applied, {\em i.e.},  the Layer 1 and 4 are handled with ASV method and Layer 2 and 3 are handled with NRC method.

\subsubsection{Experimental Comparisons between NRC,  ASV, and Carlini \emph{et al.}'s Methods.}
The experiment is performed on the {\em 3072-FCN}. We evaluate the methods on sets of 200, 1000, and 2000 dual points. These sets comprise dual points of different neurons across 4 layers. We record the runtime and accuracy, including precision and recall. Note that, due to the substantial runtime overhead of Carlini \emph{et al.}'s method,  
we only run their algorithm up to 200 dual points. For larger sets, their clustering performances are estimated by  multiplying the time of one pairwise comparison with the number of pairwise comparisons, denoted by ``$\approx$'' as shown in Table \ref{tab:clustering_comparison}.
\begin{table}\footnotesize
\centering
\caption{Comparison of clustering methods for layer 1. $T$ denotes the computational runtime in seconds; $n$ represents the number of accumulated dual points.}
\label{tab:clustering_comparison}
\setlength{\tabcolsep}{8pt} 
\begin{tabular}{clcccc}
\toprule
$n$ & Approach & $T \:(s)$ & Comparisons & Precision & Recall \\
\midrule
\multirow{2}{*}{200} & Carlini's  & $2^{15.37}$ &                                                    $2^{14.17}$ & $100\%$ & $100\%$ \\
                     & NRC & $2^{1.83}$ &
                     $2^{14.17}$ & $100\%$ & $100\%$ \\
                     & ASV        & $2^{-0.48}$ &
                     $2^{14.17}$ & $100\%$ & $100\%$ \\
\midrule
\multirow{2}{*}{1000}& Carlini's  & $\approx2^{19.75}$  & $\approx2^{18.55}$ & - & - \\
                     & NRC & $2^{6.15}$ &
                     $2^{18.55}$ & $100\%$ & $100\%$ \\
                     & ASV        & $2^{3.96}$ &
                     $2^{18.55}$ & $100\%$ & $99.54\%$ \\
\midrule 
\multirow{2}{*}{2000}& Carlini's  & $\approx2^{21.54}$  & $\approx2^{20.34}$ & - & - \\
                     & NRC & $2^{8.05}$ &
                     $2^{20.35}$ & $100\%$ & $100\%$ \\
                     & ASV        & $2^{6.09}$ &
                     $2^{20.34}$ & $100\%$ & $99.90\%$ \\
\bottomrule
\end{tabular}
\end{table}


\paragraph{Remarks: Refining Clusters after NRC or ASV Clustering.}
\begin{enumerate}
    \item {\em False Positive.} 
    Although the dual points are partitioned into distinct clusters, each cluster may contain a minor fraction of false positives. This case also happens to 
    Carlini {\em et al.} \cite{DBLP:conf/eurocrypt/CarliniCHRS25}, and they apply a secondary filter that evaluates triples of supposedly consistent dual points, discarding the entire triplet if an inconsistency is detected. This verification strategy can also be integrated into both the NRC and ASV methods. Moreover, we propose another trick: since the number of candidate points assigned to each cluster is relatively small, we can identify potential false positives by inspecting the residuals of the initial least-squares solution to Eq.~\eqref{eqn: check consistency} when recovering the signatures. We then discard the rows with anomalous residuals and recompute $g_tA^{(k)}_t$ using the remaining equations.
    \item {\em False Negative.} In a small fraction of cases, a few dual point pairs from the same target-layer neuron exhibit relatively high minimum singular values $\sigma_{\min}$ or consistency scores $\mathcal{S}$, rendering them indistinguishable from different-neuron pairs under these metrics. We collect more dual points to ensure a sufficient number of valid samples within each target neuron cluster. {\em E.g.}, in Layer 4 of {\em 3072-FCN}, we observe that both the NRC and ASV  require about 1.2 times as many dual points as the original clustering of Carlini \emph{et al.}'s method to yield a comparable number of clustered candidate points. 
\end{enumerate}

\section{Kernel-Centric ASV in Hard-label CNN}
\label{sect:ASM in CNN Extraction}

\subsection{\emph{Dual RPCP} and \emph{Dual PSP}}
In the hard-label CNN, similar to Eq. \eqref{eq:decision boundary}, the input $X^{(1)}$ corresponding to a point on the decision boundary satisfies 
\begin{equation}
    \label{eq:CNN decision boundary}\small
    \mathcal{F}_i(X^{(1)})-\mathcal{F}_j(X^{(1)})=(G_i^{k+1}-G_j^{k+1}) P^{(k)} I^{(k)} ( A^{(k)}X^{(k)} + B^{(k)} ) + B_i^{k+1}-B_j^{k+1}=0.
\end{equation}
Then, we define the \emph{Dual RPCP} and \emph{Dual PSP} in a similar way. 
\begin{definition} 
\label{def:dual RPCP}
    {\bfseries\upshape (Dual ReLU-Pooling Critical Point, Dual RPCP)}
    Dual RPCP is in the intersection of a critical hyperplane of RPCP and the decision boundary patches on both sides. So, they meet Eq. \eqref{eq:CNN decision boundary} and Pro. \ref{pro:rpcp}. 
\end{definition}
Like Def. \ref{def:dual space} in FCNs, the Dual RPCP space is defined as the subspace spanned by Dual RPCPs of the same neuron and the same decision boundary.
Suppose that a Dual RPCP $X^{(1)}$ corresponds to the $t$-th neuron in the $k$-th layer, covered by the $u$-th pooling window. Since the value of the neuron $t$ is 0, the corresponding activation matrix can be either $I_{+}^{(k)}$ or $I_{-}^{(k)}$, differing only at the $t$-th diagonal entry ({\em e.g.}, $1$ in $I_{+}^{(k)}$ and $0$ in $I_{-}^{(k)}$). By subtracting the corresponding Eq. \eqref{eq:CNN decision boundary}, we have
\begin{equation}
    \label{eq:Dual RPCP Space}
    (G_i^{k+1}-G_j^{k+1}) P^{(k)} (I^{(k)}_{+}-I^{(k)}_{-}) ( A^{(k)}X^{(k)} + B^{(k)} )=0.
\end{equation}
For any Dual RPCP $X^{(1)}$ and $X'^{(1)}$ sampled from the same Dual RPCP Space, by subtracting the corresponding Eq. \eqref{eq:Dual RPCP Space}, we have
\begin{equation}
    \label{eq:Dual RPCP difference}
    (G_i^{k+1}-G_j^{k+1}) P^{(k)} (I^{(k)}_{+}-I^{(k)}_{-}) A^{(k)} F^{k-1} (X^{(1)}-X'^{(1)}) =0. 
\end{equation}
According to Eq.~\eqref{eqn:P_matrix} in {\sf Supp.} \ref{sect:def_notations},
if the neuron $t$ is selected by the $u$-th pooling window, the $u$-th row vector $P^{(k)}_u$ of the pooling matrix $P^{(k)}$ contains a ``1'' at its $t$-th position and ``0''s elsewhere. The term $I_{+}^{(k)}-I_{-}^{(k)}$ yields a diagonal matrix with a single non-zero entry (value 1) at the entry $(t,t)$, effectively acting as a selection operator. Consequently, the product $P^{(k)}(I_{+}^{(k)} - I_{-}^{(k)})$ extracts the $t$-th column of the pooling matrix $P^{(k)}$. Since neuron $t$ is covered by exactly one pooling window, this $t$-th column contains a single ``1'' at its $u$-th row. Let the row vector $G_i^{k+1}-G_j^{k+1} = (g_1, g_2, \cdots, g_{d^{(k+1)}})$. Then, we have:
\begin{equation}
\label{eq:gr At}
    (G_i^{k+1}-G_j^{k+1}) P^{(k)} (I^{(k)}_{+}-I^{(k)}_{-}) A^{(k)}=g_u A^{(k)}_t.
\end{equation}
Let $\vec{\delta}=X^{(1)}-X'^{(1)}$, Eq. \eqref{eq:Dual RPCP difference} can be simplified as: $g_u A^{(k)}_t F^{k-1} \vec{\delta} = 0$.  
Sample $m$ directions 
$S=({\vec{\delta}_{0}}, {\vec{\delta}_{1}},\cdots, {\vec{\delta}_{m-1}})^{\sf T}$ from the same Dual RPCP Space to construct a system and recover $g_u A_{t}^{(k)}$. Specifically,
\begin{align}\label{eqn:rpcp_At}
   S {F^{k-1}}^{\sf T} g_u {A_{t}^{(k)}}^{\sf T} = 0.
\end{align}

\begin{definition}
    \label{def:Dual PSP}
    {\bfseries\upshape (Dual Pooling Switching Point, Dual PSP)}
    The Dual PSP is in the intersection of a critical hyperplane of PSP and the decision boundary patches on both sides. Thus, they fulfill Eq. \eqref{eq:CNN decision boundary} and Pro. \ref{pro:psp}. 
\end{definition}

Similarly, Dual PSP Space is defined as the subspace spanned by Dual PSPs corresponding to the same two competing neurons and the same decision boundary.
Assume that the $u$-th ($0\leq u\leq d^{(k+1)}-1$) pooling window includes neurons $p$ and $q$ ($p<q$) whose input values are $Y^{(k)}[p]$ and $Y^{(k)}[q]$ respectively, and that they share the same maximum value.
According to Eq.~\eqref{eqn:P_matrix} in {\sf Supp.} \ref{sect:def_notations}, each row $P^{(k)}_u$ of the pooling matrix $P^{(k)}$ contains a single ``1'' to indicate the selection of the maximum value for the $u$-th pooling window. Specifically, if the $p$-th neuron is selected, the row is $(0\cdots 0, 1_p, 0 \cdots0)$. Since $Y^{(k)}[p] = Y^{(k)}[q]$, the max pooling can arbitrarily select either neuron $p$ or neuron $q$ without altering its output. 
We can represent these two equivalent selecting operations using two distinct pooling matrices, $P_{+}^{(k)}$ and $P_{-}^{(k)}$, which differ only in their $u$-th row. Specifically, in $P_{+}^{(k)}$, this row is $(0\cdots 0, 1_p, 0 \cdots0, 0_q,0\cdots0)$, while in $P_{-}^{(k)}$, the same row becomes $(0 \cdots0, 0_p, 0\cdots 0,$ $ 1_q, 0 \cdots 0)$. Consequently, the pooling matrix can be either $P_{+}^{(k)}$ or $P_{-}^{(k)}$. By substituting $P_{+}^{(k)}$ and $P_{-}^{(k)}$ into Eq.~\eqref{eq:CNN decision boundary} respectively and subtracting the two resulting equations, we have
\begin{align}
    \label{eq:Dual PSP space}
  (G_i^{k+1} - G_j^{k+1})(P_{+}^{(k)} - P_{-}^{(k)}) I^{(k)} ( A^{(k)}X^{(k)} + b^{(k)} )=0. 
\end{align}
For any Dual PSP $X^{(1)}$ and $X'^{(1)}$ sampled from the same Dual PSP space, like Eq. \eqref{eq:dual space difference} in DNN, subtracting the corresponding Eq. \eqref{eq:Dual PSP space}, we have
\begin{equation}
    \label{eq:Dual PSP difference}
    (G_i^{k+1}-G_j^{k+1}) (P^{(k)}_{+} - P^{(k)}_{-})I^{(k)} A^{(k)} F^{k-1} (X^{(1)}-X'^{(1)}) =0.
\end{equation}
Note that the $u$-th row of  $P_{+}^{(k)} - P_{-}^{(k)}$ is $(0 \cdots0, 1_p, 0\cdots0, -1_q,0\cdots0)$, and the other rows are zero vectors; let  $G_i^{k+1}-G_j^{k+1}=$$(g_1, g_2, \cdots,$ $ g_{d^{(k+1)}})$, then
\begin{align}
\label{eq:gr Ap-Aq}
    (G^{k+1}_i - G^{k+1}_j) (P_{+}^{(k)} - P_{-}^{(k)}) I^{(k)} A^{(k)}
    =g_u (A_{p}^{(k)} - A_{q}^{(k)}),
\end{align}
Let $\vec{\delta}=X^{(1)}-X'^{(1)}$, Eq. \eqref{eq:Dual PSP difference} can be simplified as: $g_u (A^{(k)}_p-A^{(k)}_q) F^{k-1} \vec{\delta} = 0$. 
Sample $n$ distinct directions 
$S=({\vec{\delta}_0}, {\vec{\delta}_1},\cdots, {\vec{\delta}_{n-1}})^{\sf T}$ from the same Dual PSP space to construct a system of equations and recover $g_u (A_{p}^{(k)} - A_{q}^{(k)})$: 
\begin{align}\label{eqn:psp_Aij}
   S {F^{k-1}}^{\sf T} \cdot g_u (A_{p}^{(k)} - A_{q}^{(k)}) = 0.
\end{align}
We refer to Dual RPCPs and Dual PSPs as \emph{Dual Points}, and denote their corresponding Dual RPCP and Dual PSP Spaces simply as the \emph{Dual Spaces}.

\subsection{Kernel-Centric ASV Method}
\label{sect:Improved ASM}

\subsubsection{Limitations of {\em Neuron-Centric Clustering} for CNN.} 
Recall from Eq. \eqref{eqn: check consistency} in  the {\em Clustering step}  of Sect. \ref{subsec:Carlini's algebraic view on DNN} that the clustering principle is to ensure  Eq. \eqref{eqn: check consistency} admits a unique non-trivial solution $g_t {A_{t}^{(k)}}^{\mathsf{T}}$. Similarly clustering strategy applied to CNN is to cluster dual points and build similar system of Eq. \eqref{eqn: check consistency} from  Eq.~\eqref{eqn:rpcp_At} for Dual RPCPs or from Eq.~\eqref{eqn:psp_Aij} for Dual PSPs, and to ensure the system admits a unique non-trivial solution. Hence, 
one must collect at least two dual points belonging to the same category to make the system solvable, {\em i.e.}, the {\em neuron-centric clustering}.  
Specifically, it requires at least two Dual RPCPs associated with the same $t$-th neuron to recover $g_u A^{(k)}_t$ of Eq. \eqref{eqn:rpcp_At}, or at least two Dual PSPs corresponding to the same competing pair of neurons $(p, q)$ within the same $u$-th pooling window to recover $g_u (A^{(k)}_p - A^{(k)}_q)$ of Eq. \eqref{eqn:psp_Aij}.

Taking a toy layer with all symbols given in 
Fig. \ref{fig:symbol_cnn_supp} of  {\sf Supp.} \ref{supp:cnn_detail} as an example\footnote{The readers are suggested to read  {\sf Supp.} \ref{sec:toy_example_cnn} to better understand all the symbols.}: 
there are $d^{(k)}_{f}=36$ distinct types of Dual RPCPs corresponding to $36$ neurons. For Dual PSPs, since each $2 \times 2$ pooling window contains $\binom{4}{2}= 6$ possible competing neuron pairs, there are in total $d^{(k+1)} \times \binom{h_m^{(k)}\cdot w_m^{(k)}}{2} = 54$ distinct types of Dual PSPs in the $k$-th layer. 
If the clustering process is  {\em neuron-centric}---treating each of these $36 + 54 = 90$ categories as completely independent---then according to the Pigeonhole Principle, in the worst case, 
one might collect at least $91$ dual points to guarantee the construction of one  solvable linear system from Eq. \eqref{eqn:rpcp_At} or Eq. \eqref{eqn:psp_Aij} 
to solve $A^{(k)}_t$ or $A^{(k)}_p - A^{(k)}_q$. 
Although all 91 dual points share the {\em same} convolutional kernel, isolated linear systems fail to jointly utilize them trivially, thereby wasting most dual points. Note that, the extraction processes introduced by Carlini {\em et al.} \cite{DBLP:conf/eurocrypt/CarliniCHRS25} on FCN and Sun {\em et al.} \cite{cnn_average_pooling} on CNN are both {\em neuron-centric}. This severe inefficiency naturally raises a critical question: 
        \textit{Can we establish a ``kernel-centric'' clustering and recovery framework that exploit the weight sharing property in CNN, rather than treating each neuron independently?}

\subsubsection{Weight Sharing Property in CNN.} 
According to Eq. \eqref{eqn:A_row} in {\sf Supp}.~\ref{sect:def_notations},  $A_{i}^{(k)}$ is a sparse vector which only contains non-zero entries at the indices where the convolutional kernel is applied.  This sparsity allows us to explicitly extract the positional indices of the convolutional kernel's local receptive field (LRF-C)---defined as the specific sub-matrix of the input matrix $IN^{(k)}$, which is multiplied by the same kernel matrix $C^{(k)}$ according to {\sf Supp}.~\ref{sect:def_notations}. 
Denote the vector form of $C^{(k)}$ as $\vec{C}^{(k)}=(C^{(k)}_0, C^{(k)}_1,\dots,C^{(k)}_{h_{c}^{(k)}-1})^{\sf{T}}$. Thus, the length of $\vec{C}^{(k)}$ is $\ell_{c}^{(k)} = h^{(k)}_{c} w^{(k)}_{c}$. Assume  the stride $s_c^{(k)}=1$, and let $\mathcal{K}^{(k)}_t \subset \{0,1,\dots, d^{(k)}-1\}$ denote the set of indices belonging to the LRF-C of the $t$-th neuron ($0\leq t\leq d^{(k)}_f-1$), then
\begin{equation}  \label{eq:indices K}\small
    \mathcal{K}_t^{(k)} \mathrel{:=} \left\{w_{in}^{(k)} \cdot \left\lfloor \frac{t}{w^{(k)}_o} \right\rfloor + t \bmod w^{(k)}_o 
   + w_{in}^{(k)} \cdot \left\lfloor \frac{s}{w^{(k)}_{c}} \right\rfloor + s \bmod w^{(k)}_{c},\ 0 \le s < \ell_{c}\right\}.
\end{equation}
For Dual RPCPs, recall from Eq. \eqref{eqn:rpcp_At} that $g_u A^{(k)}_t F^{k-1} \vec{\delta} = 0$, the set $\mathcal{K}^{(k)}_t$ picks out the nonzero entries of $A^{(k)}_t$ according to Eq. \eqref{eqn:A_row}. 
Then,
\begin{align}
\label{eq:Dual RPCP difference to equation}
    g_uA_{t}^{(k)} (F^{k-1} \vec{\delta}) = g_u\vec{C}^{(k)} (F^{k-1} \vec{\delta})[\mathcal{K}^{(k)}_t]=0. 
\end{align}
Denote an empty set $\mathcal{K}^{(k)}_{\emptyset}=\emptyset$, then Eq. \eqref{eq:Dual RPCP difference to equation} can be rewritten as:
\begin{equation}
\label{eq:Dual RPCP difference 2}
    g_uA_{t}^{(k)} (F^{k-1} \vec{\delta}) = g_u\vec{C}^{(k)} ((F^{k-1} \vec{\delta})[\mathcal{K}^{(k)}_t]-(F^{k-1} \vec{\delta})[\mathcal{K}^{(k)}_{\emptyset}])=0. 
\end{equation}
For Dual PSPs, recall from Eq. \eqref{eqn:psp_Aij} that $g_u (A^{(k)}_p-A^{(k)}_q) F^{k-1} \vec{\delta} = 0$, the set $\mathcal{K}^{(k)}_p$ and $\mathcal{K}^{(k)}_q$ picks out the nonzero entries of $A^{(k)}_p$ and $A^{(k)}_q$ respectively.
Then,
\begin{align}
    \label{eq:Dual PSP difference to equation}
    g_u(A^{(k)}_p-A^{(k)}_q) (F^{k-1} \vec{\delta}) = g_u\vec{C}^{(k)} \cdot ((F^{k-1} \vec{\delta})[\mathcal{K}^{(k)}_p]-(F^{k-1} \vec{\delta})[\mathcal{K}^{(k)}_q])=0. 
\end{align}
Therefore, Eq. \eqref{eq:Dual RPCP difference 2} for  Dual RPCPs and Eq. \eqref{eq:Dual PSP difference to equation} for Dual PSPs share the same unknown vectors $g_u\vec{C}^{(k)}$ due the weight sharing property of CNN. 
From a collection of dual points, if we can recognize Dual RPCPs and Dual PSPs, and identify
the critical neuron $t$ for Dual RPCPs and the competing neurons $p$ and $q$ for Dual PSPs, then the $\mathcal{K}^{(k)}_t$, $\mathcal{K}^{(k)}_p$, and $\mathcal{K}^{(k)}_q$ will be determined.  
Hence, Eqs. \eqref{eq:Dual RPCP difference 2} and \eqref{eq:Dual PSP difference to equation} can be combined with the same unknown $g_u\vec{C}^{(k)}$, which are solved  using the SVD method to recover $\vec{C}^{(k)}$ up to a scalar multiple.

\subsubsection{The ASV Method to Identify the Dual RPCPs and Dual PSPs.}
Chen {\em et al.} introduced the identification method for RPCPs and PSPs in a raw-output CNN \cite{DBLP:journals/iacr/ChenTGSQD26}, based on the structural sparsity of $\vec{\omega} = A^{(k)}_t$ (or $\vec{\omega}=A^{(k)}_p-A^{(k)}_q$), according to Eq.~\eqref{eqn:A_row} in {\sf Supp.} \ref{sect:def_notations}. In hard-label CNN, we introduce the ASV method to identify different dual points. 

\begin{enumerate}
    \item For Dual RPCPs, denote row vectors
    \begin{equation} \label{eq:RPCP G_vectors}\small
    \begin{aligned}
    \vec{G}_{\ell} = (G_i^{k+1} - G_j^{k+1}) P^{(k)} I_{+}^{(k)} A^{(k)}, 
    \quad
    \vec{G}_r = (G_i^{k+1} - G_j^{k+1}) P^{(k)} I_{-}^{(k)} A^{(k)}.
    \end{aligned}
    \end{equation}
    Recall from Eq. \eqref{eq:gr At} that, the scaled row vector $g_u A^{(k)}_t$ of the convolutional matrix $A^{(k)}$, is the difference between $\vec{G}_{\ell}$ and $\vec{G}_r$, {\em i.e.}, $g_u A^{(k)}_t=\vec{G}_{\ell}-\vec{G}_r$.
    
    \item Similarly, for Dual PSPs, denote row vectors
    \begin{equation} \label{eq:PSP G_vectors}\small
    \begin{aligned}
    \vec{G}_{\ell} = (G_i^{k+1} - G_j^{k+1}) P_{+}^{(k)} I^{(k)} A^{(k)}, 
    \quad
    \vec{G}_r = (G_i^{k+1} - G_j^{k+1}) P_{-}^{(k)} I^{(k)} A^{(k)}. 
    \end{aligned}
    \end{equation}
    Recall from Eq. \eqref{eq:gr Ap-Aq} that, the difference of the scaled row vectors $g_u (A^{(k)}_p-A^{(k)}_q)$, is the difference between $\vec{G}_{\ell}$ and $\vec{G}_r$, {\em i.e.}, $g_u (A^{(k)}_p-A^{(k)}_q)=\vec{G}_{\ell}-\vec{G}_r$.
\end{enumerate}

For any dual point in a hard-label CNN (including both Dual RPCPs and Dual PSPs),  compute the unit normal vectors of the adjacent decision boundary patches as $\vec{n}_{\ell}=\frac{\vec{G}_{\ell}}{\lVert \vec{G}_{\ell} \rVert}$ and $\vec{n}_r=\frac{\vec{G}_r}{\lVert \vec{G}_r \rVert}$. The ASV in the hard-label CNN is then  similarly defined by Eq. \eqref{eq:signature vector}. 
Let $\vec{\omega} = \vec{G}_{\ell} - \vec{G}_r$ be the neuron weight ($g_u A^{(k)}_t$ for Dual RPCPs, or $g_u (A^{(k)}_p-A^{(k)}_q)$ for Dual PSPs). Then, Eq.~\eqref{eq:signature vector} becomes $\vec{v}= \frac{1}{\lVert \vec{G}_{\ell} \rVert}(\vec{\omega}-(\vec{\omega} \cdot \vec{n}_r^{\sf T})\vec{n}_r)$. Similarly, we measure the directional distance {\sf DGap}($\vec{v}$,$\vec{\omega}$)  between $\vec{v}$ and $\vec{\omega}$ by Eq. \eqref{eq:direction_gap}.  
According to our experiments given in {\sf Supp.} \ref{sect:experiment_cnn_asv_w}, the  {\sf DGap}($\vec{v}$,$\vec{\omega}$) is close to 0. 
Moreover, $\vec{v}$ strictly preserves the structural sparsity due to the localized connectivity of CNNs. This allows us to verify dual points and identify the corresponding neurons by examining the consistency of their sparsity. An experiment to verify the sparsity alignment between $\vec{v}$ and $\vec{\omega}$ is given in {\sf Supp}. \ref{sect:cnn_v_w_test}.

With known architecture of a CNN, we could iteratively construct $A^{(k)}_t$ for each $0 \leq t \leq d^{(k)}_f-1$ according to Eq.~\eqref{eqn:A_row} in $\sf Supp$. \ref{sect:def_notations}, thereby determining the locations of zero and non-zero values. Then, for Dual RPCPs, verify if $A^{(k)}_t$ and $\vec{v}$ are of the same sparsity consistency:  

\begin{enumerate}
    \item \textbf{Non-zero-Entry Consistency}: At any index where $A_{t}^{(k)}$ is non-zero, the corresponding element in the ASV $\vec{v}$ should be either non-zero or {\sf NaN}.
    
    \item \textbf{Zero-Entry Consistency}: At any index where $A_{t}^{(k)}$ is zero, the corresponding element in $\vec{v}$ should be either $0$ or {\sf NaN}.
\end{enumerate}
Similarly, for Dual PSPs, build $A^{(k)}_p$ and $A^{(k)}_q$ with Eq.~\eqref{eqn:A_row} by enumerating all possible competing neuron pairs $(p, q)$, and 
verify whether $A^{(k)}_p-A^{(k)}_q$ and $\vec{v}$ share the same sparsity consistency:
\begin{enumerate}
    \item \textbf{Union Non-zero Consistency}: At any index where either $A_{p}^{(k)}$ or $A_{q}^{(k)}$ is non-zero, the corresponding element in $\vec{v}$ should be either non-zero or {\sf NaN}.
    
    \item \textbf{Common Zero Consistency}: At any index where both $A_{p}^{(k)}$ and $A_{q}^{(k)}$ are zero, the corresponding element in $\vec{v}$ should be either $0$ or {\sf NaN}.
\end{enumerate}
In practice, the consistency checks are performed for Dual RPCPs before proceeding to Dual PSPs. At last, a dual point is identified as Dual RPCP or Dual PSP, together with the deduced    $\mathcal{K}^{(k)}_t$, or  $\mathcal{K}^{(k)}_p$, $\mathcal{K}^{(k)}_q$. Hence, Eqs. \eqref{eq:Dual RPCP difference 2} and \eqref{eq:Dual PSP difference to equation} can be combined and solved to determine $\vec{C}^{(k)}$ up to a scalar multiple.

\section{Black-box Experiments On Signature Recovery}
\label{Experiments}


\subsubsection{Description of the Implementation.}

In practice, our signature recovery attack\footnote{Our paper focuses on the signature recovery phase, the sign recovery method is the same to Carlini {\em et al.} \cite{DBLP:conf/eurocrypt/CarliniCHRS25}.}   proceeds layer by layer. For FCNs, when targeting a specific layer, we assume all preceding layers are extracted and directly {\em use their white-box parameters}. The reason is that when replaying Carlini {\em et al.}'s sign recovery \cite{DBLP:conf/eurocrypt/CarliniCHRS25} on 64-64$\times$4-10 network, it does not completely and correctly recover all the signs in the deep layers. This issue also leaves an open question on how to invent a more robust sign recovery method. 

For CNNs, we use the actual empirical parameters we recovered from the preceding layers during our experiments. Note that for CNNs, sign recovery is inherently straightforward, as described in \cite[Sect. 4.3 and 5.3]{DBLP:journals/iacr/ChenTGSQD26}. We provide the sign recovery process in hard-label CNNs in {\sf Supp}. \ref{supp:cnn_sign_recovery}. 

The signature recovery for each layer includes 3 phases: dual point finding, clustering, signature computing. In Carlini {\em et al.}'s proof-of-concept experiment \cite{DBLP:conf/eurocrypt/CarliniCHRS25}, they use white-box gradient access for both dual point finding and  signature computing phases, and rely on white-box neuron indices to cluster the dual points. 
In contrast, our signature recoveries on the 64-64$\times$4-10 FCN and all CNNs are black-box, while only the dual point finding for the {\em 3072-FCN} uses white-box gradient access (thus partial black-box). 
Most experiments are conducted on a server equipped with an NVIDIA RTX 5080 GPU (16 GB memory). 




\subsubsection{Hard-label FCN Extraction By the ASV-NRC Combined Clustering.}
The FCN with a 64-64$\times$4-10 architecture is trained on MNIST dataset. 
As shown in Table \ref{tab:Total Experiments} in Sect. \ref{sect:Introduction}, to extract the 1st layer, about 3000 dual points are needed to cluster. Carlini \emph{et al.}'s method \cite{DBLP:conf/eurocrypt/CarliniCHRS25} requires about $0.93$ hours to extract the 1st layer. In contrast, the extraction time based on our ASV method is only $0.05$ hours. 
The difference of time complexities is even more pronounced when the number of dual points increases. For example, extracting the 2nd layer requires clustering roughly $300,000$ dual points. Our NRC method  completes this in 4.22 hours, while Carlini \emph{et al.}'s method \cite{DBLP:conf/eurocrypt/CarliniCHRS25} cannot give a result within a reasonable time (over 1 month). We estimate their runtime as follows: 
We simulate the clustering process for all $300,000$ dual points using white-box neuron indices, assuming each SVD-based consistency check is correct. This full-scale clustering requires a total of $2,653,799,211$ comparison. We then evaluate the actual time cost on a smaller subset of $3,000$ dual points, which takes $3,585.770$ seconds to perform $1,790,440$ comparison. According to Sect.~\ref{subsect:Carlini's limitation}, the estimated runtime for Carlini \emph{et al.}'s method to cluster the entire dual points set is about $\frac{3,585.770\text{ s}}{1,790,440 \times 3,600 \text{ s/h}} \times 2,653,799,211= 1,476$ hours. In Table \ref{tab:Total Experiments} and \ref{tab:Total_3072FCN_Experiments}, the runtime of Carlini \emph{et al.}'s method is similarly estimated for the 64-64$\times$4-10 FCN's layer $F_3$ and  {\em 3072-FCN}'s layers $F_1$ and $F_2$. 



\subsubsection{Black-box Hard-label CNN Extraction with Kernel-Centric ASV.}
The LeNet-5, first proposed by LeCun {\em et al.} in 1998 \cite{Lecun1998}, is a foundational work in CNNs. Modern implementations typically update its classic (2+2) architecture by incorporating ReLU activations and max-pooling layers. Our practical attacks target a (2+1) CNN, and this modern LeNet-5 (specified in \textsf{Supp.} \ref{supp:model structure}) in hard-label setting with fully black-box access. 
The results 
are given in Table \ref{tab:Total Experiments}.

 \section{Conclusion}
  \label{sect:Conclusions}
This paper resolves a computation bottleneck in  Carlini {\em et al.}'s hard-label attacks on DNNs, {\em i.e.}, the dual points clustering step,  by
proposing the {\em Normal Rank Check (NRC)} and {\em Approximate Signature Vector (ASV)} to replace  Carlini {\em et al.}'s heavy SVD‑driven clustering, drastically lowering the computational overhead. Besides, for hard-label max-pooling CNNs, we present a highly efficient {\em kernel‑centric} methodology, namely {\em Kernel‑Centric ASV}, to overcome the limitations of previous {\em neuron‑centric} extraction strategies.
Experimental results on FCNs and max-pooling LeNet‑5 confirm that our method achieves high clustering precision and accelerates the extraction by several orders of magnitude.

\bibliographystyle{plain}
\bibliography{bib}

\newpage
\appendix
\renewcommand{\theHsection}{Appendix.\arabic{section}}
\section*{\centering \textsf{Supplementary Material}}


\section{Detailed Notations of CNN}
\label{supp:cnn_detail}
\subsection{Basic Definitions and Notations}\label{sect:def_notations}
The basic definitions and notations is summarized in Table \ref{tab:notations}. 

\begin{table}[]
    \centering
    \caption{Basic Definitions and Notations}
    \begin{tabular}{p{12cm}}
    \toprule
       {\bf FCN}: Fully Connected Neural Network (FCN).\\\midrule
      {\bf CNN}: Convolutional Neural Network (CNN). \\\midrule
       {\bf ReLU}: Rectified Linear Unit. \\\midrule
         {\bf FCN layer}: Including a linear function $f^{(k)}$ called fully-connected function, and 
         a nonlinear function $\sigma$ (component-wise ReLU function). \\\midrule
         {\bf CNN layer}: Including Convolutional Function $f_c^{(k)}$, nonlinear activation 
         function $\sigma_c^{(k)}$ (ReLU), and pooling function $\rho^{(k)}$ (Max Pooling). \\\midrule
    {\bf $\mathbf{(m+n)}$-Deep Convolutional Neural Network (CNN)}: {A} CNN with $m$ 
    CNN layers and $n$ FCN layers, as well as an output layer with a  
    fully-connected linear function. \\\midrule
         $\bf(\mathsf{c}_{in}^{(k)}, \mathsf{c}_{out}^{(k)}, h_{c}^{(k)}, w_{c}^{(k)})$: Convolutional Function with  $\mathsf{c}_{in}^{(k)}$ input and  $\mathsf{c}_{out}^{(k)}$ output channels, 
         and the size of kernel matrices {is} $h_{c}^{(k)}\times w_{c}^{(k)}$. \\\midrule
         $IN^{(k)}$: input matrix of $f_c^{(k)}$, $IN^{(k)}\in \mathbb{R}^{h^{k}_{in}\times w^{(k)}_{in}}$
         \\\midrule
          $C^{(k)}$: convolutional kernel matrix of layer $k$, $C^{(k)}\in \mathbb{R}^{h^{k}_{c}\times w^{(k)}_{c}}$
         \\\midrule
         $O^{(k)}$: $h^{(k)}_{o}\times w^{(k)}_{o}$ output matrix or the feature map of $f_c^{(k)}$, $O^{(k)}\in \mathbb{R}^{h^{(k)}_{o}\times w^{(k)}_{o}}$
         \\\midrule
         $I^{(k)}$: diagonal activation indicator matrix of layer $k$, $I^{(k)}\in \mathbb{F}_2^{d^{(k)}_{f}\times d^{(k)}_{f}}$
         \\\midrule
         $\hat{O}^{(k)}$: output matrix of the $k$-th non-linear activation function  $\sigma_c^{(k)}$, $\hat{O}^{(k)}\in \mathbb{R}^{h^{(k)}_{o}\times w^{(k)}_{o}}$
         \\\midrule
         $M^{(k),i,j}$: Boolean selection matrix for the $(i,j)$-th local receptive field of max pooling function in layer $k$, $M^{(k),i,j}\in \mathbb{F}_2^{h^{(k)}_{m}\times w^{(k)}_{m}}$
         \\\midrule
         $P^{(k)}$: Boolean pooling matrix of layer $k$, $P^{(k)}\in \mathbb{R}^{d^{(k+1)}\times d^{(k)}_{f}}$
         \\\midrule
 ${X^{(k)},Y^{(k)},Z^{(k)}}$:  Input vectors of the operations $f_c^{(k)}$, $\sigma_c^{(k)}$, $\rho^{(k)}$ in the $k$-th 
 CNN layer, $k\geq 1$.  $X^{(k)}\in \mathbb{R}^{d^{(k)}},Y^{(k)},  Z^{(k)} \in \mathbb{R}^{d_f^{(k)}}$, since $\sigma_c^{(k)}$ does not
 change the dimension. $X^{(k)}[i]$ is the $i$-th ($0\leq i < d^{(k)}$) entry of  $X^{(k)}$. \\\midrule
   ${A^{(k)}}$: Matrix whose entry in  $i$-th row and $j$-th is $A^{(k)}_{i,j}$, $i,j$ {start} from 0. \\\midrule
 {\bf RPCP, PSP, FCP}: ReLU-Pooling Critical Point, Pooling Switching Point, 
 Critical points in the Fully-connected Block, respectively.\\\midrule
 {\bf LRF-C}, Local Receptive Field of Convolutional Function: A sub-matrix of $IN^{(k)}$, 
 which is multiplied by {\em Convolutional Kernel Matrix} $C^{(k)}$ to produce each entry in $Y^{(k)}$. 
 The LRF-C will be different for different {elements} of $Y^{(k)}$. Let $Y^{(k)}[t]$ {correspond}
 to the $t$-th ($0\leq t\leq d^{(k)}_f-1$) LRF-C. \\\midrule
 {\bf LRF-P}, Local Receptive Field of Pooling function: A sub-matrix of $Z^{(k)}$, which is processed by a pooling function to produce each entry of $X^{(k+1)}$. The LRF-P will be different for different {elements} of $X^{(k+1)}$. Let $X^{(k+1)}[t]$ {correspond} to the $t$-th ($0\leq t\leq d^{(k+1)}-1$) LRF-P. \\\midrule
 $s_c^{(k)}:$ The convolution stride,  which is by default set to 1 in this paper.  \\
 \bottomrule
\end{tabular}\label{tab:notations}
\end{table}

\begin{figure}
	\centering
        \includegraphics[width=0.93\linewidth]{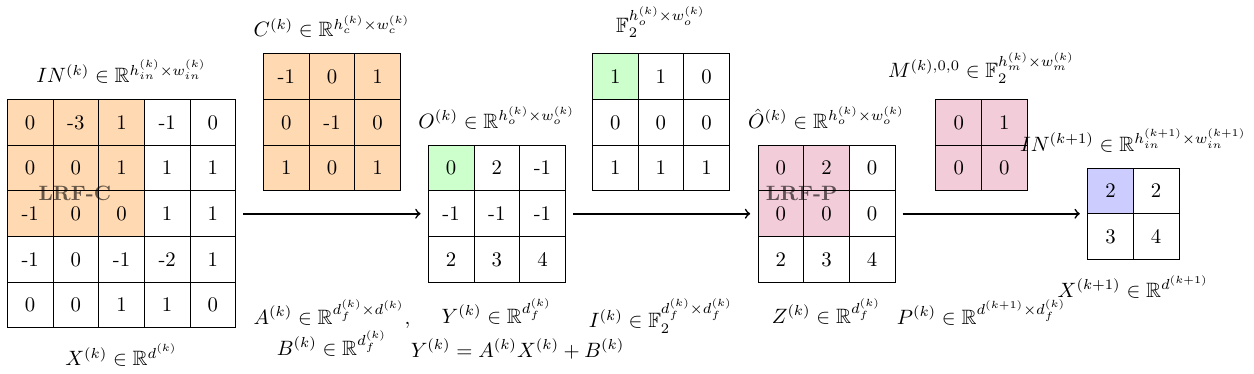}
	\caption{Symbolic View of a CNN Layer}
	\label{fig:symbol_cnn_supp}
\end{figure}
The { CNN layer} includes   a Convolutional Function $f_c^{(k)}$, a nonlinear activation 
         function $\sigma_c^{(k)}$ (ReLU), and pooling function $\rho^{(k)}$ (Max Pooling). 
\subsubsection{Convolutional Function.} The CNN's inputs are formulated as matrices. The convolution operation of the convolutional function  $f_c^{(k)}$ employs a sliding window mechanism, where a small matrix of learnable parameters (the {\em Convolutional Kernel Matrix}) traverses the input from left to right, top to bottom. At each position, the sum of element-wise products between the {\em Convolutional Kernel Matrix} and the overlapping local input patch ({\em Local Receptive Field of convolutional function}, abbreviated as {\em LRF-C}), is computed to generate a single output value. A shared bias is then added to the result at each position to constitute the output matrix ({\em Feature Map}) of the convolutional function. 

In the algebraic view in Figure \ref{fig:symbol_cnn_supp},  the input matrix of $f_c^{(k)}$ is denoted as $IN^{(k)}$  with dimension $h_{in}^{(k)} \times w_{in}^{(k)}$, so $d^{(k)} = h_{in}^{(k)} \cdot w_{in}^{(k)}$. The convolutional kernel matrix is defined by the shape $(\mathsf{c}_{in}^{(k)}, \mathsf{c}_{out}^{(k)}, h_{c}^{(k)}, w_{c}^{(k)})$, where $\mathsf{c}_{in}^{(k)}$ and $\mathsf{c}_{out}^{(k)}$ denote the number of input and output channels, and $h_{c}^{(k)}, w_{c}^{(k)}$ denote the height and width of the convolutional kernel matrix $C^{(k)}$, respectively.

For briefness, we will only describe  the simple case of single input and output channel, {\em i.e.}, $\mathsf{c}_{in}^{(k)}=\mathsf{c}_{out}^{(k)}=1$. For the case of multiple channels, please refer to \cite{DBLP:journals/iacr/ChenTGSQD26}.
Given a convolution stride $s_c^{(k)}$, the $h^{(k)}_{o} \times w^{(k)}_{o}$ output matrix or the {\em feature map} of $f_c^{(k)}$ is denoted as $O^{(k)}$, where 
    \begin{align}
        h^{(k)}_{o} = \left\lfloor \frac{h^{(k)}_{in} - h^{(k)}_{c}}{s^{(k)}_c} \right\rfloor + 1, \quad 
        w^{(k)}_{o} = \left\lfloor \frac{w^{(k)}_{in} - w^{(k)}_{c}}{s^{(k)}_c} \right\rfloor + 1.
    \end{align}

For briefness, we let the convolution stride $s_c^{(k)}=1$. The input of $f_c^{(k)}$ can be a $h_{in}^{(k)} \times w_{in}^{(k)}$ $IN^{(k)}$ or a $d^{(k)}$-dimension vector $X^{(k)}$. Denote 
$IN^{(k)}={[a^{(k)}_{i,j}]}_{h_{in}^{(k)} \times w_{in}^{(k)}}$ and the $i$-th ($0\leq i\leq h_{in}^{(k)}-1$) row $IN^{(k)}_i=(a^{(k)}_{i,0},a^{(k)}_{i,1}, \cdots, a^{(k)}_{i,w_{in}^{(k)}-1})$. Then the vector $X^{(k)} = (IN^{(k)}_0, IN^{(k)}_1,\cdots,IN^{(k)}_{h_{in}^{(k)}-1})^{\mathsf{T}}$.

Denote the kernel matrix $C^{(k)} = [c^{(k)}_{i,j}]_{{h_{c}^{(k)}} \times w_{c}^{(k)}}$, where $C^{(k)}_i=(c^{(k)}_{i,0},c^{(k)}_{i,1}, \cdots, $ $c^{(k)}_{i,w_{c}^{(k)}-1})$. 
Then convolution operation computes the output matrix $O^{(k)}$ as 
\begin{equation}
    O^{(k)}_{i,j}=
    \sum\limits _{p=0} ^{h^{(k)}_{c}-1} 
    \sum\limits _{q=0} ^{w^{(k)}_{c}-1} 
    IN^{(k)}_{i+p,j+q} C^{(k)}_{p,q} +b^{(k)}, 
\end{equation}
where the bias $b^{(k)}$ is the same for all the outputs.
In vector form, the output matrix  
$O^{(k)}$ will be a ($d^{(k)}_{f} = h^{(k)}_{o}\cdot w^{(k)}_{o}$)-dimension vector $Y^{(k)}=(O^{(k)}_{0},\cdots,O^{(k)}_{h^{(k)}_{o}-1})^{\mathsf{T}}$.

\begin{definition} \label{def:conv_matrix}
    {\bfseries\upshape (Convolutional Matrix)}
    Let $\mathsf{c}_{in}^{(k)}=\mathsf{c}_{out}^{(k)}=s_c^{(k)}=1$.    
    In the $k$-th convolutional function, given a convolution kernal matrix $C^{(k)}$, the function $f^{(k)}_c:\mathbb{R}^{d^{(k)}} \rightarrow \mathbb{R}^{d^{(k)}_{f}}$, is defined as 
    $f^{(k)}_c(X^{(k)})=A^{(k)} X^{(k)} + B^{(k)}$, where $X^{(k)} \in \mathbb{R}^{d^{(k)}}$ is the input vector.    $A^{(k)} = (A^{(k){\mathsf{T}}}_0, A^{(k){\mathsf{T}}}_1,\cdots,A^{(k){\mathsf{T}}}_{d_{f}^{(k)}-1})^{\mathsf{T}} \in \mathbb{R}^{{d_{f}^{(k)}} \times d^{(k)}}$, whose $i$-th row ($0\leq i\leq d^{(k)}_{f}-1$) is:
        \begin{equation}\label{eqn:A_row}\small
            A^{(k)}_i = (\underbrace{0, \cdots, 0}_{w^{(k)}_{in}\cdot \left\lfloor \frac{i}{w^{(k)}_{o}} \right\rfloor}, \underbrace{
        \underbrace{0, \cdots, 0}_{i\bmod w^{(k)}_o}, C^{(k)}_{0},     \underbrace{0, \dots, 0}_{padding}}_{w_{in}^{(k)}}, 
     \cdots, 
        \underbrace{
        \underbrace{0, \cdots, 0}_{i \bmod w^{(k)}_o}, C^{(k)}_{h_{c}^{(k)}-1} ,     \underbrace{0, \dots, 0}_{padding}}_{w_{in}^{(k)}}, 
        \underbrace{0, \cdots, 0}_{padding}),
        \end{equation}
     and $B^{(k)}=(b^{(k)},b^{(k)},\cdots,b^{(k)})^T \in \mathbb{R}^{d^{(k)}_{f}}$ (the bias for each neuron is the same). 
\end{definition}

\subsubsection{Nonlinear Activation Function.} 
    The $k$-th activation function $\sigma_c^{(k)}$ of a CNN consists of a set of parallel $d^{(k)}_f$ nonlinear activation  functions $\mbox{ReLU}(x_i) = \max\{x_i, 0\}$ for $i=0,\cdots, d^{(k)}_f-1$. 

The input of 
$\sigma_c^{(k)}$ function is the $h_o^{(k)}\times w_o^{(k)}$ matrix $O^{(k)}$, and we denote its output as $\hat{O}^{(k)}$ with the same dimension. In vector form, the input is denoted as $Y^{(k)}=(y_0,y_1,\cdots, y_{d_f^{(k)}-1})^{\mathsf{T}}$ and then $\sigma_c^{(k)}$ function can be expressed as a matrix $I^{(k)}$ multiplied by $Y^{(k)}$, {\em i.e.}, $Z^{(k)} =(\hat{O}^{(k)}_0,\cdots,\hat{O}^{(k)}_{h_o^{(k)}-1})^{\mathsf{T}} = \sigma_{c}^{(k)}(Y)=I^{(k)}Y^{(k)}$, where
    \begin{align}\label{eqn:sigma_I}
        I^{(k)} = 
        \begin{pmatrix}
            \tau^{(k)}_0 &0&\cdots &0\\
            0&\tau^{(k)}_1&\cdots &0\\
            \vdots &\vdots & &\vdots \\
            0&0&\cdots &\tau^{(k)}_{d^{(k)}_f-1}
        \end{pmatrix} ,~
        0\leq i \leq d^{(k)}_f-1, ~
        \tau^{(k)}_{i} = \left\{ \begin{array}{l}
        1, ~~\mbox{if} ~~~y_i>0,\\
        0, ~~\mbox{if}~~~ y_i\leq 0.
        \end{array} \right.
    \end{align}

\subsubsection{Pooling Function.}
In the pooling function,  a small fixed-size window (analogous to a convolutional kernel, and referred to as a pooling kernel) slides across a feature map (input matrix $\hat{O}^{(k)}$) with a predefined stride.  The region of the input matrix covered by the pooling window is also referred to as the {\em Local Receptive Field} of Pooling (LRF-P for short). 
For each window position, it applies a pooling function (usually use the max or average function) to compress the contents in the LRF-P into a single value, producing a dimension reduced output matrix (denoted as $IN^{(k+1)}$). Unlike convolutional kernels, pooling kernels are parameter-free (no learnable weights and biases) — their behavior is determined solely by size, stride, and compressing rule.

In the pooling function $\rho^{(k)}$, we assume that the max pooling kernel is applied to an LRF-P of size  $h^{(k)}_{m} \times w^{(k)}_{m}$ with stride $s^{(k)}_{\rho}$. The input is $h_o^{(k)}\times w_o^{(k)}$ matrix $\hat{O}^{(k)}$ in Figure \ref{fig:symbol_cnn_supp}.  The output is denoted as a $h^{(k)}_{\rho} \times w^{(k)}_{\rho}$ matrix $IN^{(k+1)}$, where:
    \begin{align*}
        h^{(k)}_{\rho} = \left\lfloor \frac{h^{(k)}_{o} - h^{(k)}_{m}}{s^{(k)}_{\rho}} \right\rfloor + 1, \quad 
        w^{(k)}_{\rho} = \left\lfloor \frac{w^{(k)}_{o} - w^{(k)}_{m}}{s^{(k)}_{\rho}} \right\rfloor + 1. 
    \end{align*} 
In particular, when the pooling kernel
 size is $h^{(k)}_{m} \times w^{(k)}_{m} = 2 \times 2$ and the stride is $s_{\rho}^{(k)}=2$, the windows partition the input into disjoint LRF-Ps, and the calculation of the output size reduces to: 
    \begin{align*}
        h^{(k)}_{\rho} = \left\lfloor \frac{h^{(k)}_{o}}{2} \right\rfloor, \quad
        w^{(k)}_{\rho} = \left\lfloor \frac{w^{(k)}_{o}}{2} \right\rfloor.
    \end{align*}
The entry of the $i$-th row and $j$-th column ($0\leq i\leq h^{(k)}_{\rho}-1$, $0\leq j\leq w^{(k)}_{\rho}-1$) of the output matrix $IN^{(k+1)}$ takes the maximum value of the elements in the $(i\cdot w^{(k)}_{\rho}+j)$-th LRF-P of $\hat{O}^{(k)}$, whose indexes are denoted as  $\Omega_{i,j}={\{(i\cdot s^{(k)}_{\rho}+ h,~ j\cdot s^{(k)}_{\rho}+ w), ~0\leq h\leq h_m^{(k)}-1, 0\leq w\leq w_m^{(k)}-1\}}$. Then, 
  
\begin{equation}\label{eqn:max_pool_1}
    IN^{(k+1)}_{i,j} = \max\{\hat{O}^{(k)}_{\Omega_{i,j}}\}. 
\end{equation}

When the input matrix $\hat{O}^{(k)}$ is given, and the max pooling operation can be expressed as multiplying a series of Boolean matrices  $M^{(k),i,j}\in \mathbb{F}_2^{h_m^{(k)}\times w_m^{(k)}}$ ($0\leq i\leq h^{(k)}_{\rho}-1$, $0\leq j\leq w^{(k)}_{\rho}-1$) by different $h^{(k)}_{m} \times w^{(k)}_{m}$  LRF-P of $\hat{O}^{(k)}$,  
\begin{equation}\label{eqn:max_pool_2}
    IN^{(k+1)}_{i,j} = 
\sum\limits_{0\leq h\leq h_m^{(k)}-1,
0\leq w\leq w_m^{(k)}-1}M^{(k),i,j}_{\{h,w\}}\cdot \hat{O}^{(k)}_{\{i\cdot s^{(k)}_{\rho}+h,j\cdot s^{(k)}_{\rho}+w\}},
\end{equation}

In the $(i\cdot w^{(k)}_{\rho}+j)$-th LRF-P of $\hat{O}^{(k)}$, let $\Lambda_{i,j}$ be the set of all index pairs that achieve the maximum value:
\begin{align}
\label{eq:maximum_index_Lamda}
\Lambda_{i,j} = \left\{ (h,w) \mid \hat{O}^{(k)}_{\{i\cdot s^{(k)}_{\rho}+h, j\cdot s^{(k)}_{\rho}+w\}} = \max\{\hat{O}^{(k)}_{\Omega_{i,j}}\} \right\}. 
\end{align}
To ensure exactly one neuron is selected from the LRF-P, let $(h^*, w^*) \in \Lambda_{i,j}$ be an arbitrarily chosen index pair. The elements of $M^{(k),i,j}$  are defined as:
\begin{align}
\label{eq:pooling_selection_M}
M^{(k),i,j}_{\{h,w\}}=\begin{cases} 
1, & \text{if } (h,w) = (h^*, w^*), \\ 
0, & \text{otherwise}, 
\end{cases} 
\end{align}
with $0\leq h\leq h_m^{(k)}-1$, $0\leq w\leq w_m^{(k)}-1$. 
Note that any arbitrary choice of $(h^*, w^*) \in \Lambda_{i,j}$ yields a functionally equivalent matrix $M^{(k),i,j}$ that produces the exact same pooling output $IN^{(k+1)}_{i,j}$.

In vector form, the input and output vector of  pooling function $\rho^{(k)}$ are $Z^{(k)} =(\hat{O}^{(k)}_0,\cdots,\hat{O}^{(k)}_{h_o^{(k)}-1})^{\mathsf{T}}\in \mathbb{R}^{d^{(k)}_f=h_o^{(k)}\cdot w_o^{(k)}}$ and $X^{(k+1)} = (IN^{(k+1)}_0,IN^{(k+1)}_1,\cdots, $ $ IN^{(k+1)}_{h_{\rho}^{(k)}-1})^{\mathsf{T}}\in \mathbb{R}^{d^{(k+1)}},~d^{(k+1)}=h^{(k)}_{\rho} \cdot w^{(k)}_{\rho}$. 

\begin{definition}\label{def:pooling_matrix}{\bfseries\upshape (Boolean Pooling Matrix)}    
There exists a $d^{(k+1)}\times d^{(k)}_{f}$ Boolean pooling matrix $P^{(k)}$  so that $\rho^{(k)}:\mathbb{R}^{d^{(k)}_{f}} \rightarrow \mathbb{R}^{d^{(k+1)}}$ satisfying  $X^{(k+1)} = \rho^{(k)}(Z^{(k)}) = P^{(k)}\cdot Z^{(k)}$. Assuming the LRF-Ps are disjoint and square ({\em i.e.}, $s^{(k)}_{\rho}=w_m^{(k)}=h_m^{(k)}$), then the max pooling operation on the $r$-th LRF-P can be defined as the $r$-th row ($0\leq r\leq d^{(k+1)}-1$) of $P^{(k)}$:  
\begin{equation}\label{eqn:P_matrix}
    \resizebox{1.0\hsize}{!}{$%
    P^{(k)}_r = (\underbrace{0, \cdots, 0}_{w^{(k)}_{o}\cdot i \cdot s^{(k)}_{\rho}}, \underbrace{
        \underbrace{0, \cdots, 0}_{j\cdot s^{(k)}_{\rho}} , M^{(k),i,j}_{0},     \underbrace{0, \dots, 0}_{padding}}_{w_{o}^{(k)}}, 
     \cdots, 
        \underbrace{
        \underbrace{0, \cdots, 0}_{j\cdot s^{(k)}_{\rho}}, M^{(k),i,j}_{h_m^{(k)}-1},     \underbrace{0, \dots, 0}_{padding}}_{w_{o}^{(k)}}, 
        \underbrace{0, \cdots, 0}_{padding}),
    $}%
\end{equation}
where $i=\left\lfloor \frac{r}{w^{(k)}_{\rho}} \right\rfloor$, $j=r \bmod w_{\rho}^{(k)}$. 
\end{definition}

Obviously, similar to the matrix of $I^{(k)}$ for $\sigma_c^{(k)}$ function, the pooling matrix $P^{(k)}$ varies with the input $X^{(1)}$ due to Eq. \eqref{eq:pooling_selection_M}. 
For example, the input matrix $\hat{O}^{(k)}$ has dimensions $h_o^{(k)}\times w_o^{(k)} = 4 \times 4$, i.e., $d_{f}^{(k)}=16$, the max pooling kernel is of $h_m^{(k)}\times w_m^{(k)} = 2 \times 2$ with stride $s_{\rho}^{(k)}=2$. The output matrix $IN^{(k+1)}$ will have dimension $h_{\rho}^{(k)}\times w_{\rho}^{(k)} = 2 \times 2$, i.e., $d^{(k+1)}=4$. The $d^{(k+1)}\times d^{(k)}_{f}=4\times 16$  pooling matrix $
    P^{(k)} = 
    \begin{pmatrix}
        &P^{(k)}_{0,0}   &{\bf 0}_{2 \times 8}  \\
        &{\bf 0}_{2 \times 8}  &P^{(k)}_{1,1} 
    \end{pmatrix}
$, where ${\bf 0}_{2\times 8}$ is $2\times8$ zero matrix, and
$$
    P^{(k)}_{0,0} =
    \left(
    \begin{array}{*{8}{c}} 
        M^{(k),0,0}_{0,0} &M^{(k),0,0}_{0,1} &0   &0   &M^{(k),0,0}_{1,0} &M^{(k),0,0}_{1,1} &0   &0  \\
        0   &0   &M^{(k),0,1}_{0,0} &M^{(k),0,1}_{0,1}
        &0   &0   &M^{(k),0,1}_{1,0} &M^{(k),0,1}_{1,1} 
    \end{array}
    \right),
$$
$$
    P^{(k)}_{1,1} =
    \left(
    \begin{array}{*{8}{c}} 
        M^{(k),1,0}_{0,0} &M^{(k),1,0}_{0,1} &0   &0   &M^{(k),1,0}_{1,0} &M^{(k),1,0}_{1,1} &0   &0   \\
        0  &0  &M^{(k),1,1}_{0,0} &M^{(k),1,1}_{0,1} 
        &0  &0  &M^{(k),1,1}_{1,0} &M^{(k),1,1}_{1,1} 
    \end{array}
    \right).
$$

\begin{definition} 
    {\bfseries\upshape (Fully-connected Function)}
    The $k$-th fully-connected function within the fully-connected block of the CNN is a function $f^{(k)}:\mathbb{R}^{d^{(k)}} \rightarrow \mathbb{R}^{d^{(k+1)}}$ given by an affine transformation:
        $$f^{(k)}(X)=A^{(k)}X^{(k)}+B^{(k)},$$ 
    where $X^{(k)}\in\mathbb{R}^{d^{(k)}}$ represents the input vector. The weight matrix $A^{(k)}\in \mathbb{R}^{d^{(k+1)}\times d^{(k)}}$ and the bias vector $B^{(k)}\in \mathbb{R}^{d^{(k+1)}}$ are composed of floating-point numbers.
\end{definition}
The matrix for the nonlinear function $\sigma$ (ReLU functions) of the fully-connected block can be similarly defined by Eq. \eqref{eqn:sigma_I}. 

Given $X^{(1)}=x\in \mathbb{R}^{d^{(1)}}$, the CNN layers and FCN layers of CNN ``collapse'' into an affine transformation. Assume that the function of the $m$ CNN layers is $ \mathcal{F}_{1}$, and the function of the $n$ FCN layers is $\mathcal{F}_{2}$, then

\begin{equation}
  \begin{array}{ll}
y=\mathcal{F}_{1}(x) &= P^{(m)}(I^{(m)}(A^{(m)}  \cdots (P^{(1)}(I^{(1)}(A^{(1)}x+B^{(1)})))\cdots+B^{(m)}))\\
       & = P^{(m)}I^{(m)}A^{(m)}  \cdots P^{(2)}I^{(2)}A^{(2)}P^{(1)}I^{(1)}A^{(1)}x+ \beta_1\\
       & = \Gamma_1 \cdot x+\beta_1
  \end{array}
\end{equation}
According to \cite[Section 3.1]{DBLP:conf/eurocrypt/CanalesMartinezCHRSS24}, given $y$, $\mathcal{F}_{2}$ collapses into 
\begin{equation}\label{eqn:affine_transformation_FCN}
  \begin{array}{ll}
\mathcal{F}_{2}(y) =\Gamma_2 \cdot y+ \beta_2.
  \end{array}
\end{equation}
Consequently, the full CNN ``collapses'' into
\begin{equation}\label{eqn:affine_transformation}
  \begin{array}{ll}
\mathcal{F}(x) =\Gamma_2  (\Gamma_1 \cdot x+\beta_1)+ \beta_2 = \Gamma_2 \cdot \Gamma_1
\cdot  x + \Gamma_2 \beta_1 + \beta_2.
\end{array}
\end{equation}
\begin{definition}{\bfseries\upshape(Linear Neighborhood of CNN)}
    Given an input  $x\in  \mathbb{R}^{d^{(1)}}$ with the corresponding  output of the CNN computed by Eq. \eqref{eqn:affine_transformation}, the linear neighborhood of $x$ is defined as the set 
\begin{equation*}
       \{u\in  \mathbb{R}^{d^{(1)}}| \mathcal{F}(u)=  \Gamma_2 \cdot \Gamma_1
\cdot  u + \Gamma_2 \beta_1 + \beta_2\}, 
\end{equation*}
i.e., the same affine transformation is used for $u$ and $x$ to compute the output of the CNN. 
\end{definition}
If we make a change of $\Delta$ to the input $x$,  and $x+\Delta$ remains within the linear neighborhood of $x$,  we can observe the corresponding change of the output 
\begin{equation*}
    \mathcal{F}(x+\Delta)-\mathcal{F}(x) =  \Gamma_2 \cdot \Gamma_1
\cdot  (x+\Delta) + \Gamma_2 \beta_1 + \beta_2 - (\Gamma_2 \cdot \Gamma_1
\cdot  x + \Gamma_2 \beta_1 + \beta_2) =\Gamma_2 \cdot \Gamma_1
\cdot  \Delta.
\end{equation*}

\begin{definition} \label{def:layer_merging}
    {\bfseries\upshape (Layer Merging)}
    We focus on the convolutional block of the $(m+n)$ network.  
   We assume that we have complete knowledge of the first $k-1$ layers of the convolutional block, and we are currently recovering layer $k$. Let $F_x^{k-1}$ and $G_x^{k+1}$ represent, respectively, the fully recovered and non-recovered parts of the CNN.
        $$\mathcal{F} = \underbrace{f^{(n+1)} \circ \cdots \circ \sigma^{(1)}\circ f^{(1)} \circ 
        \rho^{(m)} \circ \cdots \circ f_c^{(k+1)}}_{G_x^{k+1}}
        \circ \rho^{(k)} \circ \sigma_c^{(k)} \circ f_c^{(k)} \circ 
        \underbrace{\rho^{(k-1)} \circ \cdots \circ f_c^{(1)}}_{F_x^{k-1}}.$$
   Given $X^{(1)} \in  \mathbb{R}^{d^{(1)}}$, $G_x^{k+1}$ and $F_x^{k-1}$ become $G_x^{k+1}(X^{(k+1)})=
    G^{k+1}X^{(k+1)}+B^{k+1}$ and $F_x^{k-1}(X^{(1)})=F^{k-1}X^{(1)}+B^{k-1}$, where the matrices $F^{k-1}\in \mathbb{R}^{d^{(k)}\times d^{(1)}}$, $B^{k-1}\in \mathbb{R}^{d^{(k)}}$, $G^{k+1}\in \mathbb{R}^{d^{(n+2)}\times d^{(k+1)}}$, $B^{k+1}\in \mathbb{R}^{d^{(n+2)}}$,   respectively.
\end{definition}

\subsection{A Toy Example of CNN}\label{sec:toy_example_cnn}

To facilitate a clearer understanding of the algebraic notations introduced in Section \ref{sect:def_notations}, this appendix provides a concrete, step-by-step toy example. We explicitly demonstrate the complete computational pipeline --- including Convolution, ReLU activation, and Max Pooling --- for a single-input and single-output channel scenario ({\em i.e.}, $\mathsf{c}_{in}^{(k)}=\mathsf{c}_{out}^{(k)}=1$). 

\begin{sidewaysfigure}
    \centering
\includegraphics[width=\textwidth]{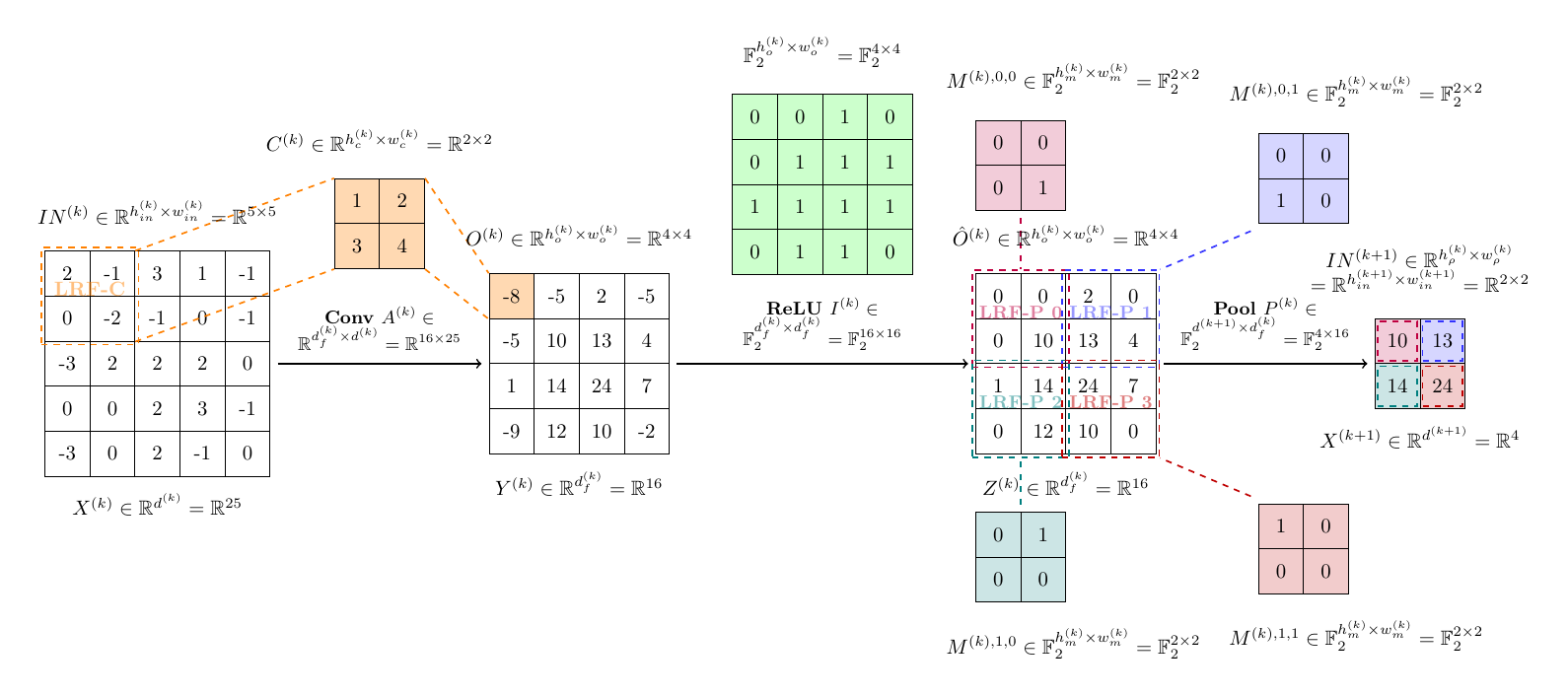} 
    \caption{Overall visual illustration of one CNN layer.}
    \label{fig:toy_example_diagram}
\end{sidewaysfigure}

Figure \ref{fig:toy_example_diagram} visualizes the spatial mapping mechanism of this process. In the following subsections, we formulate this exact process algebraically by completely expanding the equivalent transformation matrices without any truncation.

\subsubsection{Setup and Vector Flattening.}
Consider an input matrix $IN^{(k)} \in \mathbb{R}^{h_{in}^{(k)} \times w_{in}^{(k)}} = \mathbb{R}^{5 \times 5}$. We apply a Convolutional Function $f_c^{(k)}$ with a kernel $C^{(k)} \in \mathbb{R}^{h_c^{(k)} \times w_c^{(k)}} = \mathbb{R}^{2 \times 2}$, a stride of $s_c^{(k)} = 1$, and valid padding, yielding an intermediate output matrix $O^{(k)}$. According to the definitions in Section \ref{sect:def_notations}, the spatial dimensions of $O^{(k)}$ are calculated as follows:
\begin{align}
    h^{(k)}_{o} &= \left\lfloor \frac{h^{(k)}_{in} - h^{(k)}_{c}}{s^{(k)}_c} \right\rfloor + 1 = \left\lfloor \frac{5 - 2}{1} \right\rfloor + 1 = 4, \\
    w^{(k)}_{o} &= \left\lfloor \frac{w^{(k)}_{in} - w^{(k)}_{c}}{s^{(k)}_c} \right\rfloor + 1 = \left\lfloor \frac{5 - 2}{1} \right\rfloor + 1 = 4.
\end{align}
Consequently, the resulting intermediate output matrix is $O^{(k)} \in \mathbb{R}^{4 \times 4}$. Subsequently, a ReLU activation function $\sigma_c^{(k)}$ is applied to $O^{(k)}$, generating an activated output matrix $\hat{O}^{(k)} \in \mathbb{R}^{4 \times 4}$. 

Following the activation, a max pooling function $\rho^{(k)}$ with a window size of $h_m^{(k)} \times w_m^{(k)} = 2 \times 2$ and a stride of $s_\rho^{(k)} = 2$ processes $\hat{O}^{(k)}$. The spatial dimensions of the final output matrix $IN^{(k+1)}$ for the subsequent layer are determined by:
\begin{align}
    h^{(k)}_{\rho} &= \left\lfloor \frac{h^{(k)}_{o} - h^{(k)}_{m}}{s^{(k)}_{\rho}} \right\rfloor + 1 = \left\lfloor \frac{4 - 2}{2} \right\rfloor + 1 = 2, \\
    w^{(k)}_{\rho} &= \left\lfloor \frac{w^{(k)}_{o} - w^{(k)}_{m}}{s^{(k)}_{\rho}} \right\rfloor + 1 = \left\lfloor \frac{4 - 2}{2} \right\rfloor + 1 = 2.
\end{align}
This yields the matrix $IN^{(k+1)} \in \mathbb{R}^{2 \times 2}$.

To represent these operations algebraically, we flatten the spatial matrices into vectors using row-major ordering. For the input matrix $IN^{(k)}$, its $i$-th row ($0 \leq i \leq h_{in}^{(k)}-1$) is denoted as $IN^{(k)}_i=(a^{(k)}_{i,0}, a^{(k)}_{i,1}, \dots, a^{(k)}_{i,w_{in}^{(k)}-1})$. Based on the specific numerical values from Figure \ref{fig:toy_example_diagram}, we instantiate the first and the last row vectors as follows:
\begin{equation}
\begin{aligned}
    IN^{(k)}_0 &= (2, -1, 3, 1, -1), \\
    &\ \ \vdots \\
    IN^{(k)}_4 &= (-3, 0, 2, -1, 0).
\end{aligned}
\end{equation}
The flattened input vector $X^{(k)} \in \mathbb{R}^{d^{(k)}} = \mathbb{R}^{h_{in}^{(k)} \cdot w_{in}^{(k)}}= \mathbb{R}^{25}  $ is constructed by concatenating these row vectors as $X^{(k)} = (IN^{(k)}_0, IN^{(k)}_1, \dots, IN^{(k)}_{4})^{\mathsf{T}}$, which explicitly evaluates to:
\begin{equation}
    \small
    X^{(k)} = [2, -1, 3, 1, -1, 0, -2, -1, 0, -1, -3, 2, 2, 2, 0, 0, 0, 2, 3, -1, -3, 0, 2, -1, 0]^{\mathsf{T}}.
\end{equation}

\subsubsection{The Convolutional Matrix ($A^{(k)}$).}
\begin{figure}
	\centering    \includegraphics[width=0.8\linewidth]{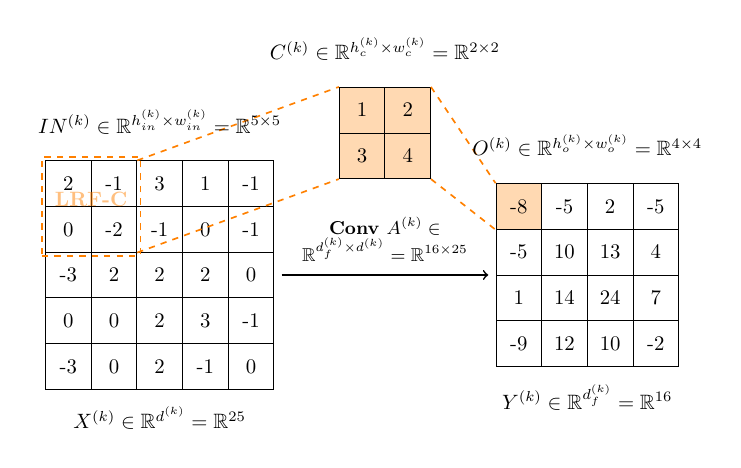}
    \caption{Local view of the convolution operation $f^{(k)}_c$}
	\label{fig:toy_conv}
\end{figure}
Let the $2 \times 2$ convolutional kernel matrix be $C^{(k)} = \left[\begin{smallmatrix} 1 & 2 \\ 3 & 4 \end{smallmatrix}\right]$. For simplicity in this example, we assume the bias is $b^{(k)}=0$, yielding an all-zero bias vector $B^{(k)} = \mathbf{0}$. Therefore, the convolution operation translates into a linear matrix multiplication $Y^{(k)} = A^{(k)} X^{(k)}$, where $Y^{(k)} \in \mathbb{R}^{d_{f}^{(k)}} = \mathbb{R}^{h^{(k)}_{o} \cdot w^{(k)}_{o} } = \mathbb{R}^{16}$ is the flattened output vector and $A^{(k)} \in \mathbb{R}^{d_{f}^{(k)} \times d^{(k)} }  = \mathbb{R}^{16 \times 25}$ is the convolutional matrix. 

As illustrated by the orange dashed lines in Figure \ref{fig:toy_conv}, the first element of the feature map $O^{(k)}_{0,0}$ (which corresponds to $Y^{(k)}_0$) is computed by the sum of the element-wise multiplications between $C^{(k)}$ and the top-left LRF-C. Formally, this is calculated as:
\begin{equation}
\begin{aligned}
    O^{(k)}_{0,0} &= \sum_{p=0}^{h_{c}^{(k)}-1} \sum_{q=0}^{w_{c}^{(k)}-1} IN^{(k)}_{0+p,0+q} C^{(k)}_{p,q} + b^{(k)} \\
    &= IN^{(k)}_{0,0} C^{(k)}_{0,0} + IN^{(k)}_{0,1} C^{(k)}_{0,1} + IN^{(k)}_{1,0} C^{(k)}_{1,0} + IN^{(k)}_{1,1} C^{(k)}_{1,1} + b^{(k)} \\
    &= (2)(1) + (-1)(2) + (0)(3) + (-2)(4) + 0 = -8.
\end{aligned}
\end{equation}

According to Eq. \eqref{eqn:A_row}, the first row of the convolutional matrix ($i=0$) is systematically constructed by substituting the kernel rows into the padded structure. Given  $w^{(k)}_{in}=5$ and $w^{(k)}_o=4$, the substitution and expansion explicitly demonstrate how the kernel parameters are positioned within $A^{(k)}_0$:
\begin{equation}
\begin{aligned}
    A^{(k)}_0 &= (\underbrace{0, \cdots, 0}_{w^{(k)}_{in}\cdot \lfloor 0/w^{(k)}_{o} \rfloor}, \underbrace{
        \underbrace{0, \cdots, 0}_{0 \bmod w^{(k)}_o}, C^{(k)}_{0}, \underbrace{0, \dots, 0}_{padding}}_{w_{in}^{(k)}}, 
        \underbrace{
        \underbrace{0, \cdots, 0}_{0 \bmod w^{(k)}_o}, C^{(k)}_{1}, \underbrace{0, \dots, 0}_{padding}}_{w_{in}^{(k)}}, 
        \underbrace{0, \cdots, 0}_{padding}) \\
    &= (\underbrace{}_{0}, \underbrace{\underbrace{}_{0}, \mathbf{1, 2}, \underbrace{0, 0, 0}_{3}}_{5}, \underbrace{\underbrace{}_{0}, \mathbf{3, 4}, \underbrace{0, 0, 0}_{3}}_{5}, \underbrace{0, \cdots, 0}_{15}) \\
    &= (1, 2, 0, 0, 0, 3, 4, 0, 0, 0, 0, 0, 0, 0, 0, 0, 0, 0, 0, 0, 0, 0, 0, 0, 0).
\end{aligned}
\end{equation}

By representing the exhaustive sliding window operations over the 2D input matrix, the entire equivalent matrix $A^{(k)}$ is expanded below. We highlight the non-zero elements corresponding to the convolutional kernel in each row in bold:

\begin{equation}
\setlength{\arraycolsep}{4pt} 
\resizebox{0.8\textwidth}{!}{$
A^{(k)} = \begin{bmatrix}
\mathbf{1} & \mathbf{2} & 0 & 0 & 0 & \mathbf{3} & \mathbf{4} & 0 & 0 & 0 & 0 & 0 & 0 & 0 & 0 & 0 & 0 & 0 & 0 & 0 & 0 & 0 & 0 & 0 & 0 \\
0 & \mathbf{1} & \mathbf{2} & 0 & 0 & 0 & \mathbf{3} & \mathbf{4} & 0 & 0 & 0 & 0 & 0 & 0 & 0 & 0 & 0 & 0 & 0 & 0 & 0 & 0 & 0 & 0 & 0 \\
0 & 0 & \mathbf{1} & \mathbf{2} & 0 & 0 & 0 & \mathbf{3} & \mathbf{4} & 0 & 0 & 0 & 0 & 0 & 0 & 0 & 0 & 0 & 0 & 0 & 0 & 0 & 0 & 0 & 0 \\
0 & 0 & 0 & \mathbf{1} & \mathbf{2} & 0 & 0 & 0 & \mathbf{3} & \mathbf{4} & 0 & 0 & 0 & 0 & 0 & 0 & 0 & 0 & 0 & 0 & 0 & 0 & 0 & 0 & 0 \\
0 & 0 & 0 & 0 & 0 & \mathbf{1} & \mathbf{2} & 0 & 0 & 0 & \mathbf{3} & \mathbf{4} & 0 & 0 & 0 & 0 & 0 & 0 & 0 & 0 & 0 & 0 & 0 & 0 & 0 \\
0 & 0 & 0 & 0 & 0 & 0 & \mathbf{1} & \mathbf{2} & 0 & 0 & 0 & \mathbf{3} & \mathbf{4} & 0 & 0 & 0 & 0 & 0 & 0 & 0 & 0 & 0 & 0 & 0 & 0 \\
0 & 0 & 0 & 0 & 0 & 0 & 0 & \mathbf{1} & \mathbf{2} & 0 & 0 & 0 & \mathbf{3} & \mathbf{4} & 0 & 0 & 0 & 0 & 0 & 0 & 0 & 0 & 0 & 0 & 0 \\
0 & 0 & 0 & 0 & 0 & 0 & 0 & 0 & \mathbf{1} & \mathbf{2} & 0 & 0 & 0 & \mathbf{3} & \mathbf{4} & 0 & 0 & 0 & 0 & 0 & 0 & 0 & 0 & 0 & 0 \\
0 & 0 & 0 & 0 & 0 & 0 & 0 & 0 & 0 & 0 & \mathbf{1} & \mathbf{2} & 0 & 0 & 0 & \mathbf{3} & \mathbf{4} & 0 & 0 & 0 & 0 & 0 & 0 & 0 & 0 \\
0 & 0 & 0 & 0 & 0 & 0 & 0 & 0 & 0 & 0 & 0 & \mathbf{1} & \mathbf{2} & 0 & 0 & 0 & \mathbf{3} & \mathbf{4} & 0 & 0 & 0 & 0 & 0 & 0 & 0 \\
0 & 0 & 0 & 0 & 0 & 0 & 0 & 0 & 0 & 0 & 0 & 0 & \mathbf{1} & \mathbf{2} & 0 & 0 & 0 & \mathbf{3} & \mathbf{4} & 0 & 0 & 0 & 0 & 0 & 0 \\
0 & 0 & 0 & 0 & 0 & 0 & 0 & 0 & 0 & 0 & 0 & 0 & 0 & \mathbf{1} & \mathbf{2} & 0 & 0 & 0 & \mathbf{3} & \mathbf{4} & 0 & 0 & 0 & 0 & 0 \\
0 & 0 & 0 & 0 & 0 & 0 & 0 & 0 & 0 & 0 & 0 & 0 & 0 & 0 & 0 & \mathbf{1} & \mathbf{2} & 0 & 0 & 0 & \mathbf{3} & \mathbf{4} & 0 & 0 & 0 \\
0 & 0 & 0 & 0 & 0 & 0 & 0 & 0 & 0 & 0 & 0 & 0 & 0 & 0 & 0 & 0 & \mathbf{1} & \mathbf{2} & 0 & 0 & 0 & \mathbf{3} & \mathbf{4} & 0 & 0 \\
0 & 0 & 0 & 0 & 0 & 0 & 0 & 0 & 0 & 0 & 0 & 0 & 0 & 0 & 0 & 0 & 0 & \mathbf{1} & \mathbf{2} & 0 & 0 & 0 & \mathbf{3} & \mathbf{4} & 0 \\
0 & 0 & 0 & 0 & 0 & 0 & 0 & 0 & 0 & 0 & 0 & 0 & 0 & 0 & 0 & 0 & 0 & 0 & \mathbf{1} & \mathbf{2} & 0 & 0 & 0 & \mathbf{3} & \mathbf{4}
\end{bmatrix}
$}
\label{eq:appendix_A}
\end{equation}

Performing the multiplication yields the pre-activation vector $Y^{(k)}$:
\begin{equation}
    \small
    Y^{(k)} = A^{(k)} X^{(k)} = [-8, -5, 2, -5, -5, 10, 13, 4, 1, 14, 24, 7, -9, 12, 10, -2]^{\mathsf{T}}.
\end{equation}

\subsubsection{The ReLU Activation Matrix ($I^{(k)}$)}

\begin{figure}
	\centering    \includegraphics[width=0.85\linewidth]{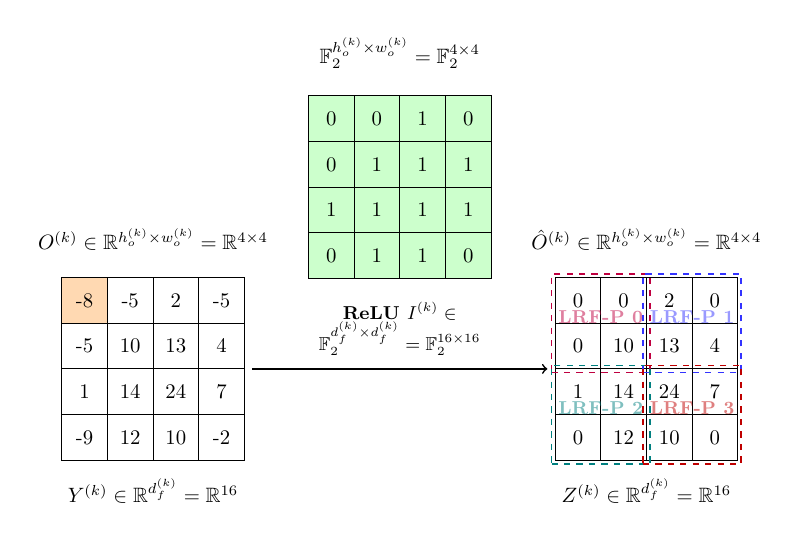}
    \caption{Local view of the ReLU activation $\sigma_c^{(k)}$}
	\label{fig:toy_relu}
\end{figure}

Fig. \ref{fig:toy_relu} shows the operation of ReLU activation.
Algebraically, it is represented as $Z^{(k)} = I^{(k)} Y^{(k)}$, where $I^{(k)} \in \mathbb{F}_2^{d_{f}^{(k)}\times d_{f}^{(k)}}  = \mathbb{F}_2^{16 \times 16}$ is a diagonal matrix. According to Eq. \eqref{eqn:sigma_I}, the $i$-th diagonal entry of $I^{(k)}$ is set to 1 if the corresponding element $Y^{(k)}_i$ is strictly positive, and 0 otherwise. 
Based on the numerical values of $Y^{(k)}$ calculated previously, the fully expanded activation matrix is constructed as follows. We highlight the entire main diagonal in bold:

\begin{equation}
\setlength{\arraycolsep}{4pt}
\resizebox{0.6\textwidth}{!}{$
I^{(k)} = \begin{bmatrix}
\mathbf{0} & 0 & 0 & 0 & 0 & 0 & 0 & 0 & 0 & 0 & 0 & 0 & 0 & 0 & 0 & 0 \\
0 & \mathbf{0} & 0 & 0 & 0 & 0 & 0 & 0 & 0 & 0 & 0 & 0 & 0 & 0 & 0 & 0 \\
0 & 0 & \mathbf{1} & 0 & 0 & 0 & 0 & 0 & 0 & 0 & 0 & 0 & 0 & 0 & 0 & 0 \\
0 & 0 & 0 & \mathbf{0} & 0 & 0 & 0 & 0 & 0 & 0 & 0 & 0 & 0 & 0 & 0 & 0 \\
0 & 0 & 0 & 0 & \mathbf{0} & 0 & 0 & 0 & 0 & 0 & 0 & 0 & 0 & 0 & 0 & 0 \\
0 & 0 & 0 & 0 & 0 & \mathbf{1} & 0 & 0 & 0 & 0 & 0 & 0 & 0 & 0 & 0 & 0 \\
0 & 0 & 0 & 0 & 0 & 0 & \mathbf{1} & 0 & 0 & 0 & 0 & 0 & 0 & 0 & 0 & 0 \\
0 & 0 & 0 & 0 & 0 & 0 & 0 & \mathbf{1} & 0 & 0 & 0 & 0 & 0 & 0 & 0 & 0 \\
0 & 0 & 0 & 0 & 0 & 0 & 0 & 0 & \mathbf{1} & 0 & 0 & 0 & 0 & 0 & 0 & 0 \\
0 & 0 & 0 & 0 & 0 & 0 & 0 & 0 & 0 & \mathbf{1} & 0 & 0 & 0 & 0 & 0 & 0 \\
0 & 0 & 0 & 0 & 0 & 0 & 0 & 0 & 0 & 0 & \mathbf{1} & 0 & 0 & 0 & 0 & 0 \\
0 & 0 & 0 & 0 & 0 & 0 & 0 & 0 & 0 & 0 & 0 & \mathbf{1} & 0 & 0 & 0 & 0 \\
0 & 0 & 0 & 0 & 0 & 0 & 0 & 0 & 0 & 0 & 0 & 0 & \mathbf{0} & 0 & 0 & 0 \\
0 & 0 & 0 & 0 & 0 & 0 & 0 & 0 & 0 & 0 & 0 & 0 & 0 & \mathbf{1} & 0 & 0 \\
0 & 0 & 0 & 0 & 0 & 0 & 0 & 0 & 0 & 0 & 0 & 0 & 0 & 0 & \mathbf{1} & 0 \\
0 & 0 & 0 & 0 & 0 & 0 & 0 & 0 & 0 & 0 & 0 & 0 & 0 & 0 & 0 & \mathbf{0}
\end{bmatrix}
$}
\label{eq:appendix_I}
\end{equation}

Applying this masking matrix to $Y^{(k)}$ zeros out its non-positive components, yielding the activated vector $Z^{(k)}$:
\begin{equation}
    \small
    Z^{(k)} = I^{(k)} Y^{(k)} = [0, 0, 2, 0, 0, 10, 13, 4, 1, 14, 24, 7, 0, 12, 10, 0]^{\mathsf{T}}.
\end{equation}

\subsubsection{The Max Pooling Matrix ($P^{(k)}$)}
In the pooling function $\rho^{(k)}$, the $2 \times 2$ max pooling operation with stride 2 partitions the $4 \times 4$ intermediate feature map $\hat{O}^{(k)}$ into 4 disjoint Local Receptive Fields of Pooling (LRF-Ps). For each LRF-P, the pooling function selects the maximum value. 
\begin{figure}
	\centering    \includegraphics[width=0.8\linewidth]{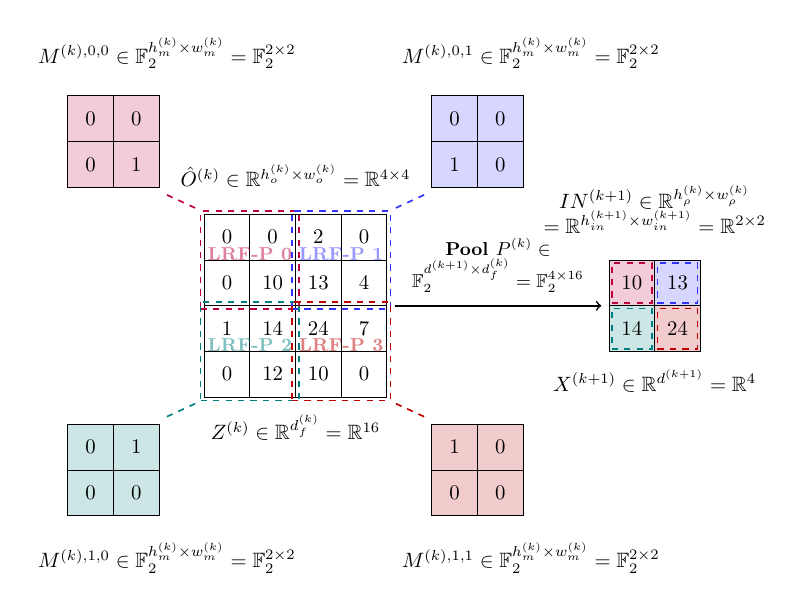}
    \caption{Local view of the max pooling $\rho^{(k)}$}
	\label{fig:toy_pooling}
\end{figure}
Fig. \ref{fig:toy_pooling} illustrates the operation of max pooling. According to Eq. \eqref{eqn:max_pool_1}, the first element of the pooled output is the maximum value within the $(i\cdot w^{(k)}_{\rho}+j)=(0\cdot w^{(k)}_{\rho}+0)=0$-th LRF-P denoted as $\Omega_{0,0}$:
\begin{equation}
    IN^{(k+1)}_{0,0} = \max\{\hat{O}^{(k)}_{\Omega_{0,0}}\} = \max(\hat{O}^{(k)}_{0,0}, \hat{O}^{(k)}_{0,1}, \hat{O}^{(k)}_{1,0}, \hat{O}^{(k)}_{1,1}) = \max(0, 0, 0, 10) = 10.
\end{equation}

Since the maximum value ``$10$'' in $\Omega_{0,0}$ is located at local index $(h^*=1, w^*=1)$, the Boolean matrix $M^{(k),0,0}$ assigns $1$ exclusively to this position, making its row vectors $M^{(k),0,0}_0 = (0, 0)$ and $M^{(k),0,0}_1 = (0, 1)$. According to Eq.  \eqref{eqn:P_matrix}, the first row ($r=0$) of the pooling matrix  
is constructed by mapping these Boolean rows of $M^{(k),0,0}$ into the padded structure. Given $w^{(k)}_o=4$ and $s^{(k)}_\rho=2$, the substitution explicitly demonstrates how the selected bits are positioned within $P^{(k)}_0$:
\begin{equation}
\begin{aligned}
    P^{(k)}_0 &= (\underbrace{0, \cdots, 0}_{w^{(k)}_{o}\cdot 0 \cdot s^{(k)}_{\rho}}, \underbrace{
        \underbrace{0, \cdots, 0}_{0 \cdot s^{(k)}_{\rho}}, M^{(k),0,0}_{0}, \underbrace{0, \dots, 0}_{padding}}_{w_{o}^{(k)}}, 
        \underbrace{
        \underbrace{0, \cdots, 0}_{0 \cdot s^{(k)}_{\rho}}, M^{(k),0,0}_{1}, \underbrace{0, \dots, 0}_{padding}}_{w_{o}^{(k)}}, 
        \underbrace{0, \cdots, 0}_{padding}) \\
    &= (\underbrace{}_{0}, \underbrace{ \underbrace{}_{0}, \mathbf{0, 0}, \underbrace{0, 0}_{2}}_{4}, \underbrace{ \underbrace{}_{0}, \mathbf{0, 1}, \underbrace{0, 0}_{2}}_{4}, \underbrace{0, \cdots, 0}_{8}) \\
    &= (0, 0, 0, 0, 0, 1, 0, 0, 0, 0, 0, 0, 0, 0, 0, 0).
\end{aligned}
\end{equation}

By extending this construction to all four LRF-Ps, the entire equivalent matrix $P^{(k)}$ extracts exactly one element per row. We highlight these selected positions in bold:

\begin{equation}
\resizebox{0.55\textwidth}{!}{$
P^{(k)} = \begin{bmatrix}
0 & 0 & 0 & 0 & 0 & \mathbf{1} & 0 & 0 & 0 & 0 & 0 & 0 & 0 & 0 & 0 & 0 \\
0 & 0 & 0 & 0 & 0 & 0 & \mathbf{1} & 0 & 0 & 0 & 0 & 0 & 0 & 0 & 0 & 0 \\
0 & 0 & 0 & 0 & 0 & 0 & 0 & 0 & 0 & \mathbf{1} & 0 & 0 & 0 & 0 & 0 & 0 \\
0 & 0 & 0 & 0 & 0 & 0 & 0 & 0 & 0 & 0 & \mathbf{1} & 0 & 0 & 0 & 0 & 0
\end{bmatrix}
$}
\label{eq:appendix_P}
\end{equation}

The final output vector for the entire layer is therefore:
\begin{equation}
    X^{(k+1)} = P^{(k)} Z^{(k)} = [10, 13, 14, 24]^{\mathsf{T}}.
\end{equation}

\section{Sign Recovery in Hard-label CNN}
\label{supp:cnn_sign_recovery}

According to \cite[Sect. 4.3 and 5.3]{DBLP:journals/iacr/ChenTGSQD26}, for raw-output CNNs, sign recovery is inherently straightforward. The situation and the recovery procedure remain identical for hard-label CNNs.

\subsubsection{Sign Recovery from Dual RPCPs.}
For Dual RPCPs, after recovering $\hat{A}^{(k)}$ and $\hat{B}^{(k)}$ up to an unknown sign, we compute the input vector of the neurons as
$\hat{A}^{(k)} (F^{k-1} X+B^{k-1})+\hat{B}^{(k)}$, where $X$ is the Dual RPCP of the $t$-th neuron of layer $k$. Assume that neuron $t$ belongs to the $u$-th pooling window $\Omega_u$. According to Prop. \ref{pro:rpcp}, all the inputs of the neurons within  $\Omega_u$ should be $\leq 0$. Then, 
\begin{itemize}
\item if all these inputs are $\leq 0$, the sign of the extracted kernel matrix is correct;
\item if there exists a positive input, the extracted kernel matrix has an inverted sign, and we multiply the matrices $\hat{A}^{(k)}$ and $\hat{B}^{(k)}$ by ``$-1$". 
\end{itemize}

\subsubsection{Sign Recovery from Dual PSPs.}
For Dual PSPs, we reconstruct the convolutional matrix $\hat{A}^{(k)}$, and once we extract $\hat{p}$ and $\hat{q}$, the involved pooling window $\Omega_u$ ($0\leq u < d^{(k+1)}$) corresponding to the PSP could be deduced. According to Prop. \ref{pro:psp}, the outputs of the $p$-th and $q$-th neurons should be the maximum value in $\Omega_u$. Then, with $\hat{A}^{(k)}$, 
\begin{itemize}
\item if $\hat{A}^{(k)}_p(F^{k-1}X+B^{k-1}) = \hat{A}^{(k)}_q(F^{k-1}X+B^{k-1}) > \hat{A}^{(k)}_{\ell}(F^{k-1}X+B^{k-1}), \ \ell\in \Omega_u \setminus \{p, q\}$, the sign of the extracted kernel matrix is correct;
\item if $\hat{A}^{(k)}_p(F^{k-1}X+B^{k-1}) = \hat{A}^{(k)}_q(F^{k-1}X+B^{k-1}) < \hat{A}^{(k)}_{\ell}(F^{k-1}X+B^{k-1}), \ \ell\in \Omega_u \setminus \{p, q\}$, the extracted kernel matrix has an inverted sign, and we multiply the vector $\hat{A}^{(k)}_p$ by ``$-1$".
\end{itemize}


\section{White-box Experiments for verification of ${\sf DGap}(\vec{v},\vec{\omega})$ on 64-64$\times$4-10 FCN and {\em 3072-FCN}.}
\label{supp:DGap_verification_on_FCN}

For both the 64-64$\times$4-10 FCN and the {\em 3072-FCN}, we randomly sample 10,000 dual points. For each point, we compute the ${\sf DGap}$ between its ASV $\vec{v}$ in the corresponding layer and the underlying neuron weight $\vec{w}$.

\begin{table}[htbp]
    \centering

    \setlength{\tabcolsep}{9pt} 
    \caption{Layer-wise statistics of average {\sf DGap} for the 64-64$\times$4-10 FCN and 3072-FCN.}
    \label{tab:dgap_statistics}
    \resizebox{\textwidth}{!}{
    \begin{tabular}{l c c c c}
        \toprule
        \multirow{2}{*}{\textbf{Layer}} & \multicolumn{2}{c}{\textbf{64-64$\times$4-10 FCN}} & \multicolumn{2}{c}{\textbf{3072-FCN}} \\
        \cmidrule(lr){2-3} \cmidrule(lr){4-5}
        & \textbf{Valid Points} & \textbf{Average {\sf DGap}} & \textbf{Valid Points} & \textbf{Average {\sf DGap}} \\
        \midrule
        0 & 3,248 & 0.0161 & 3,582 & 0.0032 \\
        1 & 2,678 & 0.0559 & 2,389 & 0.0083 \\
        2 & 1,498 & 0.0669 & 2,695 & 0.0087 \\
        3 & 1,926 & 0.0610 & 1,181 & 0.0173 \\
        \midrule
        \textbf{Total} & \textbf{9,350} & \textbf{0.0449} & \textbf{9,847} & \textbf{0.0076} \\
        \bottomrule
    \end{tabular}
    }
\end{table}

\section{Proof of Proposition \ref{pro:expectation_dgap} 
in Unitary Balanced Networks}
\label{supp:decoupled_error_proof}

This appendix provides a proof of Proposition~\ref{pro:expectation_dgap} in unitary balanced networks, where each neuron weight is a randomly chosen unit vector in $d^{(k)}$-dimensional space. 
This assumption follows the back-of-the-envelope analysis in \cite[Appendix A]{DBLP:conf/eurocrypt/CanalesMartinezCHRSS24}. 

To compute the cosine distance
${\sf DGap}(\vec{v},\vec{\omega})= 1-\lvert \cos \varphi_{\vec{v},\vec{\omega}} \rvert= 1-\frac{\lvert \vec{v}\cdot \vec{\omega}^{\sf T}\rvert}{\|\vec{v}\|\cdot \|\vec{\omega}^{\sf T}\|}$,
  we first  compute 
\begin{align}\small
    &\vec{v} \cdot \vec{\omega}^{\sf T} 
    = \frac{1}{\|\vec{G}_{\ell}\|} \left( \vec{\omega} - (\vec{\omega} \cdot \vec{n}_r^{\sf T})\vec{n}_r \right) \cdot \vec{\omega}^{\sf T} 
    = \frac{1}{\|\vec{G}_{\ell}\|} \left( \|\vec{\omega}\|^2 - (\vec{\omega} \cdot \vec{n}_r^{\sf T})^2 \right), \\
    &\|\vec{v}\| = \sqrt{ \vec{v} \cdot \vec{v}^{\sf T} } 
    = \sqrt{ \frac{1}{\|\vec{G}_{\ell}\|^2} \left( \|\vec{\omega}\|^2 - (\vec{\omega} \cdot \vec{n}_r^{\sf T})^2 \right)}
    = \frac{1}{\|\vec{G}_{\ell}\|} \sqrt{ \|\vec{\omega}\|^2 - (\vec{\omega} \cdot \vec{n}_r^{\sf T})^2 }.
\end{align}
Then, 
\begin{equation} \label{eq:reduced_cos}
    \cos \varphi_{\vec{v},\vec{\omega}} = \frac{ \|\vec{\omega}\|^2 - (\vec{\omega} \cdot \vec{n}_r^{\sf T})^2 }{ \sqrt{ \|\vec{\omega}\|^2 - (\vec{\omega} \cdot \vec{n}_r^{\sf T})^2 } \cdot \|\vec{\omega}\| } 
    =\sqrt{ 1 - \left( \frac{\vec{\omega} \cdot \vec{n}_r^{\sf T}}{\|\vec{\omega}\|} \right)^2 }.
\end{equation}
According to Eq.~\eqref{eq:G_vectors}, we have 
   $ \vec{n}_r 
   = \frac{\vec{G}_r}{\|\vec{G}_r\|} =\frac{1}{\|\vec{G}_r\|}\sum_{s \in \Lambda^{(k)}}g_sA^{(k)}_s. $
Hence, 
\begin{equation}\small
    \label{eq:w  dot n}
    \vec{\omega} \cdot \vec{n}_r^{\sf T} = g_t A_t^{(k)} \cdot \frac{1}{\|\vec{G}_r\|}\left( \sum_{s \in \Lambda^{(k)}} g_s A_s^{(k)} \right) ^{\sf T}
    = \frac{g_t}{\|\vec{G}_r\|} \sum_{s \in \Lambda^{(k)}} g_s \left(  A_t^{(k)} \cdot  {A_s^{(k)}}^{\sf T} \right).
\end{equation}
Let $\varphi_{t,s}$ denote the angle between $A^{(k)}_t$ and $A^{(k)}_s$, {\em i.e.},  $A_t^{(k)} \cdot  {A_s^{(k)}}^{\sf T}=\|A_t^{(k)}\|\cdot\|A_s^{(k)}\|\cos\varphi_{t,s}$, then $\frac{\vec{\omega} \cdot \vec{n}_r^{\sf T}}{\|\vec{\omega}\|}$ can be rewritten as:
\begin{equation} \small
\label{eq:decoupled_error}
    \frac{\vec{\omega} \cdot \vec{n}_r^{\sf T}}{\|\vec{\omega}\|} = \frac{g_t \lVert A_t^{(k)}\rVert}{\|\vec{\omega}\|}  \frac{\sum_{s \in \Lambda^{(k)}} g_s \lVert A_s^{(k)}\rVert \cos \varphi_{t,s}}{\|
    \vec{G}_r\|}
    =
    \frac{\sum_{s \in \Lambda^{(k)}} g_s \lVert A_s^{(k)}\rVert \cos \varphi_{t,s}}
    {\lVert \sum_{s \in \Lambda^{(k)}} g_s A_s^{(k)} \rVert}.
\end{equation}

Note that in unitary balanced networks, each weight (row vectors) satisfies $\|A_s^{(k)}\|=1$. We have:
\begin{equation}
\mathbb{E}[A_t^{(k)}] = \vec{0}, \qquad 
\mathbb{E}[{A_t^{(k)}}^{\sf T}A_t^{(k)}] = \frac{I_{d^{(k)}}}{d^{(k)}},
\end{equation}
where $I_{d^{(k)}}$ is the $d^{(k)}\times d^{(k)}$ identity matrix. We can explicitly verify these moments component by component:
\begin{itemize}
    \item \textbf{First Moment:} For the target weight $A_t^{(k)}$, denote its coordinates as $(a_1, a_2, \dots, a_{d^{(k)}})$. For any $a_i$, the spherical reflection symmetry implies that $a_i$ and $-a_i$ have the same distribution. Thus, $\mathbb{E}[a_i] = 0$ for all $i$, yielding $\mathbb{E}[A_t^{(k)}] = \vec{0}$.
    \item \textbf{Second Moment:} The outer product matrix ${A_t^{(k)}}^{\sf T}A_t^{(k)}$ has entries $a_i a_j$. For the off-diagonal entries ($i \neq j$), the reflection symmetry across coordinate planes ensures that $\mathbb{E}[a_i a_j] = 0$. For the diagonal entries ($i = j$), we consider the deterministic unit length constraint:
    \begin{equation*}
        \sum_{i=1}^{d^{(k)}} a_i^2 = 1 \implies \sum_{i=1}^{d^{(k)}} \mathbb{E}[a_i^2] = 1.
    \end{equation*}
    Because the distribution is isotropic, all coordinate axes are statistically equivalent, meaning $\mathbb{E}[a_1^2] = \mathbb{E}[x_a^2] = \dots = \mathbb{E}[a_{d^{(k)}}^2]$. Consequently, each diagonal entry evaluates to $\mathbb{E}[a_i^2] = 1/d^{(k)}$.
    Combining the zero off-diagonal elements and the $1/d^{(k)}$ diagonal elements directly yields the covariance matrix $I_{d^{(k)}}/d^{(k)}$. 
    
\end{itemize}

\subsubsection{Evaluation on the Error Term.}
According to Eq.~\eqref{eq:decoupled_error}, the error term $\frac{\vec{\omega} \cdot \vec{n}_r^{\sf T}}{\|\vec{\omega}\|}$ can be written as:
\begin{equation*}
    \frac{\vec{\omega} \cdot \vec{n}_r^{\sf T}}{\|\vec{\omega}\|}
    =
    \frac{\sum_{s \in \Lambda^{(k)}} g_s \lVert A_s^{(k)}\rVert \cos \varphi_{t,s}}
    {\lVert \sum_{s \in \Lambda^{(k)}} g_s A_s^{(k)} \rVert}.
\end{equation*} 
We evaluate its expectation by first conditioning on all non-target weights $A_s^{(k)}$ ($s \neq t$). The target weight $A_t^{(k)}$ is statistically independent of other weight vectors $A_s^{(k)}$ for $s \in \Lambda^{(k)}$. Since $\cos\varphi_{t,s} = A_t^{(k)} \cdot {A_s^{(k)}}^{\sf T}$, we have the conditional expectation:
\begin{equation}
\mathbb{E}_{A_t^{(k)}}[\cos\varphi_{t,s}] 
= \mathbb{E}[A_t^{(k)}] \cdot {A_s^{(k)}}^{\sf T} = 0.
\end{equation}
Note that $t \notin \Lambda^{(k)}$. 
We adopt a local idealization in which $\lVert \sum_{s \in \Lambda^{(k)}} g_s A_s^{(k)} \rVert$ and $g_s$ are deterministic. The conditional expectation of the entire ratio vanishes:
\begin{equation}
\mathbb{E}_{A_t^{(k)}}\!\left[
\frac{\sum_{s\in\Lambda^{(k)}}g_s\cos\varphi_{t,s}}
{\left\|\sum_{s\in\Lambda^{(k)}}g_sA_s^{(k)}\right\|}
\right] = 0.
\end{equation}

For any $s,q \in \Lambda^{(k)}$ (including $s=q$), the expectation of the product of two cosine terms leverages the covariance matrix of $A_t^{(k)}$:
\begin{equation}
\begin{aligned}
\mathbb{E}_{A_t^{(k)}}[\cos\varphi_{t,s}\cos\varphi_{t,q}]
= A_s^{(k)} \mathbb{E}[{A_t^{(k)}}^{\sf T}A_t^{(k)}] {A_q^{(k)}}^{\sf T}
= \frac{A_s^{(k)}\cdot{A_q^{(k)}}^{\sf T}}{d^{(k)}}.
\end{aligned}
\end{equation}
The numerator $\left(\sum g_s\cos\varphi_{t,s}\right)^2$ naturally produces a double sum over all pairs $(s, q)$, including the cross-terms ($s \neq q$). We apply the covariance identity derived above to every pair in the double sum. This substitution transforms the expectation of the numerator into a sum of inner products $\left(A_s^{(k)}\cdot{A_q^{(k)}}^{\sf T}\right)$, perfectly mirroring the algebraic expansion of the denominator's squared norm. Explicitly, evaluating the expectation of the squared ratio yields:
\begin{equation}  \small
\begin{aligned}
\mathbb{E}_{A_t^{(k)}}\!\left[
\left(
\frac{\sum_{s\in\Lambda^{(k)}}g_s\cos\varphi_{t,s}}
{\left\|\sum_{s\in\Lambda^{(k)}}g_sA_s^{(k)}\right\|}
\right)^2
\right] 
&=
\frac{
\displaystyle
\sum_{s,q\in\Lambda^{(k)}}g_sg_q\,
\mathbb{E}_{A_t^{(k)}}[\cos\varphi_{t,s}\cos\varphi_{t,q}]
}{
\displaystyle
\left\|\sum_{s\in\Lambda^{(k)}}g_sA_s^{(k)}\right\|^2
}\\
&=
\frac{1}{d^{(k)}}
\frac{
\displaystyle
\sum_{s,q\in\Lambda^{(k)}}g_sg_q
\left(A_s^{(k)}\cdot{A_q^{(k)}}^{\sf T}\right)
}{
\displaystyle
\left\|\sum_{s\in\Lambda^{(k)}}g_sA_s^{(k)}\right\|^2
}.
\end{aligned}
\end{equation}
The double summation in the numerator is precisely the expanded form of the squared norm in the denominator:
\begin{equation}
\sum_{s,q\in\Lambda^{(k)}}g_sg_q
\left(A_s^{(k)}\cdot{A_q^{(k)}}^{\sf T}\right)
=
\left\|\sum_{s\in\Lambda^{(k)}}g_sA_s^{(k)}\right\|^2.
\end{equation}
Consequently, the terms perfectly cancel, evaluating to exactly $1/d^{(k)}$. 
Applying the definitions of $\vec{\omega}$ and $\vec{n}_r$, we conclude:
\begin{equation}
\mathbb{E}\!\left[
\left(
\frac{\vec{\omega}\cdot\vec{n}_r^{\sf T}}
{\|\vec{\omega}\|}
\right)^2
\right]
=\frac{1}{d^{(k)}}.
\end{equation}


Finally, we map this projection error back to the Directional Gap ({\sf DGap}). According to Eq.~\eqref{eq:direction_gap} and Eq.~\eqref{eq:reduced_cos}, the exact relationship between {\sf DGap} and the normalized projection term is given by:
\begin{equation}
{\sf DGap}(\vec v,\vec\omega)
=
1-\sqrt{1-\left(
\frac{\vec{\omega}\cdot\vec{n}_r^{\sf T}}
{\|\vec{\omega}\|}
\right)^2}.
\end{equation}
Since $1-\sqrt{1-z} \leq z$ for $z \in [0, 1]$, we can strictly bound the directional gap by the squared error term:
\begin{equation}
{\sf DGap}(\vec v,\vec\omega)
\leq
\left(
\frac{\vec{\omega}\cdot\vec{n}_r^{\sf T}}
{\|\vec{\omega}\|}
\right)^2.
\end{equation}
Taking the expectation of both sides and substituting our derived second moment yields a rigorous upper bound on the average directional error:
\begin{equation}\label{eq:expectation_dgap}
\mathbb{E}\bigl[{\sf DGap}(\vec{v},\vec{\omega})\bigr] 
\le 
\mathbb{E}\!\left[
\left(
\frac{\vec{\omega}\cdot\vec{n}_r^{\sf T}}
{\|\vec{\omega}\|}
\right)^2
\right] 
= \frac{1}{d^{(k)}}.
\end{equation}
This formally demonstrates that the expected directional gap gracefully converges to zero as the layer width increases:
\begin{equation*}
\lim_{d^{(k)}\to\infty} \mathbb{E}\bigl[{\sf DGap}(\vec{v},\vec{\omega})\bigr] = 0.
\end{equation*}

\section{Detailed Model Structure Information in Experiments}
\label{supp:model structure}
\paragraph{(2+1)-Deep CNN:}
\begin{enumerate}
    \item CNN layer 1 ($C_1$): 
    \begin{itemize}
        \item \textbf{Convolutional function} $f_c^{(1)}$: $32 \times 32$ Input matrix $IN^{(1)}$, convolved with $(1,1,5,5)$ convolution kernel, with stride $s_c^{(1)}=1$, padding $pad^{(1)}=0$, produces $28 \times 28$ output matrix $O^{(1)}$, followed by ReLU activation.
        \item \textbf{Max Pooling function} $\rho^{(1)}$: 
        $2 \times 2$ max pooling kernel, with stride $s_{\rho}^{(1)}=2$, produces $14 \times 14$ output matrix $IN^{(2)}$.
    \end{itemize}
    
    \item CNN layer 2 ($C_2$):
    \begin{itemize}
        \item \textbf{Convolutional function} $f_c^{(2)}$: $14 \times 14$ Input matrix $IN^{(2)}$, convolved with $(1,1,5,5)$ convolution kernel, with stride $s_c^{(2)}=1$, padding $pad^{(2)}=0$, produces $10 \times 10$ output matrix $O^{(2)}$, followed by ReLU activation.
        \item \textbf{Max Pooling function} $\rho^{(2)}$: 
        $2 \times 2$ max pooling kernel, with stride $s_{\rho}^{(2)}=2$, produces $5 \times 5$ output matrix $IN^{(3)}$.
    \end{itemize}

    \item FCN layer 1 ($F_1$): 
    \begin{itemize}
        \item Flatten the matrix $IN^{(3)}$ and get the 25-dimensional vector $X^{(1)}$.
        \item \textbf{Fully-connected function} $f^{(1)}$: $d^{(1)}=25$ input neurons and $d^{(2)}=18$ output neurons, followed by ReLU activation.
    \end{itemize}
    
    \item Last layer $f^{(2)}$: $d^{(2)}=18$ input neurons and $d^{(3)}=10$ output neurons.
\end{enumerate}

\paragraph{(2+2)-Deep LeNet-5:}
\begin{enumerate}
    \item CNN layer 1 ($C_1$):
    \begin{itemize}
        \item \textbf{Convolutional function} $f_c^{(1)}$: $32 \times 32$ Input matrix $IN^{(1)}$, convolved with $(6,1,5,5)$ convolution kernel, with stride $s_c^{(1)}=1$, padding $pad^{(1)}=0$, produces 6 output matrices  $O^{(1)}$ of size $28 \times 28$ , followed by ReLU activation.
        \item \textbf{Max Pooling function} $\rho^{(1)}$: 
        $2 \times 2$ max pooling kernel, with stride $s_{\rho}^{(1)}=2$, produces 6 output matrices $IN^{(2)}$ of size $14 \times 14$ .
    \end{itemize}
    
    \item CNN layer 2 ($C_2$):
    \begin{itemize}
        \item \textbf{Convolutional function} $f_c^{(2)}$: 6 Input matrix $IN^{(2)}$ of size $14 \times 14$, convolved with $(6,16,5,5)$ convolution kernel, with stride $s_c^{(2)}=1$, padding $pad^{(2)}=0$, produces 16 output matrix $O^{(2)}$ of size $10 \times 10$, followed by ReLU activation.
        \item \textbf{Max Pooling function} $\rho^{(2)}$: 
        $2 \times 2$ max pooling kernel, with stride $s_{\rho}^{(2)}=2$, produces 16 output matrix $IN^{(3)}$ of size $5 \times 5$.
    \end{itemize}

    \item FCN layer 1 ($F_1$): 
    \begin{itemize}
        \item Flatten the matrix $IN^{(3)}$ and get the 400-dimensional vector $X^{(1)}$.
        \item \textbf{Fully-connected function} $f^{(1)}$: $d^{(1)}=400$ input neurons and $d^{(2)}=120$ output neurons, followed by ReLU activation.
    \end{itemize}

    \item FCN layer 2 ($F_2$): 
    \begin{itemize}
        \item \textbf{Fully-connected function} $f^{(2)}$: $d^{(2)}=120$ input neurons and $d^{(3)}=84$ output neurons, followed by ReLU activation.
    \end{itemize}
    
    \item Last layer $f^{(3)}$: $d^{(3)}=84$ input neurons and $d^{(3)}=10$ output neurons.
\end{enumerate}

\section{The Experiments on the {\sf DGap} among the ASV and the weight for CNN}
\label{supp:experiment_asv_cnn}

\subsection{White-box Experiment for Verification of ${\sf DGap}(\vec{v}, \vec{\omega})\rightarrow 0$ in Hard-Label CNN} \label{sect:experiment_cnn_asv_w}
\begin{figure}[htbp]
    \centering
    \includegraphics[width=0.8\textwidth]{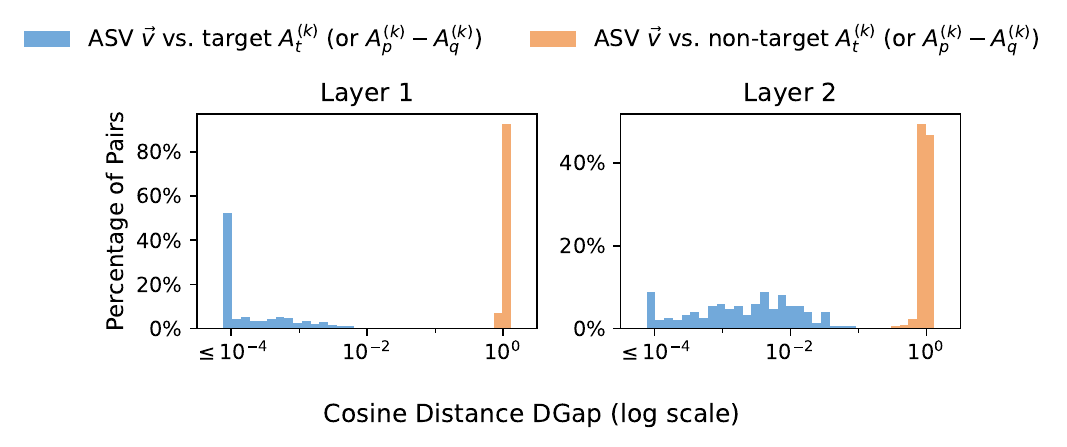}
    \caption{Distribution of the Cosine Distance (${\sf DGap}$) between ASV $\vec{v}$ and all possible pairs of row vectors across CNN layers}
    \label{fig:combined_cosine_dists_histogram_cnn}
\end{figure}

We experiment on a $(2+1)$-deep CNN with detail configurations given in {\sf Supp.}  \ref{supp:model structure}.  
We randomly sample 1000 valid dual points across the input space and compute their corresponding ASVs $\vec{v}$. For each $\vec{v}$, we then calculate the ${\sf DGap}$ with respect to all row vectors $A^{(k)}_t$, as well as all possible pairwise differences $A^{(k)}_p-A^{(k)}_q$ (neurons $p$ and $q$ are in the same pooling window) of the convolutional matrix $A^{(k)}$.
As shown in Fig. \ref{fig:combined_cosine_dists_histogram_cnn}, the distributions  exhibit a clear separation: the directional distances between ASVs and their corresponding target row vectors $A^{(k)}_t$ (or $A^{(k)}_p-A^{(k)}_q$) tightly cluster near 0, while the distances to all other non-target row vectors consistently concentrate around $1$.

\subsection{Experimental Verification of ASV-based Consistency Check} \label{sect:experiment_cnn_two_asv}
\begin{figure}[htbp]
    \centering
    \includegraphics[width=0.8\textwidth]{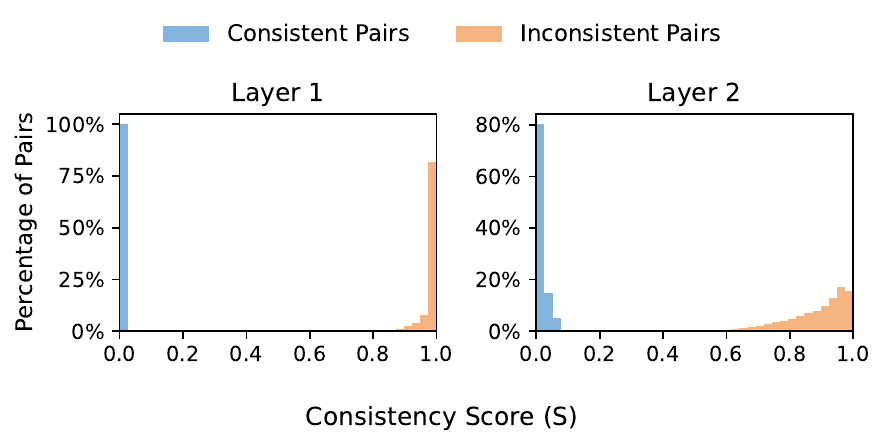}
    \caption{Distribution of Consistency Scores for pairs sharing the same target vector versus those from different target vectors across CNN layers.}
\label{fig:combined_consistency_distribution_cnn}
\end{figure}

Targeting a $(2+1)$ CNN, we evaluate the consistency scores across pairs of dual points. As shown in Fig. \ref{fig:combined_consistency_distribution_cnn}, dual points from the \emph{same} target vector $A^{(k)}_t$ (or $A^{(k)}_p-A^{(k)}_q$) yield a significantly different distribution compared to pairs from \emph{different} targets. This separation in distribution demonstrates that the ASV-based consistency score provides a reliable criterion for clustering dual points.

\section{Experiment Verification of Sparsity Alignment between $\vec{v}$ and $\vec{\omega}$ in CNNs} \label{sect:cnn_v_w_test} 
Targeting a $(2+1)$ CNN, we collect a set of $1000$ dual points and compute their ASVs~$\vec{v}$. 
If $\vec{v}$ can serve as a reliable proxy for the ground-truth vectors $\vec{\omega}$ in identifying Dual RPCPs and Dual PSPs, the positions where $\vec{\omega}$ is exactly $0$ should also remain $0$ in $\vec{v}$. This sparsity alignment guarantees that $\vec{v}$ preserves the structural identity of either the neuron weight vector ($A^{(k)}_t$) or the difference vector ($A^{(k)}_p - A^{(k)}_q$).
To quantify this zero-position fidelity, 
we extract the absolute values of our ASVs $\vec{v}$ at the indices where   $\vec{\omega}$ are $0$.
Our statistical analysis across all 1000 dual points evaluates the distribution of these theoretically inactive entries. As summarized in Table \ref{tab:theoretical zero-positions}, the element-wise noise is small enough to recognize the sparse structure across both convolutional layers.

\begin{table}
\centering
\setlength{\tabcolsep}{9pt}
\caption{Statistics of element-wise noise at the theoretical zero-positions.}
\label{tab:theoretical zero-positions}
\begin{tabular}{cccccc}
\toprule
Layer & \makecell{Theoretical Value} &Valid Duals & $\max$ & $\min$ & $\text{mean}$\\
\midrule
\multirow{2}{*}{$C_1$} & zero & \multirow{2}{*}{$807$} & $2^{-4.28}$& $2^{-47.97}$ & $2^{-11.05}$\\
& non-zero & & $2^{-1.01}$& $2^{-18.99}$ & $2^{-2.80}$\\
\midrule
\multirow{2}{*}{$C_2$} &  zero & \multirow{2}{*}{$140$}  & $2^{-3.36}$ & $2^{-44.25}$ & $2^{-7.81}$\\
& non-zero &  & $2^{-0.53}$ & $2^{-13.35}$ & $2^{-2.58}$\\

\bottomrule
\end{tabular}
\end{table}

\end{document}